\documentclass[
amsmath,amssymb,
aps, nofootinbib
rmp, floatfix, reprint,
superscriptaddress,
groupedaddress,
]{revtex4-2}
\usepackage{xcolor}
\usepackage{graphicx}
\usepackage{dcolumn}
\usepackage{bm}

\usepackage[toc,page,header]{appendix}
\makeatletter
\newcommand\appendix@section[1]{%
  \refstepcounter{section}%
  \orig@section*{Appendix \@Alph\c@section: #1}%
}
\let\orig@section\section
\g@addto@macro\appendix{\let\section\appendix@section}
\makeatother

\usepackage{amsmath,amssymb}
\usepackage{slashed}
\usepackage{graphicx}
\usepackage{dsfont}
\usepackage{bm}
\usepackage[top=1.0in,bottom=1.0in,left=1.0in,right=1.0in]{geometry}
\usepackage[colorlinks,linkcolor=blue,citecolor=blue]{hyperref}
\usepackage{CJKutf8}
\usepackage{tikz-feynhand}
\usepackage[caption=false]{subfig}
\usepackage{float}

\begin{document}
\setlength{\oddsidemargin}{0.5cm}
\setlength{\topmargin}{-0.1cm}
\setlength{\textheight}{21cm}
\setlength{\textwidth}{15cm}
\newcommand{\be}{\begin{equation}}
\newcommand{\ee}{\end{equation}}
\newcommand{\bea}{\begin{eqnarray}}
\newcommand{\eea}{\end{eqnarray}}

\newcommand{\Tr}{\mbox{Tr}\;}
\newcommand{\tr}{\mbox{tr}\;}
\newcommand{\ket}[1]{\left|#1\right\rangle}
\newcommand{\bra}[1]{\left\langle#1\right|}

\newcommand{\avg}[1]{\left\langle #1\right\rangle}
\newcommand{\vnabla}{\mathbf{\nabla}}
\newcommand{\notes}[1]{\fbox{\parbox{\columnwidth}{#1}}}


\title{Nucleon Wave Functions from QCD Vacuum Topology}

\author{Wei-Yang Liu}
\email{wyl@jlab.org}
\affiliation{Center for Nuclear Theory, Department of Physics and Astronomy, Stony Brook University, Stony Brook, New York 11794-3800, USA}
\affiliation{Theory Center, Jefferson Lab, Newport News, VA 23606, USA}

\author{Ismail Zahed }
\email{ismail.zahed@stonybrook.edu}
\affiliation{Center for Nuclear Theory, Department of Physics and Astronomy, Stony Brook University, Stony Brook, New York 11794-3800, USA}

\date{\today}


\begin{abstract}
Instantons, as topologically nontrivial gauge field configurations,
provide a microscopic mechanism for chiral symmetry breaking
and the $U_A(1)$ anomaly in QCD. Within the instanton liquid
model (ILM), instantons generate multi-fermion interactions governed
by their density and size. Using this framework, we
develop a microscopic description of the nucleon based on
instanton-induced quark--quark correlations. We derive the
diquark interaction kernels arising from single instantons
and instanton--anti-instanton molecules, revealing a strong
attraction in the scalar channel and
weaker correlations in the spin-one channels. Embedding
these interactions in a covariant, permutation-symmetric
Bethe--Salpeter--Faddeev equation yields a nucleon mass
$m_N=939.53~{\rm MeV}$, with a predominantly scalar-diquark
component ($98\%$) and a small axial-vector admixture ($2\%$) around the mass pole.
Despite its small weight, the small admixture generates substantial
spin--isospin interference, yielding an axial charge
$g_A=1.265$, close to experiment. The resulting covariant
amplitude is analytically continued to the light front to obtain the
leading three-quark wave function. The corresponding unpolarized, helicity, and transversity parton distributions, along with their QCD-evolved moments, provide a direct connection to lattice QCD calculations and global-fit analyses.
Our results show that the topology of the QCD vacuum
organizes not only the nucleon mass and spin structure,
but also its leading valence partonic content, providing
a microscopic foundation for a more complete description
including confinement and higher Fock components.
\end{abstract}

\maketitle

\section{Introduction}

Understanding the structure of hadrons directly from Quantum Chromodynamics (QCD) remains one of the central goals of modern nuclear and particle theory. While QCD is asymptotically free at large momentum transfer, enabling perturbative treatments of hard processes, its infrared behavior is governed by strong coupling where perturbation theory breaks down. The emergence of confinement, the pattern of chiral symmetry breaking, and the dynamical origin of hadron masses are deeply tied to the nonperturbative structure of the QCD vacuum. In this regime, semiclassical field configurations play a crucial role by providing a bridge between the microscopic gluon dynamics and the effective degrees of freedom describing low-lying hadrons.

Among the various nonperturbative mechanisms proposed to explain infrared QCD phenomena, instantons play a particularly central role as topologically nontrivial tunneling configurations of the gauge field. Ever since the observation that instantons induce chirality-changing fermionic zero modes and generate the well-known 't Hooft multi-fermion interaction, these pseudoparticles have offered an intuitively appealing picture of chiral symmetry breaking, the $U_{A}(1)$ anomaly, and the mass splitting in the pseudoscalar meson sector. When embedded into an ensemble of instantons and anti-instantons, this mechanism forms the basis of the instanton liquid model (ILM), which captures the dominant infrared degrees of freedom of the gluonic vacuum. Lattice gauge theory, particularly with the aid of gradient flow, now provides strong evidence that cooled gauge configurations reveal a dilute but strongly correlated ensemble of instantons and anti-instantons, supporting the ILM picture \cite{Leinweber:1999cw,Biddle:2019gke,Athenodorou:2018jwu}. 

The same gradient flow techniques using the P-vortex projection, have also shown a direct correlation between the location of the emergent branching points, sources of monopoles and anti-monopoles and the location of the instantons and anti-instantons. In retrospect, this is not surprising,  given the geometrical origin of both configurations \cite{Reinhardt:2002cm}, and their intimate relationship at finite temperature~\cite{Kraan:1998pm,Bruckmann:2009pa}. What is now emerging is the intimate relationship between chiral symmetry and confinement, where the instantons and anti-instantons are dominant in the former and in particular in the low-lying hadrons (relatively tight worldlines not prone to flux piercing), and where monopoles and anti-monopoles through center vortices are dominant for the latter, especially for the high-lying hadrons (relatively broad worldlines prone to flux piercing). The two descriptions are likely dual in the vacuum,  extending the initial observation in the thermal state~\cite{Ramamurti:2018evz}.

The physics motivation for revisiting the instanton vacuum is multifaceted.  
First, instantons naturally generate the spontaneous breaking of chiral symmetry through delocalized fermion zero modes, which serve as precursors for the formation of the constituent quark mass and the quark condensate.  
Second, instantons generate effective quark--quark interactions in specific chiral and color channels, providing a microscopic derivation of the attractive scalar diquark channel and the corresponding hierarchy of diquark correlations.  
Third, for baryons, instanton-induced interactions support a quark--diquark clustering, yet remain fully compatible with a three-body Faddeev structure. Remarkably, the strength and channel dependence of these interactions are not free parameters but determined directly by the vacuum instanton density and size distribution.

This paper is devoted to developing a unified and intuitive framework for light hadron physics directly rooted in the topological structure of the QCD vacuum. In the previous work \cite{Liu:2023yuj,Liu:2023fpj,Kock:2020frx} we established how the instanton background fully determines the isospin structure, mass scales, and couplings of light mesons. Here we extend the framework to the baryon sector. Our goals are:

\begin{itemize}
    \item to derive the instanton-induced diquark interactions for arbitrary color and flavor representations, including both single-instanton and instanton--anti-instanton molecular contributions;
    \item to determine the hierarchy and binding strengths of the scalar, pseudoscalar, vector, axial-vector, and tensor diquark channels within the ILM;
    \item to construct a fully covariant, permutation-symmetric baryon Bethe--Salpeter--Faddeev (BSF) equation rooted in these instanton-induced kernels;
    \item to determine the nucleon mass and covariant Dirac--flavor structure, and analytically continue the resulting three-quark amplitude to the light front;
    \item to calculate the axial and tensor charges and the unpolarized, helicity, and transversity parton distributions, including their perturbative evolution and comparison with lattice and phenomenological analyses.
\end{itemize}

The ILM produces a remarkably economical description: the entire baryon sector is controlled by two vacuum parameters, the instanton density and size, which also determine the constituent quark mass and quark condensate \cite{Diakonov:1985eg,Pobylitsa1989TheQP}. This provides a direct, parameter-sparse link between the microscopic gauge vacuum and the emergent low-energy hadronic world. In particular, the scalar color-antisymmetric diquark emerges as the dominant correlated pair inside the nucleon, while the axial-vector diquark provides a small but phenomenologically essential admixture. For the parameters fixed in the meson sector, we find a scalar diquark mass of about $592$ MeV and a nucleon mass of $939.53$ MeV. The axial-vector component carries only about $2\%$ of the nucleon normalization $Z_N$, yet its interference with the scalar component and between different quark pairings is essential for the axial charge $g_A=1.265$.

By embedding these diquark kernels into a fully Poincare--covariant three-body Faddeev construction, we obtain a self-consistent nucleon amplitude that respects all required $\mathds{S}_{3}$ permutation symmetries. We discuss the full structure of mixed-antisymmetric and mixed-symmetric Dirac--flavor components, their transformation under quark exchange, and the role of orbital angular momentum in shaping the covariant nucleon wave function.

Analytically continuing the covariant amplitude to the light front, we construct the leading three-quark Fock-space wave function at the intrinsic instanton scale $\mu_0\simeq0.63$ GeV. From this same wave function we evaluate the nucleon axial and tensor charges, the flavor-separated valence distributions, and the helicity and transversity distributions. Their moments and QCD-evolved distributions are compared with available lattice determinations and global analyses. These observables expose the interplay of scalar-diquark dominance, axial-vector interference, and permutation symmetry in the nucleon's spin and flavor structure.

The present calculation isolates the instanton-induced, leading-valence contribution. In particular, the absence of a bound $\Delta$ in the present kernel and the differences in some spin-dependent and sea-sensitive observables indicate the complementary roles of confinement and higher Fock components. The latter are needed to describe intrinsic sea-quark and gluon correlations beyond perturbative evolution of the valence core.

The paper is organized as follows. Section II derives the instanton and molecular diquark kernels and their spectrum. Section III constructs the covariant three-quark Bethe--Salpeter--Faddeev amplitudes, and Sec. IV continues them to the light front. Sections V and VI address the axial and tensor charges; Secs. VII--IX present the unpolarized, helicity, and transversity distributions. Section X discusses the sea-quark and gluon distributions at low resolution, and Sec. XI summarizes the results and their limitations.

The resulting framework thus provides a microscopic, vacuum-based approach rooted in the topological QCD vacuum to nucleon structure, diquark dynamics, and the emergence of hadronic spectroscopy. It complements lattice QCD and Dyson--Schwinger approaches while offering a highly intuitive physical picture of baryons as three correlated quarks interacting via instanton-induced vertices.

\section{Diquarks in Instanton Vacuum}




The spectrum and low-energy dynamics of diquarks are governed by the instanton-induced interaction. In the case of $N_c=2$, diquarks can be color-singlet and thus coincide with baryonic states. Pauli–G\"{u}rsey symmetry, an extended $SU(4)$ supersymmetry, emerges in two-color QCD with two light flavors in the chiral limit, unifying massless baryons and mesons within the same multiplet. For physical QCD with $N_c=3$, although diquarks are no longer color-singlet asymptotic states, they remain essential building blocks in both light and heavy–light baryons \cite{Shuryak:2022wtk,Schafer:1993ra}. The form of the diquark bound states is constrained by color, isospin, and Dirac matrices $C\Gamma$ where $C=i\gamma^0\gamma^2$ is the charge conjugation matrix. In the three-color QCD, the color matrices are divided into two irreducible representations for diquarks,
$$\bf3\otimes\bf3=\bf{6}\oplus\bar{\bf{3}}$$ 
where $\bf6$ is color-symmetric and  $\bar{\bf3}$ is color-antisymmetric (See Appendix~\ref{App:color}).
The isospin matrix is given by $\tau^2$ for an isosinglet diquark and by $\tau^2\tau^a$ for an isovector diquark. Due to the overall antisymmetric characteristic regarding exchanging two quarks, the diquark channels associated with $\Gamma=1,\gamma^5,\gamma_\mu\gamma^5$ are constrained to be isoscalar diquarks whereas $\Gamma=\gamma_\mu, \sigma_{\mu\nu}$ gives isovector diquarks. Note that the parity of the Dirac matrix $C\Gamma$ is opposite to that of the matrix $\Gamma$ as the internal parity of a $qq$ pair is positive while the one of the $\bar qq$ system is negative.

With this in mind, the effective diquark interaction can be constructed by transposing the ’t Hooft vertices derived in \cite{Liu:2025ldh,Liu:2023fpj,Schafer:1996wv}, followed by a Fierz transformation. The resulting effective vertices, restricted to direct fermionic bubble diagrams, fully reproduce the contributions obtained from the original instanton-induced 't Hooft Lagrangian. The leading contribution comes from single instanton, corresponding to the effective Lagrangian given by


\begin{widetext}
\begin{equation}
\begin{aligned}
\label{eq:diq}
    \mathcal L^{(\mathrm{diq})}_{I}
    =\frac{G_I}{8N_c^2}&\bigg\{\frac{1}{N_c-1}\left[(\bar{\psi}\tau^2\beta_{A}^\alpha i\gamma^5C\bar{\psi}^T)(\psi^TC\tau^2\beta_{A}^\alpha i\gamma^5\psi)-(\bar{\psi}\tau^2\beta_{A}^\alpha C\bar{\psi}^T)(\psi^TC\tau^2\beta_{A}^\alpha\psi)\right]\\
    &-\frac{1}{2(N_c+1)}(\bar{\psi}\tau^2\beta^\alpha_{S}\sigma_{\mu\nu}C\bar{\psi}^T)(\psi^TC\tau^2\beta_{S}^\alpha\sigma_{\mu\nu}\psi)\bigg\}\\
\end{aligned}
\end{equation}
\end{widetext}
where the color matrices $\beta^\alpha_{S,A}$ is defined in Appendix~\ref{App:color} and the  effective coupling for single instanton interaction is defined
\begin{equation}
    G_I=\int d\rho n(\rho)\rho^{N_f}(2\pi\rho)^{2N_f}
\end{equation}
is fixed by the effective instanton density $n(\rho)$ given in \cite{Liu:2025ldh,Vainshtein:1981wh}. The color-antisymmetric $\bar{\bf{3}}$ channel induces a strong attraction for isosinglet scalar diquark $(\psi^T Ci\gamma^5\psi)$ and a strong repulsion for isosinglet pseudoscalar diqaurk $(\psi^T C\psi)$ while color-symmetric $\bf{6}$ channel produces a weakly attraction for isovector tensor diquark.

We note that no vector or axial vector channels are included in \eqref{eq:diq}. The formation of spin-1 diquark bound states therefore requires interactions beyond the single-instanton approximation. Correlated clusters formed by a few nearby instantons and anti-instantons become relevant. The next-to-leading corrections beyond the single-instanton contribution arise from molecular configurations, namely correlated instanton–anti-instanton pairs \cite{Schafer:1994nv}




To obtain spin-1 diquark bound states, the additional effective Lagrangian from the instanton-anti-instanton pair reads
\begin{widetext}

\begin{equation}
\begin{aligned}
\label{eq:diq_2}
    \mathcal L^{(\mathrm{diq})}_{IA}
    =\frac{G_{IA}}{2N_c^2}&\bigg\{-\frac{1}{N_c-1}\left(\bar{\psi}\gamma^\mu\gamma^5\tau^2\beta^\alpha_AC\bar{\psi}^T\psi^TC\gamma_\mu\gamma^5\tau^2\beta^\alpha_A\psi+\bar{\psi}\gamma^\mu\tau^2\tau^a\beta^\alpha_A C\bar{\psi}^T\psi^TC\gamma_\mu\tau^a\tau^2\beta^\alpha_A\psi\right) \\
       &+\frac{4}{N_c-1}\left(\bar{\psi}i\gamma^5\tau^2\beta^\alpha_AC\bar{\psi}^T\psi^TCi\gamma^5\tau^2\beta^\alpha_A\psi+\bar{\psi}\tau^2\beta^\alpha_AC\bar{\psi}^T\psi^TC\tau^2\beta^\alpha_A\psi\right)\\
       &+\frac{3}{N_c+1}\left(\bar{\psi}\gamma^\mu\tau^2\beta^\alpha_S C\bar{\psi}^T\psi^TC\gamma_\mu\tau^2\beta^\alpha_S\psi+\bar{\psi}\gamma^\mu\gamma^5\tau^2\tau^a\beta^\alpha_SC\bar{\psi}^T\psi^TC\gamma_\mu\gamma^5\tau^a\tau^2\beta^\alpha_S\psi\right)\bigg\}\\
\end{aligned}
\end{equation}
The effective molecule-induced coupling is defined as
\begin{equation}
 \label{MOLX}
     G_{IA}=\int d\rho_I d\rho_{A} ~ \frac{1}{8}(4\pi^2\rho^2_I)(4\pi^2\rho^2_{A})\int dud^4R|T_{IA}(u,R)|^{2N_f-2}n(\rho_I)n(\rho_{A})\rho_I^{N_f}\rho_{A}^{N_f}
\end{equation} 
\end{widetext}
where $R=z_I-z_{A}$ is the relative separation in between and $u=U_{A} U^\dagger_I$ is the relative color orientation. The pair-induced coupling in (\ref{MOLX}) is understood as the failed tunneling rate for a molecular configuration, whereby each of quark lines hopping insed the pair is removed by each division of $T_{IA}$, the overlap of the zero mode between instanton and anti-instanton.

The instanton-anti-instanton molecules in color-antisymmetric $\bar{\bf{3}}$ channel attracts for both isosinglet scalar and pseudoscalar diqaurk, particularly turning the pseudoscalar channel from repulsive to attractive, thus demonstrating the decisive importance of molecular configurations. The vector diquark and axial vector diquark channels are also weakly attracted in color-symmetric $\bar{\bf3}$ channel while in color-symmetric $\bf{6}$ channel, vector and axial vector diquark are weakly repulsive.

Those quark fields in the emergent vertices in \eqref{eq:diq} and \eqref{eq:diq_2} are modified by finite size effects of the pseudoparticles. More specifically, each interacting quark field get dressed by
\begin{align}
&\psi(k)\rightarrow\sqrt{\mathcal{F}(\rho k)}~\psi(k)
\end{align}
with non-local quark form factor $\mathcal{F}(k)$
which is essentially the profile of quark zero mode  defined by
\begin{equation}
\label{ZMform}
    \sqrt{\mathcal{F}(k)}=-z\frac{d}{dz}[I_0(z)K_0(z)-I_1(z)K_1(z)]\bigg|_{z=\frac{\rho k}{2}}
\end{equation}

Now the resulting full Lagrangian including single instanton and instanton molecules reads,

\begin{widetext}
\begin{equation}
\begin{aligned}
\mathcal{L}=&\bar{\psi}(i\slashed{\partial}-M)\psi+\frac{g_{\bar3_S}}{2N_c}\bar{\psi}i\gamma^5\tau^2\beta_{A}^\alpha C\bar{\psi}^T\psi^TCi\gamma^5\tau^2\beta_{A}^\alpha\psi+\frac{g_{\bar3_P}}{2N_c}\bar{\psi}\tau^2\beta_{A}^\alpha C\bar{\psi}^T\psi^TC\tau^2\beta_{A}^\alpha\psi\\
 &-\frac{g_{\bar3_V}}{2N_c}\bar{\psi}\gamma^\mu\gamma^5\tau^2\beta^\alpha_AC\bar{\psi}^T\psi^TC\gamma_\mu\gamma^5\tau^2\beta^\alpha_A\psi-\frac{g_{\bar3_A}}{2N_c}\bar{\psi}\gamma^\mu\tau^2\tau^a\beta^\alpha_A C\bar{\psi}^T\psi^TC\gamma_\mu\tau^a\tau^2\beta^\alpha_A\psi \\
 &-\frac{g_{6_V}}{2N_c}\bar{\psi}\gamma^\mu\gamma^5\tau^2\tau^a\beta^\alpha_SC\bar{\psi}^T\psi^TC\gamma_\mu\gamma^5\tau^a\tau^2\beta^\alpha_S\psi-\frac{g_{6_A}}{2N_c}\bar{\psi}\gamma^\mu\tau^2\beta^\alpha_S C\bar{\psi}^T\psi^TC\gamma_\mu\tau^2\beta^\alpha_S\psi\\
 &-\frac{g_{6_T}}{2N_c}\bar{\psi}\tau^2\beta^\alpha_{S}\sigma_{\mu\nu}C\bar{\psi^T}\psi^TC\tau^2\beta_{S}^\alpha\sigma_{\mu\nu}\psi
\end{aligned}
\end{equation}
where the coupling constants of each channel are fixed by two parameters $G_{I}$ and $G_{IA}$ from the instanton vacuum.

\begin{align}
\label{diqaurk_tHooft_couplings}
&g_{\bar3_S}=\left(\frac1{N_c-1}\right)\left(\frac{G_{I}}{4N_c}+\frac{4G_{IA}}{N_c}\right) && g_{\bar3_P}=\left(\frac1{N_c-1}\right)\left(-\frac{G_{I}}{4N_c}+\frac{4G_{IA}}{N_c}\right) \nonumber\\
&g_{\bar3_V}=\left(\frac1{N_c-1}\right)\frac{G_{IA}}{N_c} \nonumber && g_{\bar3_A}=\left(\frac1{N_c-1}\right)\frac{G_{IA}}{N_c} \nonumber\\
&g_{6_V}=-\left(\frac1{N_c+1}\right)\frac{3G_{IA}}{N_c} && g_{6_A}=-\left(\frac1{N_c+1}\right)\frac{3G_{IA}}{N_c} \nonumber\\
&g_{6_T}=\left(\frac1{N_c+1}\right)\frac{G_{I}}{8N_c}
\end{align}
\end{widetext}

With given constituent mass $M=410$ MeV~\cite{Liu:2023fpj} and phenomenological value of instanton size $\rho=0.33$ fm~\cite{Schafer:1996wv}, the coupling strength $G_I$ and $G_{IA}$ is fixed by physical pion mass ($m_\pi=139$ MeV) and rho mass ($m_\rho=757.2$ MeV). The values of those three parameters are presented in Table~\ref{tab:parameters_ILM}. 

The couplings in each diquark channel can also be determined from the values of $G_I$ and $G_{IA}$ given in Table~\ref{tab:parameters_ILM}

\begin{center}
\begin{align*}
&g_{\bar3_S}=76.46~ \mathrm{GeV}^{-2}  \\
&g_{\bar3_V}=g_{\bar3_A}=18.94~ \mathrm{GeV}^{-2} \\
&g_{\bar3_P}=75.04~ \mathrm{GeV}^{-2}  \\
\end{align*}
\end{center}

\begin{table}
    \centering
    \begin{tabular}{|c|c|c|c|c|}
   \hline
    & $G_I$  & $G_{IA}$ & $M$   \\ 
    \hline
  ILM & 16.98~$\mathrm{GeV}^{-2}$ & 113.63 $\mathrm{GeV}^{-2}$ &
  $410$ MeV  \\
   \hline
\end{tabular}
    \caption{Parameters in instanton liquid model with $\rho=0.33$ fm with physical pion mass ($m_\pi=139.0$ MeV) and rho mass ($m_\rho=757.2$ MeV)}
    \label{tab:parameters_ILM}
\end{table}

For completeness, we also fix the quark mass $m$ and quark condensate $\langle\bar\psi\psi\rangle=\langle\bar u u+\bar d d\rangle$ by the gap equation (See \eqref{eq:gap} in Appendix~\ref{vac_pol}) induced by chiral symmetry breaking~\cite{Liu:2025ldh}. The values are compared to the calculation from Flavour Lattice Averaging Group (FLAG)~\cite{FlavourLatticeAveragingGroupFLAG:2021npn}.

\begin{table}
    \centering
    \begin{tabular}{|c|c|c|c|}
\hline
     & $m$ &  $|\langle\bar{\psi}\psi\rangle|$ & $m|\langle\bar{\psi}\psi\rangle|$   \\[5pt] 
    \hline
    ILM & $2.13$ MeV & ($316.25$ MeV)$^3$ & $6.74\times10^{-5}$ \\
     \hline
    FLAG & $3.381(40)$ MeV & ($272(5)$ MeV)$^3$ & $6.80\times10^{-5}$ \\
   \hline
\end{tabular}
    \caption{Quark mass and chiral condensate in the ILM using the parameters listed in Table~\ref{tab:parameters_ILM} at the resolution
    $\mu_0\approx1/\rho$. We compare the ILM results to the Flavour Lattice Averaging Group (FLAG) \cite{FlavourLatticeAveragingGroupFLAG:2021npn}
    at the resolution $\mu=$2 GeV. The last renormalization-group (RG) invariant combination is in GeV$^4$.}
    \label{tab:parameters_ILM_2}
\end{table}



For general $N_c$, the diqaurk couplings are down by the additional universal color factor $1/(N_c - 1)$ compared to those of their corresponding 
meson partner \cite{Rapp:1999qa,Shuryak:2022wtk}. In the case of two-color QCD, this color factor becomes a unity, implying additional Pauli-G$\ddot{\mathrm{u}}$rsey symmetry that mixes quarks with antiquarks. As a result, diquarks (the baryons of the two-color QCD) are degenerate with the corresponding mesons, forming a supersymmetric pair. 

Since our focus is on the diquark states relevant for nucleon structure, we restrict the follwoing discussion to the color–antisymmetric $\bar{\mathbf{3}}$ channel.  

\subsection{Diquark spectrum}

The diquark correlation functions in each channel can be calculated by suming the leading $1/N_c$ $s$-channel bubble diagrams presented in Fig.~\ref{fig:diquark_bubble}. The loop integrals for each bubble diagrams are presented in Appendix~\ref{vac_pol}. 
\begin{figure*}
    \centering
    \includegraphics[width=\linewidth]{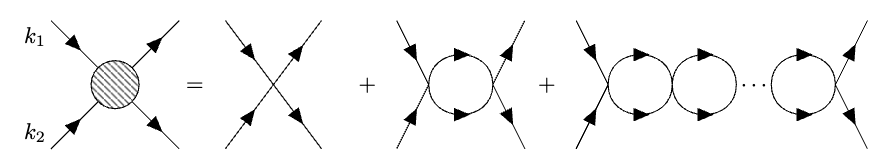}
    \caption{Feynman diagrams for the fermionic bubble resummation in the diquark channels under the random phase approximation. 
    }
    \label{fig:diquark_bubble}
\end{figure*}

In the scalar diquark channel, the attraction is strong enough to generate a diquark bound state, implying a mass pole below the two–quark threshold. The corresponding mass can be obtained from the pole condition of the resummed quark–quark correlation function, leading to a gap-like equation for the diquark mass. Since the isosinglet scalar diquark channel mixes with the longitudinal isosinglet vector diquark, by taking the mixing effect into account, the scalar diquark mass can be determined by
\begin{widetext}
\begin{equation}
\label{bound_eq_4}
    g_{\bar3_S}\Pi_{PP}(m^2_{qq_S})+g_{\bar3_V}\Pi^{(l)}_{AA}(m^2_{qq_S})-g_{\bar3_S}g_{\bar3_V}\left[\Pi_{PP}(m^2_{qq_S})\Pi^{(l)}_{AA}(m^2_{qq_S})+\Pi^2_{PA}(m^2_{qq_S})\right]=1
\end{equation}
\end{widetext}

While the single-instanton contribution leaves no interaction in the vector and axial-vector diquark channels and induces strong repulsion in pseudoscalar channel, the situation changes once molecular configurations are taken into account. The instanton–anti-instanton molecules induce overall attraction to each channel and modify the diquark dynamics. Thus, for the pseudoscalar, vector and axial diquark, similar gap-like equation can be derived from the bubble summation.
\begin{equation}
\label{bound_eq_2}
    g_{\bar3_P}\Pi_{SS}(m_{qq_P}^2)=1
\end{equation}

\begin{equation}
\label{bound_eq_3}
    g_{\bar3_V}\Pi^{(t)}_{AA}(m_{qq_V}^2)=1
\end{equation}

\begin{equation}
\label{bound_eq_1}
    g_{\bar3_A}\Pi_{VV}(m_{qq_A}^2)=1
\end{equation}

However, these equations do not have solutions below the threshold $2M$ for the present coupling fixed by physical pion and rho meson mass. The interaction in pseudoscalar, vector, and axial channels remains too weak to produce a diquark bound state below the two–quark threshold. 

\begin{figure}
    \centering
    \subfloat[]{\includegraphics[width=\linewidth]{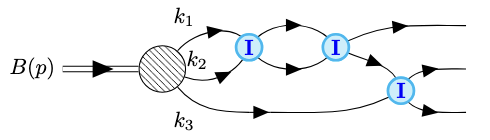}}
    \hfill
    \subfloat[]{\includegraphics[width=\linewidth]{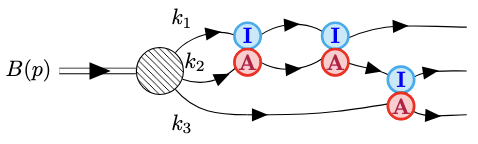}}
    \caption{Baryon bound state in instanton vacuum from (a) single instantons and (b) instanton-anti-instanton molecules.}
    \label{fig:WF_b}
\end{figure}




With the parameters given in Table~\ref{tab:parameters_ILM} together with \eqref{bound_eq_4}, \eqref{bound_eq_2}, \eqref{bound_eq_3}, and \eqref{bound_eq_1}, the diquark mass spectrum are shown in Table~\ref{tab:diq-q_couple}. The pseudoscalar $[qq]_{P}$ and vector $[qq]_{V}$ channels are found above the threshold mass, $2M = 820$ MeV. The other two channels, the scalar $[qq]_{S}$ and axial vector $[qq]_{A}$, are bounded below the threshold. The diquark masses are compared to the calculation in \cite{Roberts:2011cf} using rainbow ladder (RL) truncated Dyson-Schwinger equation (DSE) and lattice calculation \cite{Bi:2015ifa}. A lighter mass is expected as the instanton vacuum does not fully cover long-range confinement string contribution.

The color factor $1/(N_c - 1)$ in \eqref{diqaurk_tHooft_couplings} implies the real QCD with $N_c = 3$ is half-way between $N_c = 2$ (the “small $N_c$ limit” of QCD) with a relative weight of $1$, and $N_c = \infty$ (the “large $N_c$ limit” of QCD) with relative weight $0$ \cite{Shuryak:2005pk}. 

We also compute the effective quark-diquark couplings using 

\begin{equation}
\label{eq:ax_di}
\lambda^2_{qq_A}=\frac2{N_c}\left(\frac{\partial \Pi_{VV}}{\partial P^2}\right)^{-1}\Bigg|_{P^2=m_{qq_A}^2}
\end{equation}
and 





\begin{widetext}

\begin{equation}
\label{eq:sca_di}
\lambda^2_{qq_S}=\frac2{N_c}\frac{g_{\bar3_S}-g_{\bar3_V}+g_{\bar3_S}g_{\bar3_V}(\Pi_{PP}-\Pi^{(l)}_{AA})}{g_{\bar3_S}(1-g_{\bar3_V}\Pi^{(l)}_{AA})\frac{\partial \Pi_{PP}}{\partial P^2}+g_{\bar3_V}(1-g_{\bar3_S}\Pi_{PP})\frac{\partial \Pi^{(l)}_{AA}}{\partial P^2}-2g_{\bar3_S} g_{\bar3_V}\Pi_{PA}\frac{\partial}{\partial P^2}\Pi_{PA}}\Bigg|_{P^2=m_{qq_S}^2}
\end{equation}


\end{widetext}

\begin{table*}
    \centering
\begin{tabular}{|c|c|c|c|c|c|c|c|}
    \hline
    Model & $m_{qq_S}$ (MeV) & $m_{qq_P}$ (MeV) &$m_{qq_V}$ (MeV) & $m_{qq_A}$ (MeV) \\
    \hline
     ILM & $591.63$ & $-$ & $-$ & $818.5$ \\
     RL-DSE \cite{Roberts:2011cf} & $780$ & $1060$ & $930$ & $1160$ \\
      Lattice\cite{Bi:2015ifa} & $690(47)$ & $-$ & $-$ & $990(60)$ \\
      Spectral fit (ILM) \cite{Schafer:1993ra} & $420 \pm 30$ & $-$ & $940 \pm 20$ & $940 \pm 20$  \\ 
     \hline
\end{tabular}   
    \caption{The mass spectrum on color antisymmetric diquark bound states. The result is compared to the calculation in \cite{Roberts:2011cf} using RL-truncated DSE and lattice calculation \cite{Bi:2015ifa} using Laudau gauge  with a slight larger constituent quark mass $M_q(\mathrm{lattice})=427(25)$ MeV compared to our values in Table \ref{tab:parameters_ILM}.}
    \label{tab:diquark_m}
\end{table*}

The mixing angle $\theta_{SV}$ between the scalar and vector diqaurk channels can be obtained by the diagonalization of the $[qq]_S$ and $[qq]_V$ bubble diagrams, which yields

\begin{equation}
\begin{aligned}
\tan\theta_{SV}=&\frac{g_{\bar3_V}\Pi_{PA}(m^2_{qq_S})}{1-g_{\bar3_V}\Pi^{(l)}_{AA}(m^2_{qq_S})}\\
=&\frac{1-g_{\bar3_S}\Pi_{PP}(m^2_{qq_S})}{-g_{\bar3_S}\Pi_{PA}(m^2_{qq_S})}
\end{aligned}
\end{equation}

With the parameters in Table~\ref{tab:parameters_ILM}, the mixing angle between the scalar and vector diqaurk channel is $\theta_{SV}=4.17^\circ$, indicating a small mixing effect from the time-component of vector diquark channel. Thus in the following analysis, we will neglect the mixing.

\begin{table}
\begin{center}
\begin{tabular}{|c|c|c|c|c|}
   \hline
     &  $\lambda_{qq_S}$ & $\lambda_{qq_P}$ & $\lambda_{qq_V}$ & $\lambda_{qq_A}$\\ 
   \hline
   ILM & $4.33$ & $-$ & $-$ & $0.718$ \\
   \hline
\end{tabular}
\end{center}
    \caption{Effective diquark-quark couplings determined by \eqref{eq:sca_di}.}
    \label{tab:diq-q_couple}
\end{table}



\section{Covariant Baryon Wave Functions}
\label{sec:baryon}


The covariant BSF nucleon wave function can be defined as

\begin{widetext}
\begin{equation}
\begin{aligned}
\label{NBS}
&(2\pi)^4\delta^4(p-k_1-k_2-k_3)\Psi_{B,ijk}(k_1,k_2,k_3;PS)\frac1{\sqrt{N_c!}}\epsilon_{\alpha\beta\gamma}\\
=&-i\int d^4x_1d^4x_2d^4x_3 e^{ik_1x_1+ik_2\cdot x_2+ik_3\cdot x_3}\frac1{3!}\langle0|\psi_{\alpha i}(x_1)\psi_{\beta j}(x_2)\psi_{\gamma k}(x_3)|B(PS)\rangle    
\end{aligned}
\end{equation}
where $\alpha,\beta, \gamma$ denote the color indices and $i,j,k$ denote the flavor-Dirac indices. The corresponding BSF equation is written as

\begin{equation}
\begin{aligned}
\label{eq:BSF0}
    &\Psi_{B,ijk}(k_1,k_2,k_3;PS)\\
    =&S_{ii'}(k_1)S_{jj'}(k_2)\int\frac{d^4k}{(2\pi)^4} \mathcal{K}_{i'j'i''j''}(k_1-k,k_2+k,k_1,k_2)\Psi_{B,i''j''k}(k_1-k,k_2+k,k_3;PS)\\
    &+S_{jj'}(k_2)S_{kk'}(k_3)\int\frac{d^4k}{(2\pi)^4} \mathcal{K}_{j'k'j''k''}(k_2-k,k_3+k,k_2,k_3)\Psi_{B,ij''k''}(k_1,k_2-k,k_3+k;PS)\\
    &+S_{ii'}(k_1)S_{kk'}(k_3)\int\frac{d^4k}{(2\pi)^4} \mathcal{K}_{k'i'k''i''}(k_3-k,k_1+k,k_3,k_1)\Psi_{B,i''jk''}(k_1+k,k_2,k_3-k;PS)
\end{aligned}
\end{equation}
\end{widetext}
where $\mathcal{K}$ is the irreducible two-body diagram obtained by interactions in \eqref{eq:diq} and \eqref{eq:diq_2}, as shown in Fig.~\ref{fig:WF_b}.


The Pauli principle requires the full Dirac-flavor-color
structure of the BSF wave function to be antisymmetric under exchange of any two quark legs, i.e. under a combined permutation of Dirac, flavor and color indices and the corresponding exchange of momenta. The color wave function in \eqref{NBS} forms a fully antisymmetric singlet, denoted by $\epsilon_{\alpha\beta\gamma}$. As a result, the combined Dirac-flavor-momentum part must be fully symmetric under the exchange of quarks. The transformation
properties of the flavor and Dirac structure follow from Fierz identities. 

\begin{equation}
\begin{aligned}
\label{eq:BWF}
\Psi_{B,ijk}=&\frac1{\sqrt{6}}S_{ii'}(k_1)S_{jj'}(k_2)S_{kk'}(k_3)\mathbf{\Gamma}_{B,i'j'k'}\\
&+(1\leftrightarrow2\leftrightarrow3)
\end{aligned}
\end{equation}

In our case, only two-body interactions are involved in the interaction kernel and thus we can futher simplify the BSF equation \eqref{eq:BSF0} into quark-diquark form, where the nucleon and delta baryon BSF vertex functions $\mathbf{\Gamma}_{B,ijk}$ can be constructed with the effective quark-diquark two-body vertex function.

\begin{widetext}

\begin{equation}
\begin{aligned}
\label{eq:NWF}
&\mathbf{\Gamma}_{N,ijk}\rightarrow\sqrt{Z_{N}}\sqrt{\mathcal{F}(k_1)\mathcal{F}(k_2)\mathcal{F}(k_3)}\begin{pmatrix}
        i\gamma^5C\tau^2 D_{s}(k_{12}) & 0 \\
        0 & \gamma_\mu C\tau^2\tau^aD^{\mu\nu}_{a}(k_{12})
    \end{pmatrix}_{ij}\Gamma_{N,k}(P,k_3)\\
&\mathbf{\Gamma}_{\Delta,ijk}\rightarrow\sqrt{Z_{\Delta}}\sqrt{\mathcal{F}(k_1)\mathcal{F}(k_2)\mathcal{F}(k_3)}(\gamma^\mu C\tau^2\tau^a)_{ij}D_{a}^{\mu\nu}(k_{12})
    \Gamma^\nu_{\Delta,k}(P,k_3)
\end{aligned}
\end{equation}
where $k_{12}=k_1+k_2$. The scalar diquark propagator is defined as
\begin{equation}
\begin{aligned}
    D_{s}(k)=&\frac2{N_c}\frac{i\left[g_{\bar3_S}-g_{\bar3_V}+g_{\bar3_S}g_{\bar3_V}\left(\Pi_{PP}-\Pi^{(l)}_{AA}\right)\right]}{(1-g_{\bar3_S}\Pi_{PP})(1-g_{\bar3_V}\Pi^{(l)}_{AA})+g_{\bar3_S}g_{qq_V}\Pi^2_{PA}}
\end{aligned}
\end{equation}
and the axial vector diquark propagator is defined as
\begin{equation}
\begin{aligned}
    D_{a}^{\mu\nu}(k)=&\frac{2}{N_c}\frac{-ig_{qq_A}}{1-g_{qq_A}\Pi_{VV}}\left(g^{\mu\nu}-g_{qq_A}\Pi_{VV}\frac{k^\mu k^\nu}{k^2}\right)
\end{aligned}
\end{equation}
\end{widetext}

To make the diquark dynamics more tractable, we further adopt the pole approximation for the diquark propagators, with the diquark masses and effective quark–diquark couplings fixed by the corresponding two-body calculations presented in the preceding section.

\begin{equation}
\begin{aligned}
D_s(k)
&\rightarrow
\frac{-i\lambda_{qq_S}^2}{P^2-m_{qq_S}^2},
\\[2mm]
D_a^{\mu\nu}(k)
&\rightarrow
\frac{i\lambda_{qq_A}^2}{P^2-m_{qq_A}^2}
\left(
g^{\mu\nu}
-\frac{k^\mu k^\nu}{m_{qq_A}^2}
\right).
\end{aligned}
\end{equation}


The nucleon structure as a quark-diquark bound state has been discussed in many literatures \cite{Ishii:1995bu,Cloet:2005pp,Oettel:1998bk,Buck:1992wz,Cloet:2014rja,Rezaeian:2003kq,Oettel:2000jj,Hellstern:1997pg,Mineo:2002bg,Mineo:1999eq,Bloch:1999ke}. It is well established that the instanton vacuum induces strong attractive interactions in the diquark channels through the effective Lagrangian in \eqref{eq:diq} and \eqref{eq:diq_2}. As a result, a pair of tightly correlated quarks are dynamically bounded as a diquark. These diquarks subsequently interact with an additional quark via quark exchange processes, forming a baryon bound state.



With a quark-diquark construction of the nucleon state, the color structure can be determined by projection $3\otimes\bar3=8\oplus1$ as the baryon states are color-singlet. For the isospin part, as nucleon is an isospin doublet state with spin-$1/2$, it mixes the scalar diquark and axial diquark whereas the $\Delta$ baryon only consists of axial diquark due to the spin-$3/2$ and isospin quadrplet nature. 

\begin{figure}
    \subfloat[\label{color}]{\includegraphics[width=0.5\linewidth]{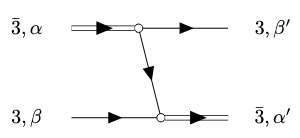}}
\hfill
    \subfloat[\label{isospin}]{\includegraphics[width=0.5\linewidth]{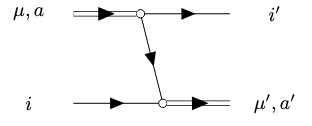}}
\caption{The quark exchange matrix in color space (a) and in flavor-spin space (b)}
\label{fig:Qkernel}
\end{figure}

To solve the full Faddeev equation for the nucleon as a quark-diquark
bound state, we first compute the quark exchange matrix $G_N$ for nucleons, which mixes the scalar diquarks and axial diquarks, as shown in Fig.~\ref{fig:Qkernel}. 

\begin{widetext}
\begin{equation}
\label{quark_exchange}
\begin{aligned}
    iG_{N}(k)=&(\beta^\alpha_A\beta^{\alpha'}_A)
    \begin{pmatrix}
        i\gamma^5C\tau^2S^T(-k)Ci\gamma^5\tau^2 & \gamma^{\mu}C\tau^2\tau^a S^T(-k)Ci\gamma^5\tau^2 \\
        i\gamma^5C\tau^2S^T(-k)C\gamma^{\mu'}\tau^{a'}\tau^2& \tau^2\tau^a\gamma^{\mu}C S^T(-k)C\gamma^{\mu'}\tau^{a'}\tau^2 \\
    \end{pmatrix}\\
    \rightarrow&-(\beta^\alpha_A\beta^{\alpha'}_A)
    \begin{pmatrix}
        \tau^2\tau^2 & \tau^2\tau^a\tau^2 \\
        \tau^2\tau^{a'}\tau^2& \tau^2\tau^a\tau^{a'}\tau^2 \\
    \end{pmatrix}
    \begin{pmatrix}
        -\gamma^5S(k)\gamma^5 & i\gamma^{\mu}S(k)\gamma^5 \\
        i\gamma^5S(k)\gamma^{\mu'}& \gamma^{\mu} S(
        k)\gamma^{\mu'} \\
    \end{pmatrix}
\end{aligned}
\end{equation}
\end{widetext}
For the $\Delta$, the isospin-$3/2$ and spin-$3/2$ baryon, can only be formed by the isovector and spin-$1$ diquark. Therefore, the quark exchange kernel reads

\begin{widetext}
\begin{equation}
    iG_{\Delta}(k)=(\beta^\alpha_A\beta^{\alpha'}_A)_{\beta'\beta}
     \tau^2\tau^a\gamma^{\mu}C S^T(-k)C\gamma^{\mu'}\tau^{a'}\tau^2 \rightarrow-(\beta^\alpha_A\beta^{\alpha'}_A)_{\beta'\beta}(\tau^2\tau^a\tau^{a'}\tau^2)\gamma^{\mu}S(k) \gamma^{\mu'} 
\end{equation}
\end{widetext}

As nucleon is color-singlet, the color matrix in the quark exchange kernel in \eqref{quark_exchange} is projected onto a color-singlet subspace and reduced to $-3$  

\begin{equation}
(\beta^\alpha_A\beta^{\alpha'}_A)_{\beta'\beta}\rightarrow-3
\end{equation}

In isospin space, the isoscalar diquark (scalar) only contributes to $I=1/2$ baryon as $\frac12\otimes0=\frac12$ while the isovector diquark (axial vector) contributes to both $I=3/2$ and $I=1/2$ baryon as $\frac12\otimes1=\frac32\oplus\frac12$. Now with this in mind, the quark exchange kernel for nucleon with color and isospin projected yields a 5 by 5 matrix

\begin{equation}
\label{G_N}
    iG_N\rightarrow3
    \begin{pmatrix}
        1 & \sqrt{3} \\
        \sqrt{3}& -1\\
    \end{pmatrix}
    \begin{pmatrix}
        -\gamma^5S(k)\gamma^5 & i\gamma^{\mu}S(k)\gamma^5 \\
        i\gamma^5S(k)\gamma^{\mu'}& \gamma^{\mu} S(k)\gamma^{\mu'} \\
    \end{pmatrix}
\end{equation}
For $\Delta$ baryon, we have a 4 by 4 matrix
\begin{equation}
\label{G_Delta}
    iG^{\mu'\mu}_\Delta\rightarrow3\times2\,\gamma^\mu S(k)\gamma^{\mu'}
\end{equation}

The remaining structure to deal with in the quark exchange kernel is the spin. The quark exchange kernel for nucleon is a tensor product of $5\times5$ mixing matrix between scalar and axial vector diquarks and $4\times4$ Dirac structure. For the $\Delta$-baryon, the quark exchange kernel is a tensor product of $4\times4$ matrix from the axial vector diquark and $4\times4$ Dirac structure. (See details in \cite{Ishii:1995bu} and Appendix~\ref{Baryon})

For convenience, we parameterize the three-body momentum using the center of momentum $P=k_1+k_2+k_3$ and the relative momenta $p$, $q$ for internal degrees of freedom, similar to the conventions used in \cite{Meoto:2025tns,Miesch:2025vas,Shuryak:2022wtk}.

\begin{equation}
\label{Jacobi}
\begin{aligned}
p &= \frac{k_2-k_3}{2}, \\
q &= \frac{k_1-k_2-k_3}{2}.
\end{aligned}
\end{equation}
Similarly, one can also construct other $p$ and $q$ by cyclic permutation on 1,2,3.

With the quark exchange matrix and wave function constructed in \eqref{eq:NWF}, the full Faddeev equation in \eqref{eq:BSF0} are reduced to an effective two-body equation for nucleon vertex function, given by~\cite{Ishii:1995bu}

\begin{widetext}

\begin{equation}
\label{eq:FdvN}
    \Gamma_{N}(P,q)=i\int \frac{d^4q'}{(2\pi)^4}G_{N}(q+q')S(P/2+q')\mathcal{F}(P/2+q')
    \begin{pmatrix}
        D_{s}(P/2-q') & 0 \\
        0 & D^{\mu\nu}_{a}(P/2-q')
    \end{pmatrix}\Gamma_{N}(P,q')
\end{equation}
and
\begin{equation}
\label{eq:FdvD}
    \Gamma^\mu_{\Delta}(P,q)=i\int \frac{d^4q'}{(2\pi)^4}G^{\mu\mu'}_{\Delta}(q+q')S(P/2+q')\mathcal{F}(P/2+q')
     D^{\mu'\nu}_{a}(P/2-q')\Gamma^\nu_{\Delta}(P,q')
\end{equation}
\end{widetext}

Within the static approximation, the quark propagator in the nucleon and $\Delta$ quark exchange kernels is replaced by the momentum-independent form $-i/M$. Thus, the quark-exchange matrix $G_{N,\Delta}$ can be taken outside the integrals in \eqref{eq:FdvN} and \eqref{eq:FdvD}. The Faddeev equations can be approximated to corresponding gap-like equations that determine the baryon mass in each channel.
\begin{equation}
\label{baryon_eq}
    \Gamma_{N,\Delta}(P)=G_{N,\Delta}\Pi_{N,\Delta}\Gamma_{N,\Delta}(P)
\end{equation}
The definition of $\Pi_{N,\Delta}$ and the details for solving the Faddeev equation can be found in Appendix \ref{Baryon}.  

With an appropriate choice of basis and projection onto positive-energy states, the nucleon vertex function is then given by
\begin{equation}
\begin{aligned}
\label{vertex_N}
    &\Gamma_N(P,S)=\begin{pmatrix}
        -i\Gamma_s \\
        \Gamma_a^\mu
    \end{pmatrix}u_N(P,S)\\
    =&
    \begin{pmatrix}
        -i\alpha_1 \\
        \left(\alpha_2-\frac{\alpha_3}{\sqrt{3}}\right) \frac{P^\mu}{m_N}\gamma^5-\frac{\alpha_3}{\sqrt{3}}\gamma^\mu\gamma^5
    \end{pmatrix}u_N(P,S)
\end{aligned}
\end{equation}
and the one for delta baryon is
\begin{equation}
\label{vertex_Delta}
    \Gamma_\Delta(P,S)=u^\mu_\Delta(P,S)
\end{equation}


The nucleon and $\Delta$ masses are presented in Table~\ref{tab:baryon_spec}. The nucleon forms a bound state with a mass of approximately 939.5 MeV, whereas no solution is found below the threshold in the $\Delta$-baryon channel. In the absence of axial diquark contributions, the nucleon mass is $948.74$ MeV, roughly $9.2$ MeV heavier than the one with the axial diquark included. 

The contribution of the scalar and axial diquark channels can be quantified by the inverse wave function renormalization $Z^{-1}_N=Z_{N_s}^{-1}+Z_{N_a}^{-1}$ of nucleon channel,

\begin{equation}
\begin{aligned}
\label{eq:SA_weight}
   Z_{N_s}^{-1}=& -\,\bar{\Gamma}_s(P)\,
\frac{\partial \Pi_{s}(P)}{\partial P^2}\,
\Gamma_s(P)\\
Z_{N_a}^{-1}=& -\,\bar{\Gamma}^\mu_a(P)\,
\frac{\partial \Pi^{\mu\nu}_{a}(P)}{\partial P^2}\,
\Gamma^\nu_a(P)\\
\end{aligned}
\end{equation}
where $\Pi_s$ and $\Pi_a^{\mu\nu}$ are defined in \eqref{eq:Pi_diq}.

With the coefficients in \eqref{vertex_N} obtained from the solution of the BSF equation, $\alpha_1=-0.290$, $\alpha_2=0.486$, $\alpha_3= 0.825$, the probabilistic weights associated with the scalar and axial-vector diquark components around the nucleon mass pole are given by

\begin{align}
\label{eq:Z}
   w_s =\frac{Z_{N_s}^{-1}}{Z_N^{-1}}&=98.0\%, & w_a= \frac{Z_{N_a}^{-1}}{Z_N^{-1}}&=2.0\% 
\end{align}


\begin{table}[]
    \centering
    \begin{tabular}{|c|c|c|}
    \hline
    Model & $m_{N}$ (MeV) & $m_{\Delta}$ (MeV) \\
    \hline
     ILM & $939.53$ & $-$ \\
     PDG \cite{ParticleDataGroup:2014cgo,ParticleDataGroup:2016lqr,ParticleDataGroup:2018ovx,ParticleDataGroup:2024} & $938.27$ & $1232$ \\
     \hline
    \end{tabular}
    \caption{The mass for nucleon and $\Delta$ using the ILM parameters in Table~\ref{tab:parameters_ILM} fixed by physical pion and rho mass}
    \label{tab:baryon_spec}
\end{table}

For completeness, these values are compared to the coefficients $(\alpha_1,\alpha_2,\alpha_3)=(-0.35,0.479,0.842)$ reported in \cite{Cloet:2005pp}. 
We also compare our estimated diquark admixtures with the Nambu–Jona-Lasinio (NJL) analysis in~\cite{Mineo:2002bg}, which found a phenomenologically reasonable axial-vector diquark admixture in the range of approximately 2\%–10\%. A similar conclusion is also obtained from ~\cite{Shuryak:2022wtk}, where the authors emphasize a strong dominance of scalar-diquark correlations in the nucleon, with an estimated scalar-diquark contribution of roughly 70\%–90\%.

Now the nucleon and delta baryon BSF wave functions reads 
\begin{widetext}

\begin{equation}
\begin{aligned}
\label{eq:BSFWF_N}
\Psi_N(k_1,k_2,k_3;PS)=&\frac{\sqrt{Z_{N}}}{\sqrt{6}}\sqrt{\mathcal{F}(k_1)\mathcal{F}(k_2)\mathcal{F}(k_3)}
       [S(k_1)\gamma^5C\tau^2S^T(k_2)]D_s(k_{12})S(k_3)\Gamma_su_N\\
      &+\frac{\sqrt{Z_{N}}}{\sqrt{6}}\sqrt{Z_{N}}\sqrt{\mathcal{F}(k_1)\mathcal{F}(k_2)\mathcal{F}(k_3)}[S(k_1) \gamma^\mu C\tau^2\tau^aS^T(k_2)]D_a^{\mu\nu}(k_{12})S(k_3)\tau^a\Gamma^\nu_au_N\\
&+ S_3\mathrm{~permutation}~(1\leftrightarrow2\leftrightarrow3)
\end{aligned}
\end{equation}
and
\begin{equation}
\begin{aligned}
\label{eq:BSFWF_D}
   \Psi_\Delta(k_1,k_2,k_3;PS)=& \frac{\sqrt{Z_{\Delta}}}{\sqrt{6}}\sqrt{\mathcal{F}(k_1)\mathcal{F}(k_2)\mathcal{F}(k_3)} \left[S(k_1)\gamma^\mu C\tau^2\tau^a S^T(k_2)\right]
       D_a^{\mu\nu}(k_{12})S(k_3)\xi^a_{\frac32}u^{\nu}_{\Delta}\\
    &+ S_3\mathrm{~permutation}~(1\leftrightarrow2\leftrightarrow3)
\end{aligned}
\end{equation}
\end{widetext}
with $k_1$, $k_2$, $k_3$ the constituent quark momenta and $P$ the total momenta of the baryon and $1/\sqrt{6}$ the symmetrization factor among permutations. $u^{\nu}_{\Delta}$ is the Rita-Schwinger spinor defined in \eqref{eq:RS_spinor} with the flavor projection $\xi^a_{\frac32}$ to isospin-$3/2$ defined by

\begin{equation}
(\xi^a_{\frac32})_{i',i}=\left\langle1 a;\frac12 i'\big|\frac32i\right\rangle 
\end{equation}

The effective quark-diquark spinor parameterization naturally emerges to characterize the scalar and axial diquark correlations.

\section{Analytic continuation to Light Front}

The light front wave function is defined by spanning in the Fock space. The $3$-body baryon light front wave function at leading Fock space reads

\begin{widetext}

\begin{equation}
    \ket{B}\approx\sum_{s_1,s_2,s_3}\int d[123]\frac1{\sqrt{6}}\Phi_B(k_1,k_2,k_3,s_1,s_2,s_3)\frac{\epsilon_{\alpha\beta\gamma}}{\sqrt{N_c!}}b_{s_1,\alpha}^\dagger(k_1) b^\dagger_{s_2,\beta}(k_2)b^\dagger_{s_3,\gamma}(k_3)\ket{0}
\end{equation}

\end{widetext}
where $\frac1{\sqrt{6}}$ is the symmetrization factor and the $3$-body phase space integral is defined as

\begin{equation}
\begin{aligned}
\label{eq:n-body}
    &d[123]
    =\prod_{i=1}^3\frac{dx_{i}d^2k_{i\perp}}{(2\pi)^32x_i}\\
    &\times(2\pi)^32\delta\left(1-\sum_{i=1}^3x_{i}\right)\delta^2\left(p_\perp-\sum_{i=1}^3k_{i\perp}\right)
\end{aligned}
\end{equation}
The solutions in light front formalism should be equivalent to the Lorentz covariant formalism. In our method we solve $3$-body Faddeev equation by summing over all quark bubble diagrams at leading $1/N_c$ limit. It is easy to show the resulting three-body BSF wave function is equivalent to leading Fock light front wave function by directly integrating the Faddeev wave function over the $k_{1,2,3}^-$ component of the momentum integrals within the physical region $\delta(k_1^-+k_2^-+k_3^-p^-)$. The higher Fock state light front wave function correspond to the diagrams that is sub-leading in $1/N_c$. 

\begin{widetext}

\begin{equation}
\begin{aligned}
    &\frac{1}{4x_1x_2x_3}\Phi_{B}(x_1,x_2,x_3,k_{1\perp},k_{2\perp},k_{3\perp},s_1,s_2,s_3)=\\
    &-i(P^+)^2\int_{-\infty}^{\infty
    }\frac{dk_1^-dk_2^-}{(2\pi)^2}\frac{[\bar{u}_{s_1}\gamma^+]_i}{2k_1^+}\frac{[\bar{u}_{s_2}\gamma^+]_j}{2k_2^+}\frac{[\bar{u}_{s_3}\gamma^+]_k}{2k_3^+}\Psi_{B,ijk}(k_1,k_2,k_3;ps)\bigg|_{k_{1,2,3}^+=x_{1,2,3}P^+}\\
\end{aligned}
\end{equation}

Using the light front integral in \cite{Liu:2023fpj}, 
\begin{equation}
\begin{aligned}
\label{LF_integral}
    &(iP^+)^2\int_{-\infty}^{\infty
    }\frac{dk_1^-dk_2^-}{(2\pi)^2}\frac{\sqrt{\mathcal{F}(k_1)\mathcal{F}(k_2)\mathcal{F}(k_3)}}{(k_1^2-M^2)(k_2^2-M^2)(k_3^2-M^2)}\frac{1}{(k_1+k_2)^2-m_{qq}^2}\bigg|_{P^2=m_N^2}\\
    \rightarrow&\frac{\theta(x_1x_3)\theta(x_2x_3)}{4x_1x_2x_3}\frac{\mathcal{F}^{3/2}\left(\frac{k_{1\perp}^2}{\lambda_B x_1}+\frac{k_{2\perp}^2}{\lambda_B x_2}+\frac{k_{3\perp}^2}{\lambda_B x_3}\right)}{\bar{x}_3\left(m_B^2-\frac{k_{1\perp}^2+M^2}{x_1}-\frac{k_{2\perp}^2+M^2}{x_2}-\frac{k_{3\perp}^2+M^2}{x_3}\right)\left(\frac{k_{1\perp}^2+M^2}{x_1}+\frac{k_{2\perp}^2+M^2}{x_2}-\frac{k_{3\perp}^2+m_{qq}^2}{\bar{x}_3}\right)}
\end{aligned}
\end{equation}
\end{widetext}
where $x_1+x_2+x_3=1$ and $k_{1\perp}+k_{2\perp}+k_{3\perp}=0$. The parameter $\lambda_B$ is of order unity and fixed such that the light-cone wave functions are properly normalized with the normalization condition,

\begin{equation}
\label{eq:norm}
    \sum_{s_1s_2s_3}\int d[123] \left|\Phi_B(k_1,k_2,k_3,s_1,s_2,s_3)\right|^2=1
\end{equation}
where $(1,2,3)$ denotes the 3 constituent quark phase space $(x_1,x_2,x_3,k_{1\perp},k_{2\perp},k_{3\perp})$. For the nucleon, the parameter is fixed to $\lambda_N=2.26$.

Therefore, the nucleon light front wavefunction reads
\begin{widetext}
\begin{equation}
\begin{aligned}
\label{eq:LCWF_N}
\Phi_N=
&\frac1{\sqrt{6}}\phi_{N_s}(x_1,x_2,x_3,k_{1\perp},k_{2\perp},k_{3\perp})\left(\bar u_{s_1}\gamma^5C i\tau^2\bar u_{s_2}^T\right)\bar u_{s_3}\Gamma_su_N\\
       &+\frac1{\sqrt{6}}\phi_{N_a}(x_1,x_2,x_3,k_{1\perp},k_{2\perp},k_{3\perp})\left(g_{\mu\nu}-\frac{(k_{12})_\mu (k_{12})_\nu}{m_{qq_A}^2}\right)\left(\bar u_{s_1}\gamma^\mu Ci\tau^2\tau^a\bar u_{s_2}^T\right)\bar u_{s_3}\tau^a\Gamma_a^\nu u_N\\
    &+ S_3\mathrm{~permutation}~(1\leftrightarrow2\leftrightarrow3)
\end{aligned}
\end{equation}
and the delta baryon light front wavefunction reads

\begin{equation}
\begin{aligned}
\label{eq:LCWF_delt}
\Phi_\Delta=&\frac1{\sqrt{6}}\phi_{\Delta}(x_1,x_2,x_3,k_{1\perp},k_{2\perp},k_{3\perp})\left(g_{\mu\nu}-\frac{(k_{12})_\mu (k_{12})_\nu}{m_{qq_A}^2}\right)\left(\bar u_{s_1}\gamma^\mu Ci\tau^2\tau^a\bar u_{s_2}^T\right)\bar u_{s_3} u^\nu_\Delta\\
&+ S_3\mathrm{~permutation}~(1\leftrightarrow2\leftrightarrow3)
\end{aligned}
\end{equation}
where $u_{s_{1,2,3}}$ denotes the light front quark $4-$spinor (see Appendix~\ref{Appx:LFspinor}) and the momentum wave function is defined by
\begin{equation}
\begin{aligned}
\label{eq:WFNs}
&\phi_{N_s}(x_1,x_2,x_3,k_{1\perp},k_{2\perp},k_{3\perp})=\\
&\frac{\sqrt{Z_{N}}\lambda_{qq_S}^2\mathcal{F}\left(\frac{k_{1\perp}^2}{\lambda_N x_1}+\frac{k_{2\perp}^2}{\lambda_N x_2}+\frac{k_{3\perp}^2}{\lambda_N x_3}\right)}{\bar{x}_3\left(\frac{k_{1\perp}^2+M^2}{x_1}+\frac{k_{2\perp}^2+M^2}{x_2}-\frac{k_{3\perp}^2+m_{qq_S}^2}{\bar{x}_3}\right)\left(m_N^2-\frac{k_{1\perp}^2+M^2}{x_1}-\frac{k_{2\perp}^2+M^2}{x_2}-\frac{k_{3\perp}^2+M^2}{x_3}\right)}
\end{aligned}
\end{equation}
and
\begin{equation}
\begin{aligned}
\label{eq:WFNa}
&\phi_{N_a}(x_1,x_2,x_3,k_{1\perp},k_{2\perp},k_{3\perp})=\\
&\frac{-\sqrt{Z_{N}}\lambda_{qq_A}^2\mathcal{F}\left(\frac{k_{1\perp}^2}{\lambda_N x_1}+\frac{k_{2\perp}^2}{\lambda_N x_2}+\frac{k_{3\perp}^2}{\lambda_N x_3}\right)}{\bar{x}_3\left(\frac{k_{1\perp}^2+M^2}{x_1}+\frac{k_{2\perp}^2+M^2}{x_2}-\frac{k_{3\perp}^2+m_{qq_A}^2}{\bar{x}_3}\right)\left(m_N^2-\frac{k_{1\perp}^2+M^2}{x_1}-\frac{k_{2\perp}^2+M^2}{x_2}-\frac{k_{3\perp}^2+M^2}{x_3}\right)}
\end{aligned}
\end{equation}
and
\begin{equation}
\begin{aligned}
\label{eq:WFD}
&\phi_{\Delta}(x_1,x_2,x_3,k_{1\perp},k_{2\perp},k_{3\perp})=\\
&\frac{-\sqrt{Z_{\Delta}}\lambda_{qq_A}^2\mathcal{F}\left(\frac{k_{1\perp}^2}{\lambda_\Delta x_1}+\frac{k_{2\perp}^2}{\lambda_\Delta x_2}+\frac{k_{3\perp}^2}{\lambda_\Delta x_3}\right)}{\bar{x}_3\left(\frac{k_{1\perp}^2+M^2}{x_1}+\frac{k_{2\perp}^2+M^2}{x_2}-\frac{k_{3\perp}^2+m_{qq_A}^2}{\bar{x}_3}\right)\left(m_\Delta^2-\frac{k_{1\perp}^2+M^2}{x_1}-\frac{k_{2\perp}^2+M^2}{x_2}-\frac{k_{3\perp}^2+M^2}{x_3}\right)}
\end{aligned}
\end{equation}
\end{widetext}
with the normalization constant $\sqrt{Z_N}$ and $\sqrt{Z_\Delta}$ related to the wave function renormalization constant in \eqref{eq:BSFWF_N} and \eqref{eq:BSFWF_D}.

The resulting nucleon light-front wave function is written as a product of momentum, spin, and flavor components, which can be organized into the following structure
\begin{equation}
\Phi_N=\frac{1}{\sqrt{6}}\left(\Phi_N^{(s)}+\Phi_N^{(a)}+\text{$S_3$ permutations}\right),
\end{equation}
with
\begin{align}
\Phi_N^{(s)}(1,2,3)
&=\,
\phi_{Ns}(1,2,3)
S^{(s)}_{s_1s_2s_3}(1,2,3)F^{(s)}_{q_1q_2q_3},
\\
\Phi_N^{(a)}(1,2,3)
&=\,
\phi_{Na}(1,2,3)
S^{(a)}_{s_1s_2s_3}(1,2,3)F^{(a)}_{q_1q_2q_3},
\end{align}
where $S^{(s,a)}_{s_1s_2s_3}$ and $F^{(s,a)}_{q_1q_2q_3}$ denote the three-quark light-front spinor and flavor structures, respectively, defined in Eq.~\eqref{eq:LCWF_N}.

The momentum dependence is encoded in the wave functions \(\phi_{N_s}(1,2,3)\), \(\phi_{N_a}(1,2,3)\), and the spinor components $S^{(s,a)}_{s_1s_2s_3}$, which together describe the distribution of the three constituent quarks in longitudinal and transverse momentum space under the associated diquark channel. The spin structure is organized according to the diquark channel as well. The scalar-diquark contribution is represented by \(S^{(s)}_{s_1s_2s_3}=\bar{u}_{s_1}\gamma^5 C\bar{u}_{s_2}^T\,\bar{u}_{s_3}\Gamma_su_N\), whereas the axial-vector diquark contribution is described by \(S^{(a)}_{s_1s_2s_3}=\bar{u}_{s_1}\gamma^\mu C\bar{u}_{s_2}^T\,\bar{u}_{s_3}\Gamma_a^\nu u_N\). The corresponding flavor structures are \(F^{(s)}_{q_1q_2q_3}=i(\tau^2)_{q_1q_2} \mathbf{1}_{q_3\,N}\) for the flavor-singlet scalar diquark and \(F^{(a)}_{q_1q_2q_3}=i(\tau^2\tau^a)_{q_1q_2} \tau^a_{q_3\,N}\) for the flavor-triplet axial-vector diquark. Similarly, the delta baryon light front wave function is composed of momentum wave function \(\phi_{\Delta}(1,2,3)\), spin wave function \(\bar{u}_{s_1}\gamma^\mu C\bar{u}_{s_2}^T\,\bar{u}_{s_3} u^\nu_\Delta\), and  flavor component \(i(\tau^2\tau^a)_{q_1q_2}(\xi^a_{\frac32})_{q_3\,\Delta}\). Together, these components specify the momentum dependence and the spin-flavor organization of the three-quark baryon state.

The explicit wave functions written in the conventional basis of light front wave functions in \cite{Ji:2003yj} can be found in Appendix~\ref{app:LCWF}. 

The calculation demonstrates that in the ILM, the proton and neutron are strongly correlated quark-diquark states, with a  tight scalar-iso-scalar diquark $[ud]_{S}$ and weaker  axial-vector flavor-triplet  diquark ($[uu]_{A}$, $[ud]_{A}$, $[dd]_{A}$)~\cite{Schafer:1996wv}. This strong correlation in the scalar channel follows from the particle-anti-particle symmetry between the spin-$0$ pion and the spin-$0$ diquark. 

The normalization in Eq.~\eqref{eq:norm} can be decomposed into

\begin{widetext}
\begin{equation}
\begin{aligned}
P^{(ss)}
&=
\sum_{s_1s_2s_3}\int d[123]\,
\Phi_N^{(s)}(1,2,3)
\Bigl[\Phi_N^{(s)}(1,2,3)+\mathrm{perm.}\Bigr]^*
\\[1mm]
P^{(sa)}
&=
\sum_{s_1s_2s_3}\int d[123]\,
\Bigl\{
\Phi_N^{(s)}(1,2,3)
\Bigl[\Phi_N^{(a)}(1,2,3)+\mathrm{perm.}\Bigr]^*
+
\Phi_N^{(a)}(1,2,3)
\Bigl[\Phi_N^{(s)}(1,2,3)+\mathrm{perm.}\Bigr]^*
\Bigr\}
\\[1mm]
P^{(aa)}
&=
\sum_{s_1s_2s_3}\int d[123]\,
\Phi_N^{(a)}(1,2,3)
\Bigl[\Phi_N^{(a)}(1,2,3)+\mathrm{perm.}\Bigr]^*.
\end{aligned}
\end{equation}
\end{widetext}

Their numerical values are listed in Table~\ref{tab:norm}. By construction,
\begin{equation}
P^{(ss)}+P^{(sa)}+P^{(aa)}=1,
\end{equation}
as required by the normalization condition.

These contributions $P^{(ss)}$, $P^{(sa)}$, and $P^{(aa)}$, listed in Table~\ref{tab:norm}, should be distinguished from the diquark weights $w_s$ and $w_a$ in \eqref{eq:Z}. The latter are defined from the scalar and axial contributions to the quark--diquark pole normalization at nucleon mass and therefore characterize the nucleon in a fixed quark--diquark basis. Although the corresponding vertices already incorporate the exchange dynamics through the BSF equation, $w_s$ and $w_a$ do not contain the explicit overlaps among different three-quark permutations. On the other hand, \(P^{(ss)}\), \(P^{(sa)}\), and \(P^{(aa)}\) are obtained from the norm of the fully symmetrized three-quark wave function and therefore include diagonal and quark-exchange contributions between different pairings. In particular, recoupling of the spin--flavor structure on light front allows scalar and axial-vector configurations associated with different permutations to interfere, giving rise to $P^{(sa)}$. Consequently, the strong scalar-diquark dominance $w_s\gg w_a$ does not imply a corresponding hierarchy among $P^{(ss)}$, $P^{(sa)}$, and $P^{(aa)}$. Their comparable magnitudes reflect the permutation and quark-exchange structure of the fully symmetrized three-quark state, which redistributes the norms among these components and can generate a sizable scalar--axial interference contribution despite the small axial-vector diquark weight.




\begin{table}[]
    \centering
    \begin{tabular}{|c|c|c|}
    \hline
       $P^{(ss)}$  & $P^{(sa)}$ & $P^{(aa)}$ \\
    \hline
        0.305 & 0.350 & 0.339 \\
    \hline
    \end{tabular}
    \caption{Decomposition of the nucleon wave function normalization into scalar–scalar ($ss$), scalar–axial interference ($sa$), and axial–axial ($aa$) diquark contributions}
    \label{tab:norm}
\end{table}

Since the conventional baryon $SU(6)$ wave functions can be recast in a quark--diquark basis~\cite{Anselmino:1992vg} (and references therein), we also construct the corresponding $SU(6)$ baryon wave functions with details given in Appendix~\ref{App:SU6}, which provide a useful reference for comparison of the $SU(6)$ approximation.  In a relativistic description, spin and momentum are no longer separable: relativistic spin rotations and spin--momentum correlations mix the spin and orbital structures of the wave function. Consequently, $SU(6)$ is not expected to remain an exact symmetry, and departures from the $SU(6)$ limit provide a measure of relativistic and dynamical symmetry-breaking effects. 





\begin{table}
\begin{center}
\begin{tabular}{|c|c|c|}
   \hline
     &  $\sqrt{Z_{N}}$ & $\sqrt{Z_{\Delta}}$ \\
   \hline
   ILM & $3.80$ & $-$ \\
   \hline
\end{tabular}
\end{center}
    \caption{The wave function renormalization of nucleon and delta baryon wave function where we have normalize the nucleon vertex mixing coefficients $\alpha_{1,2,3}$ to 1 (See Appendix~\ref{Baryon}).}
    \label{tab:WFren}
\end{table}

\section{Nucleon axial charge}\label{app:gA}

In this section, we derive the nucleon isovector axial charge from the light-front nucleon wave function in Eq.~\eqref{eq:LCWF_N}. We use the standard definition of nucleon axial charge, given by

\begin{equation}
\begin{aligned}
&\langle N(P,S)|\bar \psi\gamma_{\mu}\gamma_5\frac{\tau^a}{2}\psi|N(P,S)\rangle\\
&=g_A\,\bar u_N(P,S)
\gamma_{\mu}\gamma_5\frac{\tau^a}{2}
 u_N(P,S),
\end{aligned}
\end{equation}
with the measurement from Particle Data Group \cite{ParticleDataGroup:2018ovx} indicating 
$g_A\simeq 1.27$. 
In light front formalism, 
since the plus component of the axial current is diagonal in quark helicity,
the axial current selects particle-number-conserving transitions. As a result, the matrix element can be expressed as a diagonal overlap of light-front wave functions within the same Fock sector. For the proton with helicity $S=+\frac12$,
\begin{equation}
 g_A
 =
 \frac{1}{P^+}
 \langle P,S=\tfrac12|\bar \psi\gamma^+\gamma_5\frac{\tau^3}{2}\psi|P,S=\tfrac12\rangle .
 \label{eq:gA_def_lf}
\end{equation}

Thus, using the one-body form of the good current in the leading 3-Fock sector and the normalization of Eq.~\eqref{eq:norm} yield the isovector axial charge $g_A$

\begin{widetext}

\begin{equation}
 g_A
 =
 \sum_{j=1}^3
 \sum_{s_1s_2s_3}
 \int d[123]\,
 \bigl(2s_j\bigr)\,\bigl(2t_j^{(3)}\bigr)
 \left|\Phi_{N_\uparrow}(k_1,k_2,k_3;s_1,s_2,s_3)\right|^2,
 \label{eq:gA_master}
\end{equation}
and singlet axial charge (total quark spin) $\Delta\Sigma$
\begin{equation}
\Delta\Sigma
 =
 \sum_{j=1}^3
 \sum_{s_1s_2s_3}
 \int d[123]\,
 \bigl(2s_j\bigr)\,
 \left|\Phi_{N_\uparrow}(k_1,k_2,k_3;s_1,s_2,s_3)\right|^2,
 \label{eq:Sq_master}
\end{equation}

\end{widetext}
where $2t_j^{(3)}=\delta_{q_j u}-\delta_{q_j d}$, i.e. $+1$ for a $u$ quark and $-1$ for a $d$ quark.  


From \eqref{eq:LCWF_N}, the resulting axial charge splits into
\begin{equation}
g_A=g_A^{(ss)}+g_A^{(sa)}+g_A^{(aa)}.
\label{eq:gA_split}
\end{equation}
where the piece from pure scalar component is 

\begin{widetext}
\begin{equation}
 g_A^{(ss)}
 =
 \sum_{j=1}^3\sum_{s_1s_2s_3}
 \int d[123]\,
 \bigl(2s_j\bigr)\bigl(2t^{(3)}_j\bigr)
 \Phi_N^{(s)}(1,2,3)\Bigl[\Phi_N^{(s)}(1,2,3)+\text{perm.}\Bigr]^*.
 \label{eq:gA_ss}
\end{equation}
\end{widetext}

Since the scalar diquark is a spin- and isospin-singlet, in the idealized nonrelativistic scalar-diquark limit the proton may be written as $[ud]_{0,0}u_\uparrow$, so that the isovector axial charge is carried entirely by the spectator quark. One therefore expects $g_A^{(ss)} \simeq 1$. However in the full three-quark state, this limit is modified by interference among different pairings and by relativistic transverse-momentum effects. In particular, relativistic spin--momentum correlations rotate the quark spin relative to its light-front helicity, while permutation terms generate interference between different quark--diquark pairings. These effects dynamically reduce $g_A^{(ss)}$ below its naive scalar-diquark value of unity.

For the axial-vector diquark component, it  generates the additional contribution to $g_A$
by $g_A^{(sa)}$ and $g_A^{(aa)}$ with
\begin{widetext}

\begin{equation}
 g_A^{(aa)}=
 \sum_{j=1}^3\sum_{s_1s_2s_3}\int d[123]\,
 \bigl(2s_j\bigr)\bigl(2t^{(3)}_j\bigr)
\Phi_{N_\uparrow}^{(a)}(1,2,3)\Bigl[\Phi_{N_\uparrow}^{(a)}(1,2,3)+\text{perm.}\Bigr]^*,
\end{equation}

\begin{equation}
\begin{aligned}
\label{eq:gA_aa_sa}
 g_A^{(sa)}
 =
 \Re\,
 \sum_{j=1}^3\sum_{s_1s_2s_3}\int d[123]\,
 \bigl(2s_j\bigr)\bigl(2t^{(3)}_j\bigr)&\Bigl\{
 \Phi_{N_\uparrow}^{(a)}(1,2,3)\Bigl[\Phi_{N_\uparrow}^{(s)}(1,2,3)+\,\text{perm.}\Bigr]^*
 \\
 &\,+\,\Phi_{N_\uparrow}^{(s)}(1,2,3)\Bigl[\Phi_{N_\uparrow}^{(a)}(1,2,3)+\,\text{perm.}\Bigr]^*\Bigl\}.
\end{aligned}
\end{equation}
\end{widetext}


From Eq.~\eqref{eq:SA_weight}, the scalar and axial diquark components contribute approximately 
\(
w_s=98\%\) and 
\(
w_a=2\%
\)
respectively within the present ILM solution, indicating that the nucleon is strongly scalar-diquark dominated in the ILM. Thus, in the nonrelativistic (NR) limit, where relativistic spin--momentum mixing is absent and permutation-induced interference among different quark--diquark pairings is neglected, the relative size of the axial-charge contributions is expected to follow
\begin{equation}
g_A^{(ss)}:g_A^{(sa)}:g_A^{(aa)}|_{\text{NR}}
\approx
1:2\sqrt{w_s w_a}:w_a,
\end{equation}

However, once the interference among different diquark pairing in a fully symmetrized three-quark configuration are taken into account, this simple scaling is substantially modified. The full result is therefore obtained by numerically evaluating Eqs.~\eqref{eq:gA_ss}--\eqref{eq:gA_aa_sa} using the explicit light-front wave functions given in Eq.~\eqref{eq:LCWF_N} and Appendix~\ref{app:LCWF}. 

As shown in Table~\ref{tab:gA_dec}, this demonstrates that the empirical axial charge can be reproduced with only a {\it few-percent} axial-vector diquark admixture, provided that the axial channel contributes coherently through scalar--axial interference and through interference among different quark--diquark pairings.

 \begin{table}[]
     \centering
     \begin{tabular}{|c|c|c|c|}
     \hline
         $g_A^{(ss)}$ & $g_A^{(sa)}$ & $g_A^{(aa)}$ & $g_A$ \\
        \hline
         0.443 & 0.491 & 0.331 & 1.265\\
     \hline
     \end{tabular}
     \caption{Diquark decomposition of axial charge in ILM with $m_N=939.53$ MeV, $\sqrt{Z_N}=3.8$, and $\lambda_N=2.26$.}
     \label{tab:gA_dec}
 \end{table}
 



In Table~\ref{tab:gA}, we compare our result for $g_A$ with the PDG value extracted from neutron $\beta$ decay~\cite{Gorchtein:2021fce}, the recent ETMC lattice-QCD result obtained at the physical pion mass with a continuum extrapolation~\cite{Alexandrou:2024ozj}, and two covariant Faddeev calculations. The first three-quark Faddeev calculation in~\cite{Eichmann:2011pv} is based on a rainbow--ladder interaction, yielding $g_A\simeq0.99$ which underestimates the empirical value by approximately $20$--$25\%$. Although rainbow--ladder preserves the axial Ward--Takahashi identity and incorporates dynamical chiral symmetry breaking at the quark level, it omits important chiral dynamical effect beyond rainbow--ladder approximation, including pion-cloud contributions. In contrast, the symmetry-preserving quark--diquark Faddeev treatment in ~\cite{Chen:2020wuq}, which incorporates strong scalar and axial-vector diquark correlations, obtains $g_A=1.25(3)$, close to our ILM estimation $g_A=1.265$ and the empirical value. The substantial improvement with diquark dynamics highlights the importance of nonperturbative diquark correlations in describing the nucleon's axial structure. Such correlations arise naturally in the ILM, where the strong attraction in the scalar-diquark channel is generated by the instanton-induced interaction, while the weaker attraction in the axial-vector channel arises from instanton--anti-instanton molecule interactions.

\begin{table}[]
     \centering
     \begin{tabular}{|c|c|c|c|c|c|}
     \hline
          & PDG \cite{Gorchtein:2021fce} & ETMC \cite{Alexandrou:2024ozj} & \cite{Eichmann:2011pv} & \cite{Chen:2020wuq} & ILM \\
        \hline
         $g_A$ & 1.2754(13) & 1.250(24) & 0.99 & 1.25(3) & 1.265\\
     \hline
     \end{tabular}
     \caption{Axial charge compared with the PDG value from neutron \(\beta\) decay \cite{Gorchtein:2021fce}, the recent ETMC lattice result at the physical pion mass with continuum extrapolation \cite{Alexandrou:2024ozj}, and quark--diquark Faddeev calculations in rainbow-ladder \cite{Eichmann:2011pv} and beyond-rainbow-ladder treatments incorporating diquark dynamics for dynamical chiral-symmetry-breaking effect \cite{Chen:2020wuq}.}
     \label{tab:gA}
 \end{table}



 

\section{Nucleon tensor charge}\label{app:gT}

Similarly, we derive the nucleon isovector tensor charge $g_T$ from the light-front nucleon wave function in Eq.~\eqref{eq:LCWF_N}, given by

\begin{equation}
\begin{aligned}
&\langle N(P,S)|\bar \psi i\sigma_{\mu\nu}\gamma_5\frac{\tau^a}{2}\psi|N(P,S)\rangle\\
&=
g_T\,\bar u_N(P,S)
i\sigma_{\mu\nu}\gamma_5\frac{\tau^a}{2}
 u_N(P,S),
 \end{aligned}
\end{equation}

In light front formalism, the matrix element of the leading-twist tensor operator \(\sigma^{+j}\) is nonzero for a transversely polarized nucleon or, equivalently, between nucleon states of opposite helicity. Using the one-body form of the good current in the leading 3-Fock sector, one obtains the isovector tensor charge $g_T$

\begin{widetext}

\begin{equation}
\begin{aligned}
 g_T
 =
 \sum_{j=1}^3
 \sum_{s_1s_2s_3}
 \int d[123]\,
 \,\bigl(2t_j^{(3)}\bigr)
&\Phi^*_{N_\downarrow}(k_1,k_2,k_3;s_1,\cdots,-s_j,\cdots,s_3)\Phi_{N_\uparrow}(k_1,k_2,k_3;s_1,\cdots,s_j,\cdots,s_3),
 \label{eq:gT_master}
\end{aligned}
\end{equation}
and singlet tensor charge (total quark tensor charge) $\delta\Sigma$
\begin{equation}
\begin{aligned}
 \delta \Sigma
 =\sum_{j=1}^3
 \sum_{s_1s_2s_3}
 \int d[123]\,
&\Phi^*_{N_\downarrow}(k_1,k_2,k_3;s_1,\cdots,-s_j,\cdots,s_3)\Phi_{N_\uparrow}(k_1,k_2,k_3;s_1,\cdots,s_j,\cdots,s_3),
 \label{eq:dSq_master}
\end{aligned}
\end{equation}

\end{widetext}

As shown in Table~\ref{tab:gT_dec}, similar to the decomposition in the axial charge, we present the tensor charge decomposition. 

 \begin{table}[]
     \centering
     \begin{tabular}{|c|c|c|c|c|}
     \hline
         & $g_T^{(ss)}$ & $g_T^{(sa)}$ & $g_T^{(aa)}$ & $g_T$ \\
        \hline
        $\mu_0=0.63$ GeV & 0.477 & 0.527 & 0.395 & 1.399 \\
     \hline
     \end{tabular}
     \caption{Diquark decomposition of tensor charge in ILM with $m_N=939.53$ MeV, $\sqrt{Z_N}=3.8$, and $\lambda_N=2.26$.}
     \label{tab:gT_dec}
 \end{table}
 



Unlike the vector and axial current, the quark tensor current is not conserved and therefore exhibits a nontrivial renormalization-scale dependence. In Table~\ref{tab:gT}, the ILM tensor charges are evolved from the model scale $\mu_0=0.63~\mathrm{GeV}$ to $\mu=2~\mathrm{GeV}$ using the NNLO anomalous dimension calculated in~\cite{Gracey:2000am}. The evolved results are compared with recent lattice QCD determinations by the  FLAG~\cite{FlavourLatticeAveragingGroupFLAG:2024oxs}, ETMC~\cite{Alexandrou:2024ozj} and PNDME~\cite{Park:2025rxi} collaborations. Both calculations include continuum extrapolations, with ETMC employing ensembles at the physical pion mass and PNDME performing a combined chiral and continuum extrapolation. We also compare our results with the global analysis by the JAM collaboration~\cite{Cocuzza:2023vqs}, which incorporates experimental data from Belle~\cite{Belle:2011cur,Belle:2017rwm}, HERMES~\cite{HERMES:2008mcr}, COMPASS~\cite{COMPASS:2023cgk}, and STAR~\cite{STAR:2015jkc,STAR:2017wsi} collaborations, as well as the lattice constraints from earlier calculations by ETMC \cite{Gupta:2018lvp} and PNDME \cite{Alexandrou:2021oih}. 

\begin{table*}
     \centering
     \begin{tabular}{|c|c|c|c|c|c|c|}
     \hline
          & FLAG \cite{FlavourLatticeAveragingGroupFLAG:2024oxs} & ETMC \cite{Alexandrou:2024ozj} & PNDME \cite{Park:2025rxi} & JAMDiFF \cite{Cocuzza:2023vqs,Cocuzza:2023oam} & ILM ($\alpha_s(\mu_0)=0.87$) & ILM ($\alpha_s(\mu_0)=1.78$) \\
        \hline
         $g_T$ & 0.993(15) & 0.955(29) & 0.977(32) & 0.91(2) & 1.051 & 0.820  \\
     \hline
     \end{tabular}
     \caption{Tensor charges compared with the  global analysis from JAM Collaboration~\cite{Cocuzza:2023vqs} and recent lattice QCD determinations by ETMC~\cite{Alexandrou:2024ozj}, performed at the physical pion mass with continuum extrapolation, and~\cite{Park:2025rxi}, where the calculation supersedes earlier PNDME results \cite{Gupta:2018lvp} and incorporates combined chiral–continuum–finite-volume extrapolations.}
     \label{tab:gT}
 \end{table*}
 
\section{Unpolarized parton distributions}





Unpolarizaed parton density distribution is defined as
\begin{widetext}
\begin{equation}
\begin{aligned}
    q(x)=&\int_{-\infty}^\infty\frac{dz^-}{2\pi}e^{ix P^+z^-}\langle N(P,S)|\bar{q}(0)\frac{\gamma^+}2 W(0,\xi^-)q(z^-)|N(P,S)\rangle\\
    =&\sum_{j=1}^3\int d[123]\sum_{s_1s_2s_3}\left|\Phi_N(1,2,3,s_1,s_2,s_3)\right|^2\delta(x_j-x)\delta_{q_jq}
\end{aligned}
\end{equation}

\end{widetext}
where $q=u,d$ denotes the quark flavor. The unpolarized non-singlet distribution is normalized according to the valence-quark number sum rule,
\begin{equation}
1=\int_0^1 dx\,\Bigl[u(x)-d(x)\Bigr].
\label{eq}
\end{equation}
with $\int_0^1 dx u(x)=2$ and $\int_0^1 dx d(x)=1$ for leading Fock state.

\begin{figure}
    \centering
    \includegraphics[width=\linewidth]{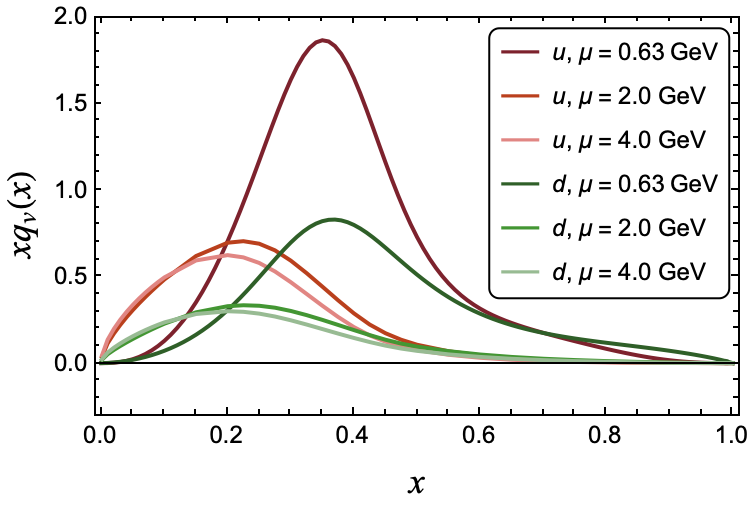}
    \caption{Comparison of the valence $u$ and $d$ quark momentum distributions at different scales, obtained using NNLO DGLAP evolution, evolved from $\mu_0=0.63~\mathrm{GeV}$ to $\mu=2~\mathrm{GeV}$.}
    \label{fig:DGLAP}
\end{figure}

In Fig.~\ref{fig:DGLAP}, we present the next-to-next-to-leading-order (NNLO) Dokshitzer–Gribov–Lipatov–Altarelli–Parisi (DGLAP) evolution of the valence $u$- and $d$-quark PDFs obtained in the ILM, from the initial scale $\mu_0=0.63~\mathrm{GeV}$, corresponding to a length scale comparable to the typical instanton size, to $\mu=2~\mathrm{GeV}$. The evolution is performed using the Higher Order Perturbative Parton Evolution Toolkit (HOPPET)~\cite{Salam:2008qg,Karlberg:2025hxk}, incorporating the charm-quark threshold and splitting functions up to three loops given in~\cite{Furmanski:1980cm,Curci:1980uw,Vogt:2004mw,Moch:2004pa}.

\begin{figure}
    \centering
\subfloat[\label{fig:u}]{\includegraphics[width=\linewidth]{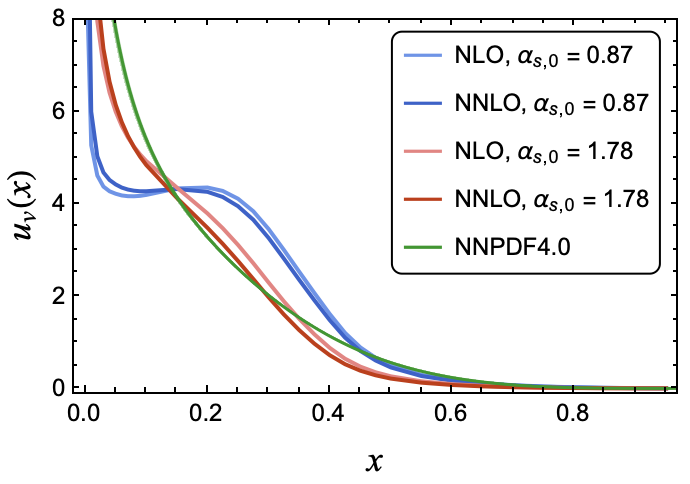}}
\hfill
\subfloat[\label{fig:xu}]{\includegraphics[width=\linewidth]{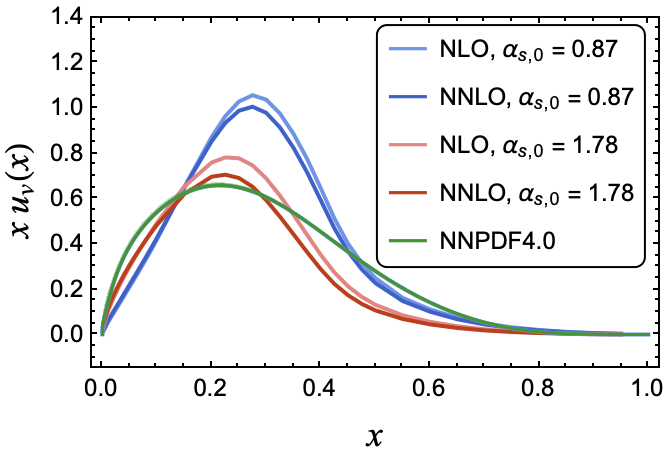}}
    \caption{ILM valence $u$ quark parton distribution $u_v=u-\bar{u}$ in proton evolved from $\mu_0=0.63$ GeV to $\mu=2$ GeV with $\alpha_{s}(\mu_0)=0.87$ (3-loop running) and $1.78$ (5-loop running), compared to NNPDF 4.0 global analysis \cite{NNPDF:2021njg}}
    \label{fig:uPDF}
\end{figure}

\begin{figure}
    \centering
\subfloat[\label{fig:d}]{\includegraphics[width=\linewidth]{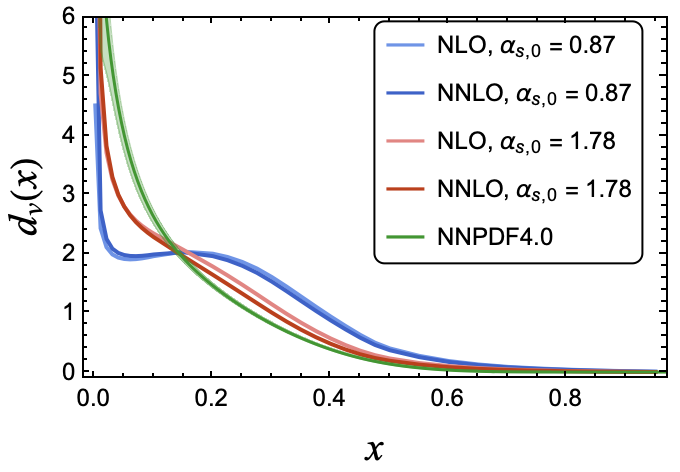}}
\hfill
\subfloat[\label{fig:xd}]{\includegraphics[width=\linewidth]{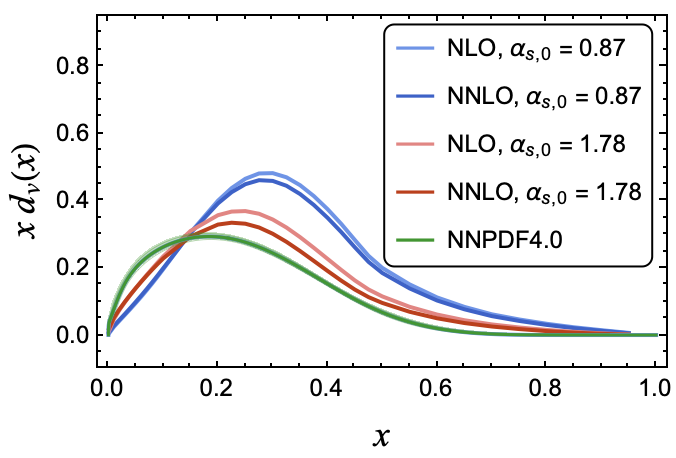}}
    \caption{ILM valence $d$ quark parton distribution in proton evolved from $\mu_0=0.63$ GeV to $\mu=2$ GeV with $\alpha_{s}(\mu_0)=0.87$ (3-loop running) and $1.78$ (5-loop running), compared to NNPDF 4.0 global analysis \cite{NNPDF:2021njg}}
    \label{fig:dPDF}
\end{figure}

In Figs.~\ref{fig:uPDF} and \ref{fig:dPDF}, we evolve the valence $u$ and $d$ quark PDFs obtained in the ILM from the initial scale $\mu_0=0.63~\mathrm{GeV}$ to the target scale $\mu=2.0~\mathrm{GeV}$. The evolved results are then compared with the global-fit valence $u$ and $d$ quark PDFs of the proton from NNPDF4.0 \cite{NNPDF:2021njg}. The evolution is performed using the HOPPET \cite{Salam:2008qg,Karlberg:2025hxk} with NLO and NNLO DGLAP evolution. To assess the sensitivity to the treatment of the running coupling at the low initial scale, we consider several values of $\alpha_s(\mu_0)$ obtained using different perturbative orders of the QCD $\beta$ function, evaluated with RunDec \cite{Herren:2017osy} using fixed $\alpha_s(M_Z)=0.118$ where $M_Z$ denotes the $Z$-boson mass. We obtain $\alpha_s(\mu_0)=0.87$ 
from three-loop running, $\alpha_s(\mu_0)=1.11$ 
from four-loop running, and $\alpha_s(\mu_0)=1.78$ 
from five-loop running. The results exhibit good convergence between the NLO and NNLO evolutions for the two choices of the running coupling, providing confidence in the reliability of the DGLAP evolution applied to the ILM distributions.

In Table~\ref{tab:moment}, we present the unpolarized second Mellin moments, corresponding to the quark gravitational charges \(A_q\), for the total \(u+\bar{u}\) and \(d+\bar{d}\) distributions given by

\begin{equation}
    \int_0^1 dx x[q(x)+\bar{q}(x)] = \langle x\rangle_q
\end{equation}

The results are evolved from the model scale \(\mu_0=0.63~\mathrm{GeV}\) to \(\mu=2~\mathrm{GeV}\) using the NNLO anomalous dimensions and fixed flavor $N_f=3$ implemented with the EKO evolution package~\cite{Candido:2022tld}. For the running coupling, we consider  \(\alpha_s(0.63~\mathrm{GeV})=0.87\) (3-loop running) and \(1.78\) (5-loop running), respectively for the initial scale. The evolved moments are compared with recent lattice-QCD results from ETMC \cite{Alexandrou:2026oks} and with global-fit results from the MSHT20 \cite{Bailey:2020ooq} and NNPDF4.0 \cite{NNPDF:2021njg} collaborations. We also compare our results with the calculation of the Dyson-Schwinger equation (DSE) based on the rainbow-ladder truncation in~\cite{Bednar:2018htv}, where a covariant Faddeev equation with scalar and axial-vector diquark is used for nucleon bound-state amplitude, supplemented by phenomenological pion-cloud contributions to estimate the rest of sea-quark and gluon momentum fractions.

\begin{table*}
    \centering
    \begin{tabular}{|c|c|c|c|}
    \hline
       & $\langle x\rangle_u$ & $\langle x\rangle_d$ \\
    \hline
    ILM ($0.63$ GeV) & 0.653 & 0.351 \\
    \hline
    ILM ($2$ GeV, $\alpha_s(\mu_0)=0.87$) & 0.416 & 0.234 \\
    \hline
    ILM ($2$ GeV, $\alpha_s(\mu_0)=1.78$) & 0.321 & 0.193 \\
    \hline
       DSE $(S+A+\pi)$ \cite{Bednar:2018htv} 
       & 0.2894 
       & 0.1739 \\
    \hline
    ETMC \cite{Alexandrou:2026oks} 
    & 0.357(17)(11)(3)
    & 0.192(16)(2)(2) \\
    \hline
    MSHT20 \cite{Bailey:2020ooq} & 
    0.347 & 
    0.193 \\
    \hline
    NNPDF4.0 \cite{NNPDF:2021njg} 
    & 0.347 
    & 0.194 \\
    \hline
    \end{tabular}
    \caption{Unpolarized second Mellin moment (gravitational charge $A_q$) of valence $u$ and $d$ quark.}
    \label{tab:moment}
\end{table*}

\section{Helicity parton distribution}

The helicity parton distribution is given by

\begin{widetext}
    \begin{equation}
\begin{aligned}
    \Delta q(x)=&\int_{-\infty}^\infty\frac{dz^-}{2\pi}e^{ix P^+z^-}\langle N(P,S)|\bar{q}(0)\frac{\gamma^+\gamma^5}2 W(0,\xi^-)q(z^-)|N(P,S)\rangle\\
    =&\sum_{j=1}^3\int d[123]\sum_{s_1s_2s_3}\left|\Phi_{N_\uparrow}(k_1,k_2,k_3,s_1,s_2,s_3)\right|^2\delta(x_j-x)\delta_{q_jq}2s_j
\end{aligned}
\end{equation}
\end{widetext}
where $2s_j=\pm1$ is the quark helicity eigenvalue.

The corresponding sum rule for polarized isovector distributions yields the axial charge,
\begin{equation}
g_A=\int_0^1 dx\,\Bigl[\Delta u(x)-\Delta d(x)\Bigr],
\label{eq:gA_Dq}
\end{equation}
which is RG invariant as a consequence of the PCAC relation.

\begin{table}[]
    \centering
    \begin{tabular}{|c|c|c|}
\hline
      & $\Delta u$   & $\Delta d$  \\
\hline
ILM ($0.63$ GeV) & 1.118 & -0.147 \\
\hline
ILM ($2$ GeV, $\alpha_s(\mu_0)=0.87$) & 1.072 & -0.193 \\
\hline
ILM ($2$ GeV, $\alpha_s(\mu_0)=1.78$) & 1.017 & -0.248 \\
\hline
ETMC \cite{Alexandrou:2024ozj} & 0.832(28) & -0.417(22) \\
\hline
NNPDFpol2.0 \cite{Cruz-Martinez:2025ahf} & 0.812 & -0.456 \\
\hline
BDSSV \cite{Borsa:2024mss} & 0.937 & -0.279 \\
\hline
    \end{tabular}
    \caption{Intrinsic quark spin given by \eqref{eq:gA_master} and \eqref{eq:Sq_master} using $\Delta u=(\Delta\Sigma+g_A)/2$ and $\Delta d=(\Delta\Sigma-g_A)/2$. The results are evolved to $\mu=2$ GeV with $\alpha_{s}(\mu_0)=0.87$ (3-loop running) and $1.78$ (5-loop running) and compared to lattice calculation from ETMC collaboration \cite{Alexandrou:2024ozj} at the physical pion mass with continuum extrapolation and global analyses from NNPDFpol2.0 \cite{Cruz-Martinez:2025ahf} and BDSSV \cite{Borsa:2024mss}.}
    \label{tab:quark spin}
\end{table}

In Table~\ref{tab:quark spin}, we present the quark helicity for both the $u$ and $d$ quarks. For each flavor, the corresponding first moment is defined as
\begin{equation}
\Delta q(\mu)
=
\int_0^1 dx\,
\left[
\Delta q(x,\mu)+\Delta \bar{q}(x,\mu)
\right],
\label{eq:Dq}
\end{equation}
and therefore includes both the quark and antiquark contributions. These quantities correspond to the flavor-diagonal axial charges and characterize the net helicity carried by quarks and antiquarks of a given flavor. The ILM results are evolved from the model scale $\mu_0=0.63~\mathrm{GeV}$ to $\mu=2~\mathrm{GeV}$ using NNLO anomalous dimensions and fixed flavor $N_f=3$ implemented within the EKO evolution package~\cite{Candido:2022tld}. Compared with phenomenological analyses and lattice results, the evolved ILM result for $\Delta u$ is slightly larger, whereas that for $\Delta d$ is somewhat smaller. A possible source of these deviations is the restriction of the present calculation to the leading Fock component of the nucleon. Higher Fock components in ILM including additional $q\bar q$ degrees of freedom provide nonperturbative sea contributions that are absent at the model scale in the present treatment. Their inclusion can modify the sea quark helicity distributions and, consequently, their first moments after QCD evolution.

\begin{figure}
    \centering
\subfloat[]{\includegraphics[width=\linewidth]{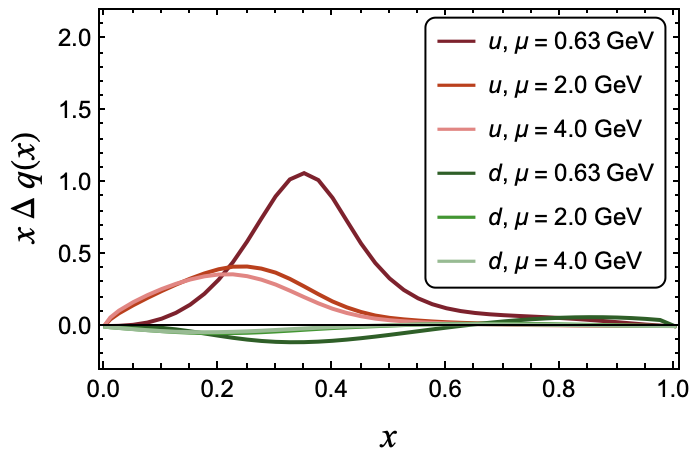}}
\hfill
\subfloat[\label{fig:xDqb}]{\includegraphics[width=\linewidth]{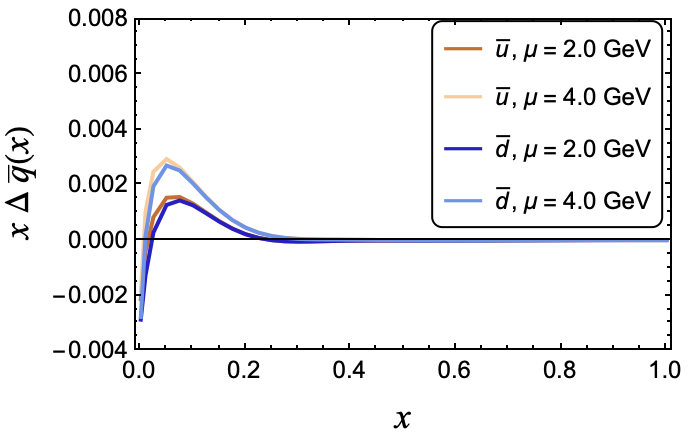}} 
    \caption{Comparison of the helicity distributions at different scales for (a) quarks, $x\Delta u$ and $x\Delta d$, and (b) antiquarks, $x\Delta\bar{u}$ and $x\Delta\bar{d}$. The distributions are evolved from the initial scale $\mu_0=0.63~\mathrm{GeV}$ to $\mu=2~\mathrm{GeV}$ using NLO DGLAP evolution implemented in HOPPET \cite{Salam:2008qg,Karlberg:2025hxk}.}
    \label{fig:helic_ILM}
\end{figure}

\begin{figure}
    \centering
\subfloat[]{\includegraphics[width=\linewidth]{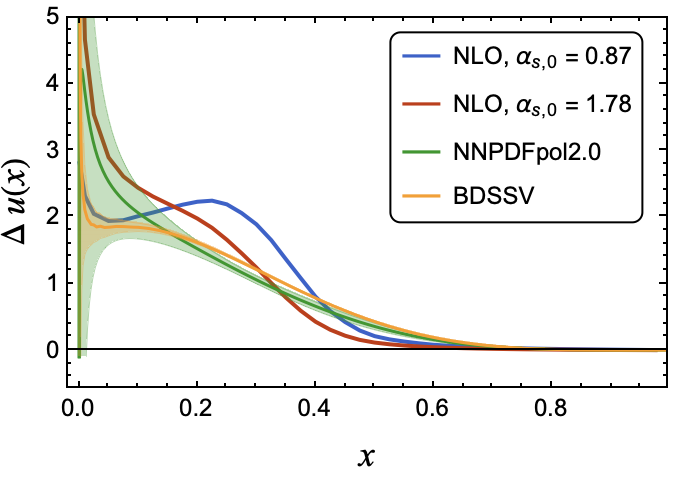}}
\hfill
\subfloat[]{\includegraphics[width=\linewidth]{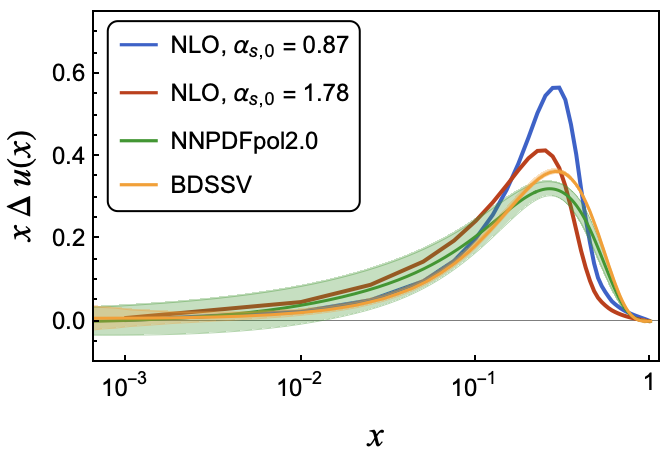}} 
    \caption{ILM predictions for the proton $u$-quark helicity distributions, (a) $\Delta u$ and (b) $x\Delta u$, evolved from the initial scale $\mu_0=0.63~\mathrm{GeV}$ to $\mu=2~\mathrm{GeV}$ with $\alpha_s(\mu_0)$ ranging from $0.87$ to $1.78$, compared with the NNPDFpol2.0 \cite{Cruz-Martinez:2025ahf} and BDSSV \cite{Borsa:2024mss} global analyses.}
    \label{fig:Du}
\end{figure}

\begin{figure}
    \centering
\subfloat[]{\includegraphics[width=\linewidth]{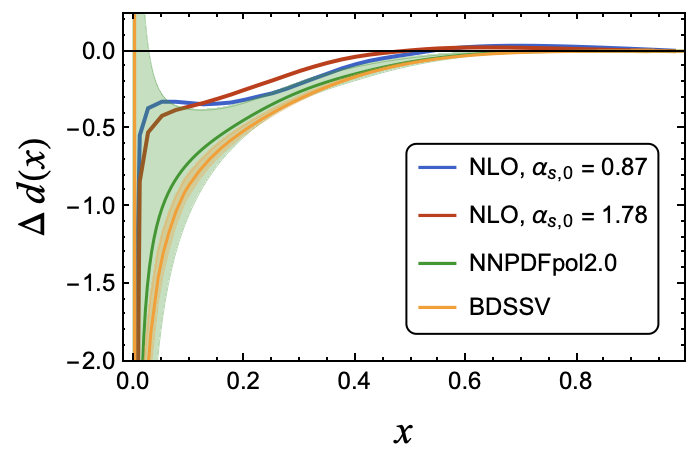}}
\hfill
\subfloat[]{\includegraphics[width=\linewidth]{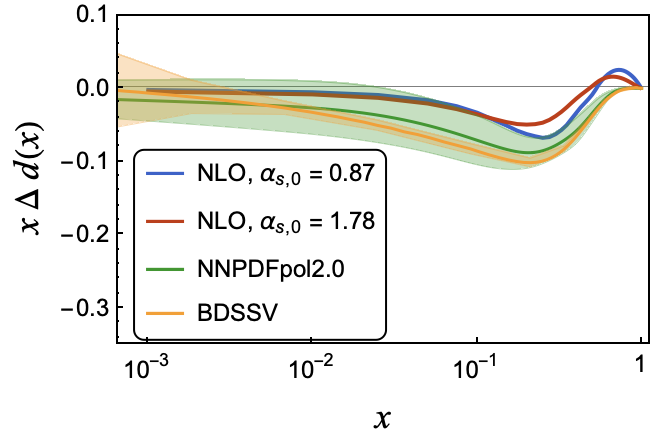}} 
    \caption{ILM $d$ quark helicity distribution in proton evolved from $\mu_0=0.63$ GeV to $\mu=2$ GeV with $\alpha_{s}(\mu_0)=0.87$ (3-loop running) and $1.78$ (5-loop running), compared to NNPDFpol2.0 \cite{Cruz-Martinez:2025ahf} and BDSSV \cite{Borsa:2024mss} global analysis.}
    \label{fig:Dd}
\end{figure}

In Fig.~\ref{fig:helic_ILM}, we present the quark and antiquark helicity PDFs for $u$ and $d$ obtained in the ILM, evolved from the initial scale $\mu_0=0.63~\mathrm{GeV}$ to $\mu=2~\mathrm{GeV}$ using the HOPPET package \cite{Salam:2008qg,Karlberg:2025hxk} with splitting functions implemented up to two loops (NLO) given in \cite{Mertig:1995ny,Vogelsang:1996im}. Although the three-loop helicity splitting functions are available in the literature \cite{Moch:2014sna}, they have not yet been implemented in HOPPET \cite{Karlberg:2025hxk}. The nearly identical $\bar u$ and $\bar d$ helicity distributions can be understood from their perturbative generation through DGLAP evolution. Since our calculation includes only the leading three-quark Fock component of the nucleon, no antiquark helicity distributions are present at the initial model scale. During evolution, the polarized sea is generated predominantly through gluon splitting (\(g\to q\bar q\)), which contributes already at leading order. The flavor independence of this splitting naturally leads to approximately equal \(\bar u\) and \(\bar d\) helicity distributions, with their rapid rise toward small \(x\) further reflecting their gluon-induced origin.


In Fig.~\ref{fig:Du} and \ref{fig:Dd}, we evolve the $u$ and $d$ quark helicity PDFs obtained in the ILM from the initial scale $\mu_0=0.63~\mathrm{GeV}$ to the target scale $\mu=2.0~\mathrm{GeV}$, using HOPPET package with NLO polarized DGLAP evolution. To assess the sensitivity of the evolution to the running coupling at the low initial scale, we consider three- and five-loop running, yielding \(\alpha_s(\mu_0)=0.87\) and \(1.78\), respectively. The evolved results are then compared with the global-fit $u$ and $d$ helicity quark PDFs of the proton from NNPDFpol2.0 \cite{Cruz-Martinez:2025ahf} and BDSSV \cite{Borsa:2024mss} with fixed $\alpha_s(M_Z)=0.118$.  Similarly, the evolved ILM helicity distributions are slightly larger for $u$ quarks and smaller for $d$ quarks, reflecting the absence of higher Fock components that include additional \(q\bar q\) pairs. These nonperturbative sea contributions could redistribute the quark helicity by reducing the \(u\) quark contribution while enhancing that of the \(d\) quark.

\begin{figure}
    \centering
\subfloat[]{\includegraphics[width=\linewidth]{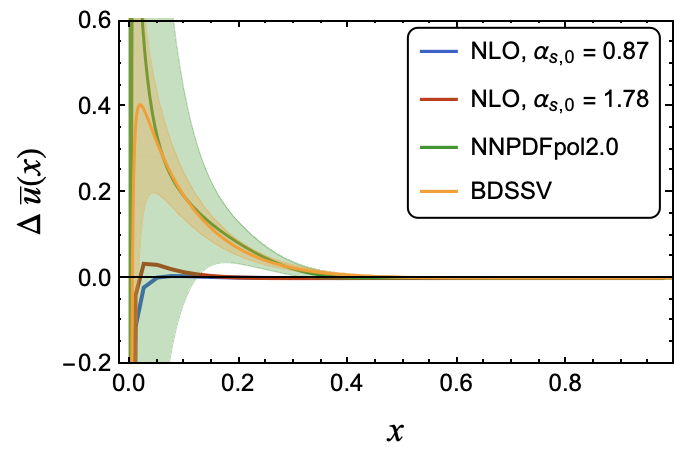}}
\hfill
\subfloat[]{\includegraphics[width=\linewidth]{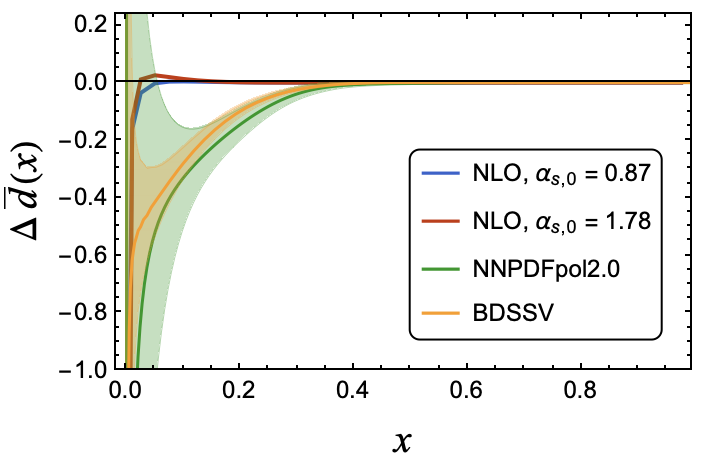}}
    \caption{ILM $\bar u$ and $\bar d$ antiquark helicity distribution in proton evolved from $\mu_0=0.63$ GeV to $\mu=2$ GeV with $\alpha_{s}(\mu_0)=0.87$ (3-loop running) and $1.78$ (5-loop running), compared to NNPDFpol2.0 \cite{Cruz-Martinez:2025ahf} and BDSSV \cite{Borsa:2024mss} global analysis.}
    \label{fig:D_antiq}
\end{figure}

In Fig.~\ref{fig:D_antiq}, we evolve the $\bar u$ and $\bar d$ antiquark helicity PDFs obtained in the ILM from the initial scale $\mu_0=0.63~\mathrm{GeV}$ to the target scale $\mu=2.0~\mathrm{GeV}$, using NLO polarized DGLAP evolution from HOPPET package with two choices of the running coupling, $\alpha_s(\mu_0)=0.87$ for three-loop running beta function and $\alpha_s(\mu_0)=1.78$ for the five-loop running beta function. In contrast to the flavor-symmetric antiquark helicity distributions generated from the leading-Fock-state nucleon in ILM through perturbative evolution as shown in Fig.~\ref{fig:xDqb}, the global analyses that are constrained in particular by flavor-sensitive polarized $W^\pm$ production data, exhibit a sizable flavor asymmetry in the polarized light-quark sea. Such a discrepancy may indicate nonperturbative sea-quark contributions that are absent in the present leading-Fock-state treatment.

The ratio between the unpolarized and helicity PDFs probes the relativistic spin-flavor structure of the nucleon. The helicity distributions satisfy the positivity bounds, \(|\Delta q(x)|\leq q(x)\) and \(|\Delta\bar q(x)|\leq\bar q(x)\), reflecting the non-negativity of quark and antiquark helicity densities. At the low resolution scale, we find the endpoint ($x\to1$) behavior
\begin{equation}
\begin{aligned}
\Delta u(x)\sim u(x)&\sim 4.397(1-x),\\
\Delta d(x)\sim d(x)&\sim 8.794(1-x),
\end{aligned}
\end{equation}
implying \(\Delta u/u\to1\) and \(\Delta d/d\to1\), which saturate the positivity bound and indicate hadron helicity conservation at large $x$ \cite{Farrar:1975yb,Brodsky:1994kg}. Furthermore, the unpolarized flavor ratio approaches \(d/u\to2\), consistent with the $SU(6)$ Clebsch–Gordan probabilities of $2/3$ and $1/3$ for the $uu$ and symmetric $ud$ axial-vector spectator configurations respectively (see Appendix~\ref{App:SU6}). Although the value deviates from pQCD prediction \cite{Farrar:1975yb,Brodsky:1994kg}, the value is on ly valid at low resolution and should change over the scale and evolution. 

These endpoint limits arise from the dominance of the axial diquark component, particularly its longitudinal polarization, despite the strong scalar-diquark attraction in the nonrelativistic regime. Although the scalar and axial radial wave functions share the same leading endpoint scaling, the spinor overlap from the scalar component is suppressed by an additional factor of \((1-x)^2\), whereas the overlap from axial component remains finite. The dominant longitudinal axial component further favors alignment of the struck-quark helicity with the nucleon. Thus, the endpoint flavor and helicity ratios reflect the combined effects of isospin coupling and relativistic spin correlations encoded in the nucleon Faddeev amplitudes. 

Similarly, the higher Fock state encodes the major information of sea quark at low resolution, which are not included in the present leading-Fock-state approximation. This has already been observed in our results for unpolarized PDF and helicity PDF. One should expect the deviation from global analysis is reduced once the higher Fock-state contributions are included. 

\begin{figure}
    \centering
    \includegraphics[width=\linewidth]{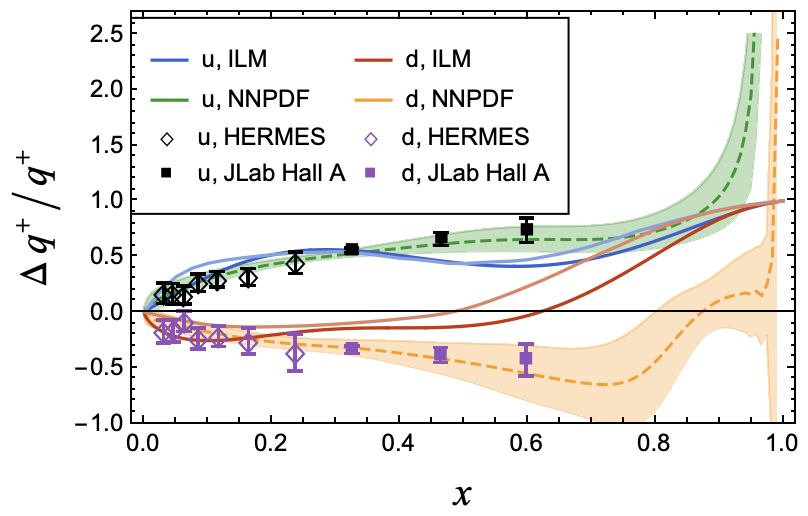}
    \caption{The ratio of the ILM polarized to unpolarized PDF evolved from $\mu_0=0.63$ GeV to $\mu=2$ GeV with $\alpha_{s}(\mu_0)=1.78$ (5-loop running), compared to experimental data from HERMES \cite{HERMES:2003gbu,Tung:2002vr} and JLab  6 GeV program from Hall A collaboration \cite{JeffersonLabHallA:2004tea}, and global analysis obtained from NNPDFpol2.0 \cite{Cruz-Martinez:2025ahf} and NNPDF4.0 \cite{NNPDF:2021njg} where $q^+=q(x)+\bar{q}(x)$ and $\Delta q^+=\Delta q(x)+\Delta\bar{q}(x)$}
    \label{fig:ratio}
\end{figure}

In Fig.~\ref{fig:ratio}, we compare our results with experimental data from HERMES (diamond) \cite{HERMES:2003gbu,Tung:2002vr} and JLab 6 GeV program from Hall A collaboration (filled square) \cite{JeffersonLabHallA:2004tea}, as well as the NNPDF global analysis obtained from NNPDFpol2.0 \cite{Cruz-Martinez:2025ahf} and NNPDF4.0 \cite{NNPDF:2021njg}. The blue and red curves represent the $u$- and $d$-quark helicity ratios respectively, with darker curves showing the distributions at the initial model scale ($\mu_0=0.63$ GeV) and lighter curves showing those evolved to ($\mu=2\,\mathrm{GeV}$). The green and orange curves correspond to the NNLO results obtained from separate NNPDF global analyses of polarized and unpolarized PDFs. Since the helicity ratios are sensitive to pQCD evolution and the corresponding renormalization scheme, ratios obtained from independently fitted polarized and unpolarized distributions are not necessarily constrained to satisfy the positivity bound. Consequently, apparent violations may arise, particularly in the poorly constrained large-$x$ region.

\section{Transversity parton distribution}

The transversity parton distribution is given by

\begin{widetext}
    \begin{equation}
\begin{aligned}
    &\delta q(x)=\int_{-\infty}^\infty\frac{dz^-}{2\pi}e^{ix P^+z^-}\langle N(P,S)|\bar{q}(0)\frac{1}2i\sigma^{+j}\gamma^5 W(0,\xi^-)q(z^-)|N(P,S)\rangle\\
    =&\sum_{j=1}^3\int d[123]\sum_{s_1s_2s_3}\Phi_{N_\uparrow}(k_1,k_2,k_3,s_1,\cdots, s_j,\cdots,s_3)\Phi^*_{N_\downarrow}(k_1,k_2,k_3,s_1,\cdots, -s_j,\cdots,s_3)\delta(x_j-x)\delta_{q_jq}
\end{aligned}
\end{equation}
\end{widetext}

The corresponding sum rule for transversely polarized isovector distributions yields the tensor charge,
\begin{equation}
g_T=\int_0^1 dx\,\Bigl[\delta u(x)-\delta d(x)\Bigr],
\label{eq:gA_Dq}
\end{equation}
which runs with renormalization scale but does not mix with gluon contribution. 

The transversity distribution is further constrained by the unpolarized and helicity PDFs through the Soffer bound \cite{Soffer:1994ww},

\begin{equation}
    |\delta q(x)| \leq\frac12\left[q(x)+\Delta q(x)\right].
\end{equation}

In Table~\ref{tab:quark transversity}, we present the quark transversity for both the $u$ and $d$ quarks. For each flavor, the corresponding first moment is defined as
\begin{equation}
\delta q(\mu)
=
\int_0^1 dx\,
\left[
\delta q(x,\mu)-\delta \bar{q}(x,\mu)
\right],
\label{eq:Dq}
\end{equation}
and therefore includes both the quark and antiquark contributions. These quantities correspond to the flavor-diagonal tensor charges and characterize the net transversity carried by quarks and antiquarks of a given flavor. The ILM results are evolved from the model scale $\mu_0=0.63~\mathrm{GeV}$ to $\mu=2~\mathrm{GeV}$ using NNLO anomalous dimensions given in \cite{Gracey:2000am}. The evolved tensor charges obtained in this work are compared with recent lattice QCD determinations with continuum extrapolation by the ETMC~\cite{Alexandrou:2024ozj} at the physical pion mass and PNDME~\cite{Park:2025rxi} using chiral extrapolation, as well as the global-fit results from the JAM Collaboration~\cite{Cocuzza:2023vqs}. The JAM analysis incorporates transversity-sensitive experimental data from Belle~\cite{Belle:2011cur,Belle:2017rwm}, HERMES~\cite{HERMES:2008mcr}, COMPASS~\cite{COMPASS:2023cgk}, and STAR~\cite{STAR:2015jkc,STAR:2017wsi}, supplemented by lattice QCD constraints on the tensor charges. Since the quark tensor currents do not mix with gluons and share the same multiplicative anomalous dimension, the ratio $\delta u/\delta d$ is RG invariant. We also compare this ratio with other determinations in Table~\ref{tab:delta_ratio}.

\begin{table}[]
    \centering
    \begin{tabular}{|c|c|c|}
\hline
      & $\delta u$   & $\delta d$  \\
\hline
ILM ($0.63$ GeV) & 1.084 & -0.316 \\
\hline
ILM ($2$ GeV, $\alpha_s(\mu_0)=0.87$) & 0.814  & -0.237 \\
\hline
ILM ($2$ GeV, $\alpha_s(\mu_0)=1.78$) & 0.635  & -0.185  \\
\hline
ETMC \cite{Alexandrou:2024ozj} & 0.756(29) & -0.196(12) \\
\hline
PNDME \cite{Park:2025rxi} & 0.782(28) & -0.195(16) \\
\hline
JAMDiFF \cite{Cocuzza:2023vqs} & 0.71(2) & -0.200(6) \\
\hline
    \end{tabular}
    \caption{Quark transversity given by \eqref{eq:gT_master} and \eqref{eq:dSq_master} using $\delta u=(\delta\Sigma+g_T)/2$ and $\delta d=(\delta\Sigma-g_T)/2$. The results are evolved to $\mu=2$ GeV with $\alpha_{s}(\mu_0)=0.87$ (3-loop running) and $1.78$ (5-loop running) and compared to the recent lattice QCD determinations by the ETMC~\cite{Alexandrou:2024ozj} and PNDME~\cite{Park:2025rxi}, as well as the $2$ GeV global-fit results from the JAM Collaboration~\cite{Cocuzza:2023vqs}.}
    \label{tab:quark transversity}
\end{table}

\begin{table}[]
    \centering
    \begin{tabular}{|c|c|c|c|}
\hline
      & ILM  & ETMC~\cite{Alexandrou:2024ozj} & PNDME~\cite{Park:2025rxi} \\
\hline
$\delta d /\delta u$ & 0.292  & 0.259(19) & 0.249(22) \\
\hline
    \end{tabular}
    \caption{RG-invariant ratio of quark transversity compared to the recent lattice QCD determinations by the ETMC~\cite{Alexandrou:2024ozj} and PNDME~\cite{Park:2025rxi}.}
    \label{tab:delta_ratio}
\end{table}

\begin{figure}
    \centering
\subfloat[]{\includegraphics[width=\linewidth]{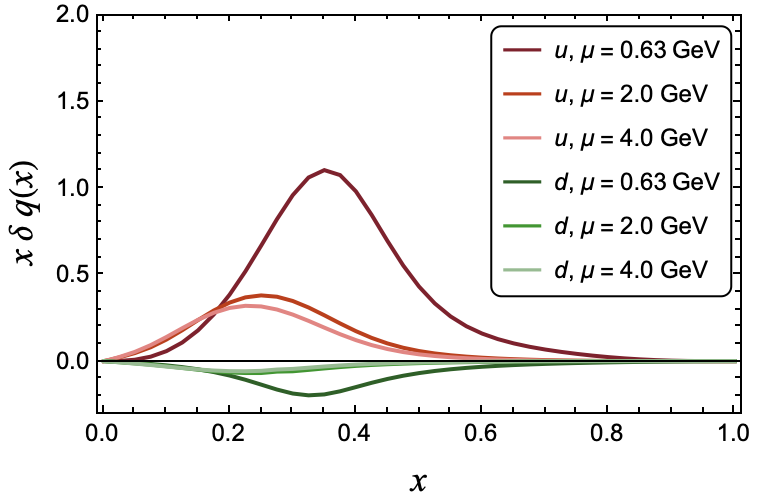}}
\hfill
\subfloat[\label{fig:xdqb}]{\includegraphics[width=\linewidth]{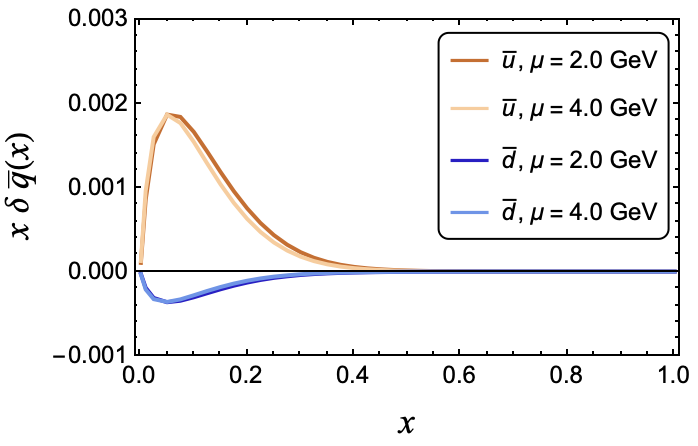}} 
    \caption{Comparison of the $x\delta u$ and $x\delta d$ quark helicity distributions at different scales, obtained using NLO DGLAP evolution, evolved from $\mu_0=0.63~\mathrm{GeV}$ to $\mu=2~\mathrm{GeV}$.}
    \label{fig:trans_ILM}
\end{figure}

In Fig.~\ref{fig:trans_ILM}, we present the NLO DGLAP evolution of the $u$ and $d$ quark and antiquark transversity PDFs obtained in the ILM, evolved from the initial scale $\mu_0=0.63~\mathrm{GeV}$ to $\mu=2~\mathrm{GeV}$, using tParton package \cite{Sha:2024thh} with fixed flavor $N_f=3$ and splitting functions given in \cite{Manashov:2024fcd}. As shown in Fig.~\ref{fig:xdqb}, transversity distributions evolve without mixing with gluons due to its chiral-odd property. In our ILM calculation, the nucleon is described by its leading three-quark Fock component, with zero antiquark distributions at the initial model scale. Thus, the nonzero antiquark transversity distributions at higher scales arise entirely from perturbative quark–antiquark spliting, which first enters at NLO. The evolved $\bar{u}$ and $\bar{d}$ transversity distributions retain the sign of their initial valence-quark distributions, reflecting the spin-flavor structure inherited from the initial ILM wave function.

\begin{figure}
    \centering
\subfloat[]{\includegraphics[width=\linewidth]{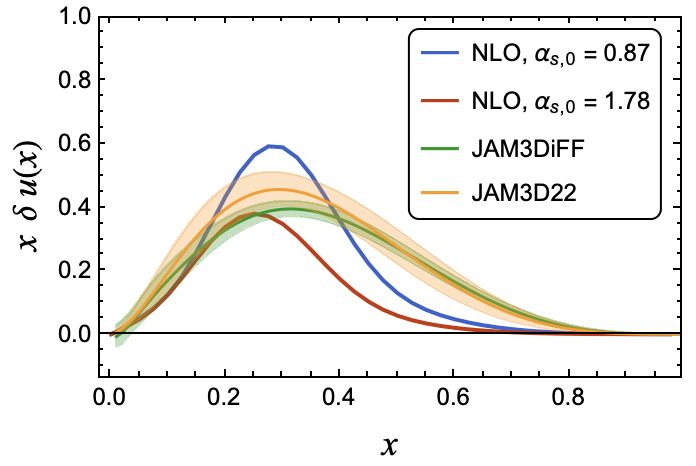}}
\hfill
\subfloat[]{\includegraphics[width=\linewidth]{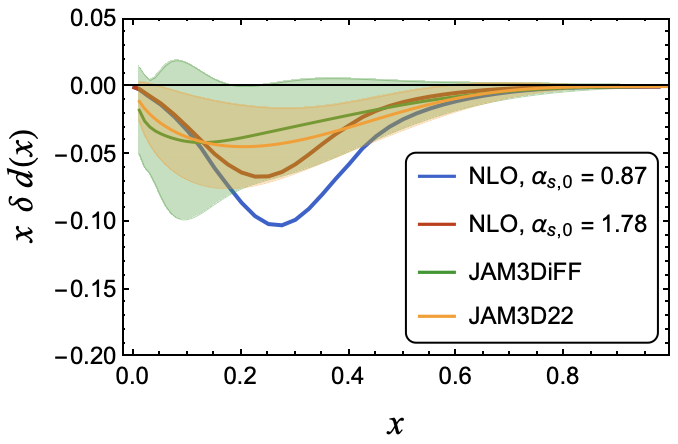}}
    \caption{ILM transversity quark distribution for (a) $u$ and (b) $d$ in proton evolved from $\mu_0=0.63$ GeV to $\mu=2$ GeV with $\alpha_{s}(\mu_0)=0.87$ (3-loop running) and $1.78$ (5-loop running), compared to JAM3D \cite{Gamberg:2022kdb} and JAMDiFF \cite{Cocuzza:2023vqs,Cocuzza:2023oam} global analyses.}
    \label{fig:dq}
\end{figure}

In Fig.~\ref{fig:dq}, we evolve the $u$ and $d$ quark transversity PDFs obtained in the ILM from the initial scale $\mu_0=0.63~\mathrm{GeV}$ to the target scale $\mu=2.0~\mathrm{GeV}$, using tParton package with NLO DGLAP evolution. We employ three- and five-loop running couplings, with \(\alpha_s(\mu_0)=0.87\) and $1.78$, respectively, to examine the sensitivity of the evolution to the low resolution scale. The evolved results are then compared with the global-fit valence $u$ and $d$ quark PDFs of the proton from JAM3D22 \cite{Gamberg:2022kdb} and JAMDiFF \cite{Cocuzza:2023vqs,Cocuzza:2023oam} global analyses. In these two global analyses,  JAM3D22 neglects antiquark transversity, whereas JAMDiFF allows for a nonzero antiquark parameterization under the assumption \(\delta\bar u_{\rm JAMDiFF}(x)=-\delta\bar d_{\rm JAMDiFF}(x)\) motivated by large-\(N_c\) QCD predictions.

\begin{figure}
    \centering
\subfloat[]{\includegraphics[width=\linewidth]{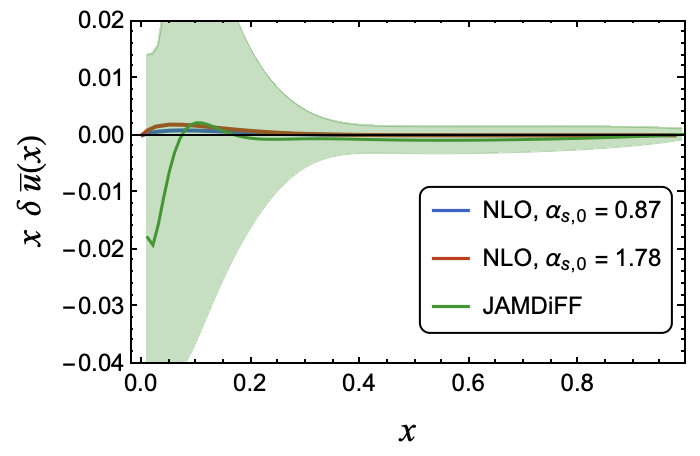}}
\hfill
\subfloat[]{\includegraphics[width=\linewidth]{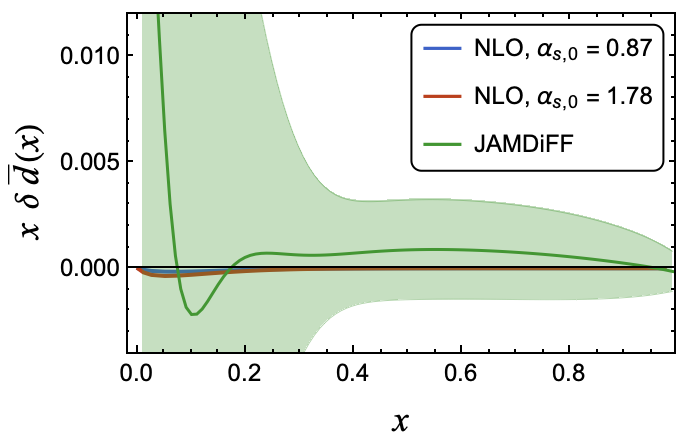}}
    \caption{ILM anti-quark transversity distribution for (a) $\bar u$ and (b) $\bar d$ in proton evolved from $\mu_0=0.63$ GeV to $\mu=2$ GeV with $\alpha_{s}(\mu_0)=0.87$ (3-loop running) and $1.78$ (5-loop running), compared to JAMDiFF \cite{Cocuzza:2023vqs,Cocuzza:2023oam} global analyses.}
    \label{fig:dqbar}
\end{figure}

In Fig.~\ref{fig:dqbar}, we evolve the $\bar u$ and $\bar d$ antiquark helicity PDFs obtained in the ILM from the initial scale $\mu_0=0.63~\mathrm{GeV}$ to the target scale $\mu=2.0~\mathrm{GeV}$, using NLO polarized DGLAP evolution from HOPPET package with three-loop running coupling $\alpha_s(\mu_0)=0.87$ and five-loop running coupling $\alpha_s(\mu_0)=1.78$. In contrast to the nearly flavor-symmetric antiquark helicity distributions shown in Fig.~\ref{fig:xDqb}, the evolved antiquark transversity distributions exhibit a pronounced flavor asymmetry in Fig.~\ref{fig:xdqb}. Unlike helicity, transversity does not couple to gluons under DGLAP evolution. Thus, antiquark transversity is generated only at NLO through quark–to-antiquark splitting. As a result, the resulting sea distributions largely retain the characteristic shapes and flavor asymmetry of the initial $u$ and $d$ quark transversity distributions.

\section{Gluon and Sea quark distribution in low scale}

\begin{figure}
    \centering
    \subfloat[]{\includegraphics[width=\linewidth]{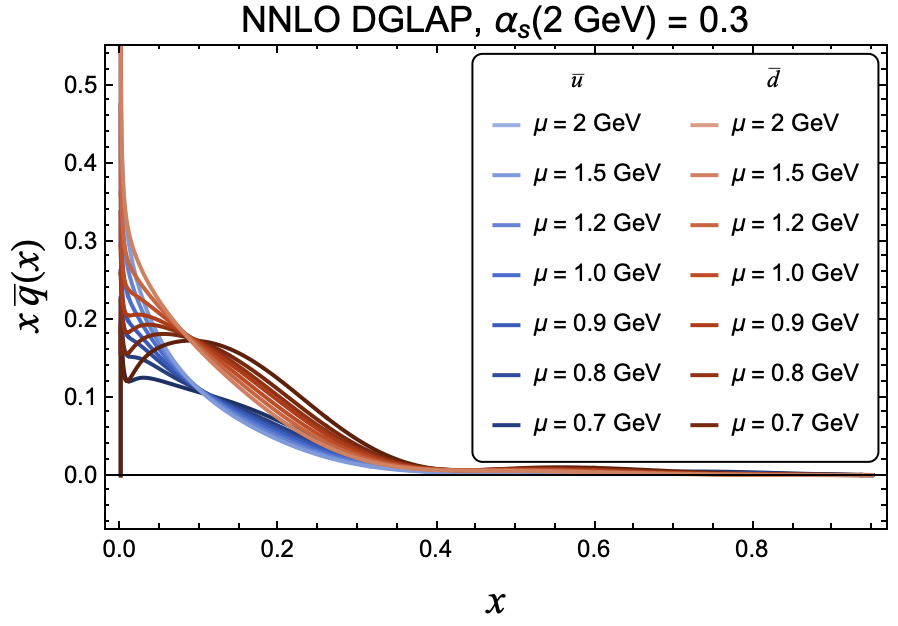}}
    \hfill
    \subfloat[]{\includegraphics[width=\linewidth]{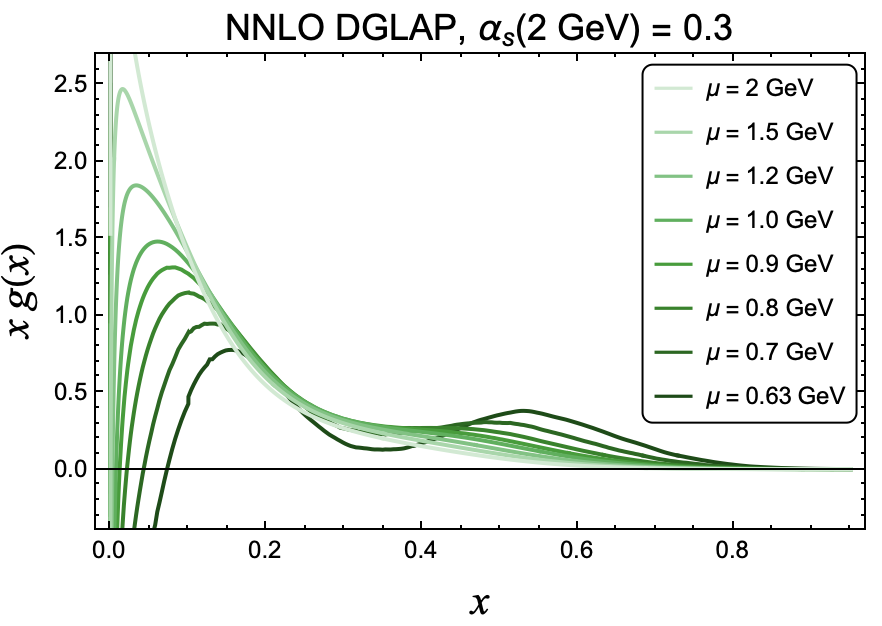}}
    \caption{Unpolarized PDFs of (a) sea quarks and (b) gluons from the NNPDF 4.0 global-fit NNLO set \cite{NNPDF:2021njg} with \(\alpha_s(M_Z)=0.118\), evolved backward from \(2~\mathrm{GeV}\) to lower scales using NNLO DGLAP evolution implemented in HOPPET, with \(\alpha_s(2~\mathrm{GeV})=0.301\).}
    \label{fig:backward}
\end{figure}

\begin{figure}
    \centering
 \includegraphics[width=\linewidth]{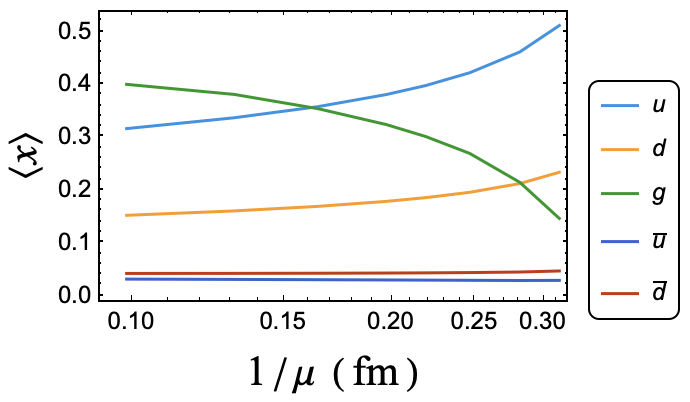}
    \caption{Second Mellin moment of PDFs of the NNPDF 4.0 global-fit NNLO set \cite{NNPDF:2021njg} with \(\alpha_s(M_Z)=0.118\), evolved backward from \(2~\mathrm{GeV}\) to lower scales using NNLO DGLAP evolution implemented in HOPPET, with \(\alpha_s(2~\mathrm{GeV})=0.301\).}
    \label{fig:mellin}
\end{figure}

In Fig.~\ref{fig:backward}, we present  the unpolarized sea-quark and gluon PDFs from the NNPDF4.0 NNLO global fit \cite{NNPDF:2021njg}, evolved backward from \(\mu=2~\mathrm{GeV}\) to lower scales using NNLO DGLAP evolution implemented in HOPPET, with \(\alpha_s(2~\mathrm{GeV})=0.301\).
Their corresponding second Mellin moments
are shown in Fig.~\ref{fig:mellin}. The backward NNLO DGLAP evolution reveals a distinct gluon enhancement at intermediate \(x\sim 0.4\,\text{–}\,0.6\), which becomes more pronounced toward lower scales, accompanied by a depletion of the small-\(x\) gluon distribution. The existence of non-negligible sea-quark distributions at lower scale suggests that higher Fock components containing additional \(q\bar q\) pairs may play an important role in describing the nonperturbative nucleon structure at the initial model scale. 

\section{Conclusion}

We have developed a microscopic description of the nucleon in which
quark--diquark correlations emerge from the topological structure of
the QCD vacuum rather than being introduced phenomenologically.
Starting from the instanton-induced 't~Hooft interaction and its
instanton--anti-instanton molecular extension, we derived the
interactions in the different diquark channels and embedded them
in a covariant Bethe--Salpeter--Faddeev description of the
three-quark bound state. In this way, the same instanton vacuum
that drives spontaneous chiral symmetry breaking also determines
the dominant short- and intermediate-distance correlations
organizing the nucleon.

The resulting hierarchy of diquark channels is particularly
transparent. A single instanton produces a strong attraction in
the color $\overline{\mathbf 3}$, isoscalar scalar channel,
while spin--one correlations arise through instanton--anti-instanton
molecules. With parameters fixed in the meson sector, the scalar
diquark is strongly bound, with $m_{qq_S}\simeq592~{\rm MeV}$,
whereas the axial-vector correlation lies close to the two-quark
threshold. The fully symmetrized Faddeev amplitude yields
$m_N=939.53~{\rm MeV}$, remarkably close to the physical nucleon
mass, with approximately $98\%$ scalar and $2\%$ axial-vector
diquark weights. Although the latter shifts the mass by only
about $9~{\rm MeV}$, its impact on spin observables is much
larger. In particular, scalar--axial interference and interference
between different quark--diquark pairings yield $g_A=1.265$,
close to experiment. This demonstrates that the physical
importance of a Faddeev component is not measured by its
probability alone: a small spin--one admixture can have a
substantial effect on spin--isospin coherence.

An important new outcome is that the same covariant nucleon
amplitude also provides its leading light-front three-quark
wave function, allowing the calculation of collinear parton
distributions without introducing an independent phenomenological
wave function. At the intrinsic instanton resolution
$\mu_0\simeq\rho^{-1}\simeq0.63~{\rm GeV}$, we obtain the
unpolarized, helicity, and transversity distributions and
their associated moments. After perturbative QCD evolution
to $2~{\rm GeV}$, the unpolarized valence distributions exhibit
the expected qualitative behavior and moments comparable in
scale to lattice calculations and global analyses.
The spin-dependent distributions provide a more sensitive test of the underlying diquark dynamics, highlighting the role of scalar--axial interference already observed in the axial charge, while exposing the limitations of a purely valence description, particularly in the individual flavor helicities. 
The resulting connection between the
nucleon spectrum, its spin structure, and its partonic
content is a central result of this work.

The present calculation also identifies the limitations
of the topological interaction alone. No $\Delta$ bound-state
pole is found below threshold, consistent with the absence
of the strongly attractive scalar diquark channel in that
state. This suggests a complementary role for confinement,
which supplies the long-distance binding and orbital dynamics
needed for a more complete baryon spectrum. In the broader
light-front hadron framework developed by Shuryak, Zahed,
and collaborators~\cite{Shuryak:2022wtk,Shuryak:2026pqt},
the instanton-induced interaction and confinement string are
therefore complementary rather than competing mechanisms:
topology selects the dominant chiral and spin--flavor
correlations, while confinement string completes the long-distance
hadronic dynamics.

There is a corresponding hierarchy in Fock space.
The three-quark wave function obtained here represents
the leading valence core at the low instanton scale,
rather than the complete physical nucleon. Higher
components, schematically
\[
 |N\rangle =
 \psi_{3q}|qqq\rangle+
 \psi_{5q}|qqq\,q\bar q\rangle+
 \psi_{3qg}|qqq\,g\rangle+\cdots ,
\]
provide the intrinsic sea-quark and gluonic degrees of
freedom that can modify flavor asymmetries, helicity
distributions, orbital angular momentum, and form factors.
Perturbative evolution generates additional partonic
radiation above the matching scale, but cannot replace
all nonperturbative correlations already present in
these higher Fock sectors. Thus, scalar-diquark dominance
and multiquark dressing are complementary: the former
organizes the valence core, while the latter completes
its partonic structure.

A natural next step is to incorporate the instanton-generated
Faddeev kernel into a confining light-front Hamiltonian and
couple the resulting three-quark state to the lowest
$qqq\,q\bar q$ and $qqq\,g$ sectors. This would allow a
unified treatment of the nucleon and $\Delta$ spectra,
orbital excitations, flavor-resolved parton distributions,
and electromagnetic and gravitational form factors.
More generally, the present results demonstrate that the
topology of the QCD vacuum already accounts quantitatively
for a substantial part of the nucleon's mass, spin, and
valence partonic structure. Confinement and higher Fock
components are then expected to complete, rather than
replace, this underlying topological configuration.

\vskip 0.5cm

\begin{acknowledgments}
This work is supported by the U.S. Department of Energy, Office of
Science, Office of Nuclear Physics under Contract No.\
DE-FG-88ER40388. This work is also supported in part by the
Quark-Gluon Tomography Topical Collaboration under Award
DE-SC0023646. The work of Wei-Yang Liu is supported in part by the U.S. Department of Energy, Office of Science, Office of Nuclear Physics under Contract No.~89243126CSC000213.
\end{acknowledgments}

\noindent 

\appendix

\section{Conventions used on the LF}
\label{Appx:LFspinor}
Throughout this paper, our conventions of the light front frame follows Kogut-Soper convention based on the Weyl chiral basis of the gamma matrices
\begin{eqnarray}
&\gamma^0=\begin{pmatrix}
0 & \mathds{1} \\
\mathds{1} & 0 \\
\end{pmatrix} ~\
&\gamma^{i}=\begin{pmatrix}
0 & \sigma^i \\
-\sigma^i  & 0 \\
\end{pmatrix}
\end{eqnarray}
The light front components are normalized to be
\begin{equation}
\gamma^\pm=\frac{\gamma^0\pm \gamma^3}{\sqrt{2}}
\end{equation}
with light front projection defined 
\begin{equation}
\frac{1}{2}\gamma^-\gamma^+=\begin{pmatrix}
1 & 0 & 0 & 0 \\
0 & 0 & 0 & 0 \\
0 & 0 & 0 & 0 \\
0 & 0 & 0 & 1 \\
\end{pmatrix}
\end{equation}
The LF spinors for the quarks are 
\begin{align}
&\frac{1\pm\gamma^5}2u_s(k)=\frac{1}{\sqrt{\sqrt{2}\,k^+}}\\
&\times
\left[
\sqrt{2}\,k^+ \frac{\mathbf{1}\pm\sigma_3}{2}
+ M\,\mathbf{1}
\pm k_L \sigma^+
\pm k_R \sigma^-
\right]\chi_s
\end{align}
where Pauli matrices $\sigma^\pm=(\sigma_x+\pm i\sigma_y)/2$, $k_{L,R}=k^x_\perp\mp ik^y_\perp$, and $M$  the constituent quark mass.

\section{Color matrices for diquarks}
\label{App:color}
For diquark bound states in $SU(3)$, the color structure are divided into two irreducible representations :
$\bf3\otimes\bf3=\bf{6}\oplus\bar{\bf{3}}$ where $\bf6$ is color-symmetric and  $\bar{\bf3}$ is color-antisymmetric. The three color generators in $\bar{\bf3}$ are defined by $(\beta_A^\alpha)_{\beta\gamma}=-i\sqrt{N_c/2}\ \epsilon_{\alpha\beta\gamma}$, with the normalization $\mathrm{tr}(\beta^\alpha\beta^\beta)=N_c\delta^{\alpha\beta}$ for $1/N_c$ counting. They are associated with the antisymmetric Gell-Mann matrices

\begin{align}
    \beta^3_{A}=&\sqrt{\frac{N_c}2}\lambda^2 \nonumber\\ 
    \beta^2_{A}=&-\sqrt{\frac{N_c}2}\lambda^5 \nonumber\\ \beta^1_{A}=&\sqrt{\frac{N_c}2}\lambda^7
\end{align}

On the other hand, the color-symmetric $\bf{6}$ generators $\beta^\alpha_{S}$ are defined by $\beta^1_S=\sqrt{N_c}\delta_{1\alpha}\delta_{1\beta}$, $\beta^2_S=\sqrt{N_c}\delta_{2\alpha}\delta_{2\beta}$, $\beta^3_S=\sqrt{N_c}\delta_{3\alpha}\delta_{3\beta}$ for the diagonal color matrices and $\beta^{4}_{S}=\sqrt{N_c/2}\left(\delta_{1\alpha}\delta_{2\beta}+\delta_{2\alpha}\delta_{1\beta}\right)$, $\beta^{5}_{S}=\sqrt{N_c/2}\left(\delta_{1\alpha}\delta_{3\beta}+\delta_{3\alpha}\delta_{1\beta}\right)$, $\beta^{6}_{S}=\sqrt{N_c/2}\left(\delta_{2\alpha}\delta_{3\beta}+\delta_{3\alpha}\delta_{2\beta}\right)$ for the non-diagonal color matrices. The former three are the linear combinations of the diagonal matrices $\lambda^{3,8}$ and the color identity matrix $\mathds{1}_c$ and the latter three correspond to non-diagonal symmetric Gell-Mann matrices $\lambda^{1,4,6}$.

\begin{align}
    \beta^1_S&=\sqrt{N_c}\left(\frac{1}{3}\mathds{1}_c+\frac{\lambda^3}{2}+\frac{\sqrt{3}}{6}\lambda^8\right) & \beta^{4}_{S}&=\sqrt{\frac{N_c}{2}}\lambda^{1} \nonumber\\
    \beta^2_S&=\sqrt{N_c}\left(\frac{1}{3}\mathds{1}_c-\frac{\lambda^3}{2}+\frac{\sqrt{3}}{6}\lambda^8\right) & \beta^{5}_{S}&=\sqrt{\frac{N_c}{2}}\lambda^{4} \nonumber\\
    \beta^3_S&=\sqrt{N_c}\left(\frac{1}{3}\mathds{1}_c-\frac{\sqrt{3}}{3}\lambda^8\right) & \beta^{6}_{S}&=\sqrt{\frac{N_c}{2}}\lambda^{6}
\end{align}

\section{Fiertz identity in diquark channels}

A useful identity to transform the $SU(2)$ flavor basis between $s$-exchange and $t$-exchange reads
\begin{equation}
    \delta_{ij}\delta_{kl}-\tau^a_{ij}\tau^a_{kl}=-\left(\delta_{il}\delta_{kj}-\tau^a_{il}\tau^a_{kj}\right)
\end{equation}
and
\begin{equation}
    \delta_{ij}\delta_{kl}+\tau^a_{ij}\tau^a_{kl}=2\delta_{il}\delta_{kj}
\end{equation}

In diquark channels, the identity can be rearranged as
\begin{equation}
    \delta_{ij}\delta_{kl}-\tau^a_{ij}\tau^a_{kl}=2\tau^2_{ik}\tau^2_{lj}
\end{equation}
and
\begin{equation}
    2\delta_{ij}\delta_{lk}=\tau^2_{ik}\tau^2_{lj}+(\tau^2\tau^a)_{ik}(\tau^a\tau^2)_{lj}
\end{equation}
with the identities

\begin{equation}
    \tau^2_{ik}\tau^2_{lj} = \delta_{ij}\delta_{kl} - \delta_{il}\delta_{kj},
\end{equation}
and
\begin{equation}
(\tau^2 \tau^a)_{ik}(\tau^a\tau^2)_{lj} = \delta_{ij}\delta_{kl} + \delta_{il}\delta_{kj}    
\end{equation}

Similar identity can also be derived for color basis in $SU(N_c)$. 
\begin{equation}
    \delta_{\alpha\beta}\delta_{\gamma\delta}=\frac{1}{N_c}\delta_{\alpha\delta}\delta_{\gamma\beta}+\frac{1}{2}\lambda^\alpha_{\alpha\delta}\lambda^\alpha_{\gamma\beta}
\end{equation}

In diquark channels, the color identity can be decomposed into antisymmetric part and symmetric part. 
\begin{equation}
    \delta_{\alpha\beta}\delta_{\gamma\delta}=\frac{1}{2}\beta^a_{A\alpha\gamma}\beta^a_{A\delta\beta}+\frac{1}{2}\beta^a_{S\alpha\gamma}\beta^a_{S\delta\beta}
\end{equation}
with the identities $$(\beta^a_{A})_{\alpha\gamma}(\beta^a_{A})_{\delta\beta}=N_c\left(\delta_{\alpha\beta}\delta_{\gamma\delta}-\delta_{\alpha\delta}\delta_{\gamma\beta}\right)/2$$ and  $$(\beta^a_{S})_{\alpha\gamma}(\beta^a_{S})_{\delta\beta}=N_c\left(\delta_{\alpha\beta}\delta_{\gamma\delta}+\delta_{\alpha\delta}\delta_{\gamma\beta}\right)/2$$
where $\beta_A^a$ is the color-antisymmetric $\bar{\bf{3}}$ generators and $\beta^a_{S}$ is the color-symmetric $\bf{6}$ generators. 

\section{Dyson-Schwinger and Faddeev equation}

In this Appendix, we summarize the connection between the Green’s function, the transfer matrix, and the Dyson-Schwinger (DS) equation, in close analogy with the relation between the covariant bound-state wave function and the Bethe-Salpeter-Faddeev (BSF) equation. For brevity, the equations are presented in quantum operator formalism in Hilbert space.

The full Green’s function $G$ is related to the free propagator $G_0$ and the transfer matrix $T$ through
\begin{equation}
    G = G_0 + G_0 \, T \, G_0 \, .
\end{equation}
The corresponding DS equation for the scattering matrix is
\begin{equation}
    T = K + K \, G_0 \, T \, ,
\end{equation}
where $K$ denotes the irreducible interaction kernel. In ILM, this kernel is specified by the effective multi-quark interaction induced by the instantons and anti-instantons in the vacuum.

When the interaction is sufficiently attractive to generate a bound state, the transfer matrix develops a pole at the mass of a hadron bound-state $X$,
\begin{equation}
    T \;\xrightarrow[P^2\to m_X^2]{}\; \frac{-i\,\Gamma\,\bar{\Gamma}}{P^2 - m_X^2} \, ,
\end{equation}
with total momentum $P$, the sum of all incoming external momenta. Here $\Gamma$ is the amputated BSF vertex function, while the corresponding covariant BSF wave function is obtained by attaching the constituent free propagators,
\begin{equation}
    \Psi = G_0 \, \Gamma \, ,
\end{equation}

Substituting the pole structure into the DS equation yields the homogeneous bound-state equation
\begin{equation}
    \Gamma = K \, G_0 \, \Gamma \, ,
\end{equation}
which is the BSF equation for the hadron vertex. Solving the equation simultaneously produces a gap-like equation to determine the hadron mass and the form of the vertex function in terms of Dirac matrices.

For an $n$-body Green's function where $n>2$, the dynamics can be reduced to a hierarchy of coupled two-body equations if the interaction kernel $K$ contains only two-body interactions. In this case, repeated interactions within a given pair of any two quark lines can be resummed into an effective two-body transfer matrix that satisfys its own DS equation. The full transfer matrix can then be expressed in terms of these reduced matrix with quark-exchange kernel that permutes quark lines between successive pairs. For more details, see \cite{Ishii:1995bu} (references therein)

\section{SU(6) baryon wave function}
\label{App:SU6}

The $SU(6)$ nucleon favor-spin wave functions can be obtained by combining the mixed anti-symmetric basis and symmetric basis equally. 

\begin{widetext}

\bea
\label{SU6}
&|p,+1/2\rangle=&\frac 1{\sqrt {18}}
 \bigg(3[ud]_S u_\uparrow + 2[uu]^+_A d_\downarrow 
 -\sqrt 2[uu]^0_A d_\uparrow-\sqrt 2[ud]^+_A u_\downarrow + [ud]^0_A u_\uparrow\bigg)\nonumber\\
&|n,+1/2\rangle=&\frac 1{\sqrt {18}}
 \bigg(3[ud]_S d_\uparrow + 2[dd]^+_A u_\downarrow 
 -\sqrt 2[dd]^0_A u_\uparrow-\sqrt 2[ud]^+_A d_\downarrow + [ud]^0_A d_\uparrow\bigg)
 \eea
The diquark states appearing above are defined in terms of two-quark combinations with definite spin and isospin. The scalar (isoscalar, spin-0) diquark is antisymmetric in both flavor and spin,
\begin{equation}
[ud]_S \equiv \frac{1}{\sqrt{2}}\left( u^\uparrow d^\downarrow - u^\downarrow d^\uparrow - d^\uparrow u^\downarrow + d^\downarrow u^\uparrow \right),
\end{equation}
which can be written compactly as an antisymmetric combination in flavor and spin.
The axial-vector (isovector, spin-1) diquarks are symmetric in spin and carry isospin components labeled by the superscript $+ , 0 , -$. They are given by
\begin{align}
[uu]^+_A &\equiv u^\uparrow u^\uparrow, \\
[dd]^+_A &\equiv d^\uparrow d^\uparrow, \\
[uu]^0_A &\equiv \frac{1}{\sqrt{2}}\left( u^\uparrow u^\downarrow + u^\downarrow u^\uparrow \right), \\
[dd]^0_A &\equiv \frac{1}{\sqrt{2}}\left( d^\uparrow d^\downarrow + d^\downarrow d^\uparrow \right), \\
[ud]^+_A &\equiv \frac{1}{\sqrt{2}}\left( u^\uparrow d^\uparrow + d^\uparrow u^\uparrow \right), \\
[ud]^0_A &\equiv \frac{1}{2}\left( u^\uparrow d^\downarrow + u^\downarrow d^\uparrow + d^\uparrow u^\downarrow + d^\downarrow u^\uparrow \right), \\
[ud]^-_A &\equiv \frac{1}{\sqrt{2}}\left( u^\downarrow d^\downarrow + d^\downarrow u^\downarrow \right).
\end{align}
These definitions ensure that the scalar diquark is fully antisymmetric, while the axial-vector diquarks form a symmetric spin-triplet and isospin-triplet representation, consistent with the Pauli principle when combined with the antisymmetric color structure.

 Similarly for delta, we can construct the flavor-spin wave functions in quark-diquark form. The $SU(6)$ quark-diquark wave functions for a helcity $\lambda=+1/2$ state thus read
\begin{equation}
    \begin{aligned}
\label{SU6}
        & |\Delta^{++},+1/2\rangle=\frac1{\sqrt{3}}\bigg([uu]_A^+ u_\downarrow+\sqrt{2}[uu]_A^0 u_\uparrow\bigg)\\
& |\Delta^{+},+1/2\rangle=\frac{1}{\sqrt{9}}\bigg([uu]_A^+ d_\downarrow+\sqrt{2}[uu]_A^0 d_\uparrow+\sqrt{2}[ud]_A^+ u_\downarrow+2[ud]_A^0 u_\uparrow\bigg)\\
& |\Delta^{0},+1/2\rangle=\frac{1}{\sqrt{9}}\bigg([dd]_A^+ u_\downarrow+\sqrt{2}[dd]_A^0 u_\uparrow+\sqrt{2}[ud]_A^+ d_\downarrow+2[ud]_A^0 d_\uparrow\bigg)\\
& |\Delta^{-},+1/2\rangle=\frac1{\sqrt{3}}\bigg([dd]_A^+ d_\downarrow+\sqrt{2}[dd]_A^0 d_\uparrow\bigg)
    \end{aligned}
\end{equation}

\end{widetext}
and for helicity $\lambda=+3/2$ state
\begin{equation}
    \begin{aligned}
\label{SU6_1}
        & |\Delta^{++},+3/2\rangle=[uu]_A^+ u_\uparrow
    \end{aligned}
\end{equation}

\begin{equation}
    \begin{aligned}
\label{SU6_2}
& |\Delta^{-},+3/2\rangle=[dd]_A^+ d_\uparrow
    \end{aligned}
\end{equation}

\begin{equation}
    \begin{aligned}
\label{SU6_3}
& |\Delta^{+},+3/2\rangle=\frac{1}{\sqrt{3}}\bigg([uu]_A^+ u_\uparrow+\sqrt{2}[ud]_A^+ u_\uparrow\bigg)
    \end{aligned}
\end{equation}

\begin{equation}
    \begin{aligned}
\label{SU6_4}
& |\Delta^{0},+3/2\rangle=\frac{1}{\sqrt{3}}\bigg([dd]_A^+ u_\uparrow+\sqrt{2}[ud]_A^+ d_\uparrow\bigg)
    \end{aligned}
\end{equation}

These baryon flavor and spin basis can be rewritten as covariant quark-diquark 4-spinor form after Lorentz boost. For the mixed antisymmetric flavor-spin combination, the result reads

\begin{equation}
\begin{aligned}
    &\begin{pmatrix}
   |udu\rangle-|duu\rangle \\
        |udd\rangle-|dud\rangle
    \end{pmatrix}\rightarrow i\tau^2\otimes 1
\end{aligned}
\end{equation}

and
\begin{equation}
    \begin{pmatrix}
   |\uparrow\downarrow\uparrow\rangle-|\downarrow\uparrow\uparrow\rangle \\
        |\uparrow\downarrow\downarrow\rangle-|\downarrow\uparrow\downarrow\rangle
    \end{pmatrix}\rightarrow \frac1{2M^{3/2}}(u_{s_1}^TC\gamma^5u_{s_2})u_{s_3}
\end{equation}

For the mixed symmetric combination, the result reads

\begin{equation}
\begin{aligned}
    &\begin{pmatrix}
   2|uud\rangle-|udu\rangle-|duu\rangle \\
    -2|ddu\rangle+|dud\rangle+|udd\rangle
    \end{pmatrix}\rightarrow i\tau^2\tau^a\otimes \tau^a
\end{aligned}
\end{equation}

and
\begin{equation}
\begin{aligned}
    &\begin{pmatrix}
   2|\uparrow\uparrow\downarrow\rangle-|\uparrow\downarrow\uparrow\rangle-|\downarrow\uparrow\uparrow\rangle \\
        -2|\downarrow\downarrow\uparrow\rangle+|\downarrow\uparrow\downarrow\rangle+|\uparrow\downarrow\downarrow\rangle
    \end{pmatrix}\\
    &\rightarrow -\frac1{2M^{3/2}}(u_{s_1}^TC\gamma^\mu u_{s_2})\left(\gamma_\mu+\frac{P_\mu}{m_N}\right)\gamma^5 u_{s_3}
\end{aligned}
\end{equation}
where the 4-spinor normalization is $\bar{u}_su_s=2M$.


Using these identities, the static $SU(6)$ wave functions can be boosted to a relativistic frame and expressed in terms of light-front spinors, allowing us to assess the validity of the $SU(6)$ approximation for baryons in a light-front setting.
 
\begin{equation}
\begin{aligned}
&\Phi_N^{SU(6)}=\frac1{\sqrt{6}}\bigg[\left(\bar u_{s_1}\gamma^5C i\tau^2\bar u_{s_2}^T\right)\bar u_{s_3}u_N\\
&-\frac13\left(\bar u_{s_1}\gamma^\mu Ci\tau^2\tau^a\bar u_{s_2}^T\right)\bar u_{s_3}\tau^a\left(\gamma_\mu+\frac{P_\mu}{m_N}\right)\gamma^5 u_N\bigg]\\
&+S_3~\mathrm{permutation}
\end{aligned}
\end{equation}

\section{Light-cone wave functions}
\label{app:LCWF}

In this Appendix, we present the explicit forms of the light-cone wave functions following the framework of \cite{Ji:2003yj} using fermion creation operator $q^\dagger_{s,\alpha}(i)$ creating single quark state of flavor $q$ with spin $s$, color $\alpha$, and momentum $k^\mu_i=\left(x_i P^+,\frac{k_{i\perp}^2+M^2}{2x_iP^+},k_{i\perp}\right)$. The construction is carried out such that angular momentum conservation is preserved upon projection onto the light front. In particular, the total orbital angular momentum $l_z=\sum l_{zi}$ and the total parton helicity  $s_z=\sum s_{zi}$, combine to yield the hadron helicity, $\lambda=l_z+s_z$. Our formulation and classification are consistent with the general structure summarized in \cite{Ji:2003yj}


\subsection{Proton}

We consider the proton $p^+$ as an illustrative example. Only the state with positive helicity will be discussed as the negative helicity state can be
obtained simply from the parity transformation. The leading Fock-space component of the proton wave function is given by

\begin{widetext}
\begin{equation}
\begin{aligned}
|p\uparrow\rangle^{l_z=0}
&= \int d[123]\,
\left(
\psi_N^{(1)}(1,2,3)
+ i \epsilon_{\alpha\beta} k^\alpha_{1\perp} k_{2\perp}^\beta\,\psi^{(2)}_N(1,2,3)
\right)
\nonumber \\
&\quad \times
\frac{\epsilon^{\alpha\beta\gamma}}{\sqrt{N_c!}}\,
u^\dagger_{\alpha\uparrow}(1)
\left(
u^\dagger_{\beta\downarrow}(2)\,d^\dagger_{\gamma\uparrow}(3)
-
d^\dagger_{\beta\downarrow}(2)\,u^\dagger_{\gamma\uparrow}(3)
\right)
|0\rangle ,
\end{aligned}
\end{equation}

\begin{equation}
\begin{aligned}
|p\uparrow\rangle^{l_z=1}
&= \int d[123]\,
\left(
k_{1R}\,\psi^{(3)}_N(1,2,3)
+
k_{2R}\,\psi^{(4)}_N(1,2,3)
\right)
\nonumber \\
&\quad \times
\frac{\epsilon^{\alpha\beta\gamma}}{\sqrt{N_c!}}\,
\left(
u^\dagger_{\alpha\uparrow}(1)\,u^\dagger_{\beta\downarrow}(2)\,d^\dagger_{\gamma\downarrow}(3)
-
d^\dagger_{\alpha\uparrow}(1)\,u^\dagger_{\beta\downarrow}(2)\,u^\dagger_{\gamma\downarrow}(3)
\right)
|0\rangle .
\end{aligned}
\end{equation}

\begin{equation}
\begin{aligned}
|p\uparrow\rangle^{l_z=-1}
&= \int d[123]\,
k_{2L}\,\psi^{(5)}_N(1,2,3)
\nonumber \\
&\quad \times
\frac{\epsilon^{\alpha\beta\gamma}}{\sqrt{N_c!}}\,
u^\dagger_{\alpha\uparrow}(1)
\left(
u^\dagger_{\beta\uparrow}(2)\,d^\dagger_{\gamma\uparrow}(3)
-
d^\dagger_{\beta\uparrow}(2)\,u^\dagger_{\gamma\uparrow}(3)
\right)
|0\rangle ,
\\[1em]
|p\uparrow\rangle^{l_z=2}
&= \int d[123]\,
\left(k_{1R}\,k_{3R}\,\psi^{(6)}_N(1,2,3)+k_{2R}\,k_{3R}\,\psi^{(6)'}_N(1,2,3)\right)
\nonumber \\
&\quad \times
\frac{\epsilon^{\alpha\beta\gamma}}{\sqrt{N_c!}}\,
u^\dagger_{\alpha\downarrow}(1)
\left(
u^\dagger_{\beta\downarrow}(2)\,d^\dagger_{\gamma\downarrow}(3)-d^\dagger_{\beta\downarrow}(2)\,u^\dagger_{\gamma\downarrow}(3)
\right)
|0\rangle .
\end{aligned}
\end{equation}
where $k_{L,R}=k^x_\perp\mp ik^y_\perp$ and  $d[123]$ is the 3-body phase space integral defined in \eqref{eq:n-body}. With the useful identity $i\epsilon_{\alpha\beta} k^\alpha_{1\perp} k_{2\perp}^\beta k_{1R}=k_{1\perp}^2k_{2R}-k_{1\perp}\cdot k_{2\perp}k_{1R}$ and momentum conservation, we can rewrite the wave functions in \eqref{eq:LCWF_N} and \eqref{eq:LCWF_delt} into these independent basis.

\begin{equation}
\begin{aligned}
    &\psi^{(1)}_N(1,2,3)=-
\frac{
2 \, M(M + m_N x_1)(x_2 + x_3)
}{\sqrt{x_1 x_2 x_3}}\alpha_1\, \phi_{N_s}(2,3,1)
+
\frac{
2 x_1 k_{2\perp}^2  + 2k_{1\perp}\cdot k_{2\perp} (x_1 + x_3)
}
{\sqrt{x_1 x_2 x_3}}\alpha_1\, \phi_{N_s}(3,1,2)\\
&+\frac{4\left(M - m_N x_3\right)\left(k_{1\perp}\cdot k_{2\perp} - M^2 + m_N^2 x_1 x_2\right)}
{m_N \sqrt{x_1 x_2 x_3}}\left(\alpha_2-\frac{\alpha_3}{\sqrt{3}}\right)\phi_{N_a}(1,2,3)
\\
&+
\frac{2(M - m_N x_1)
\left(
k_{1\perp}\cdot k_{2\perp}
+ k^2_{2\perp}
+ M^2
- m_N^2 x_2 x_3
\right)}{m_N \sqrt{x_1 x_2 x_3}}
\left(\alpha_2-\frac{\alpha_3}{\sqrt{3}}\right)\phi_{N_a}(2,3,1)\\
&+
\frac{2M k^2_{2\perp}
}{m_N \sqrt{x_1 x_2 x_3}}\left(\alpha_2-\frac{\alpha_3}{\sqrt{3}}\right)\, \phi_{N_a}(3,1,2)\\
&-\frac{8\left(x_1k^2_{2\perp} + (x_1 + x_3)k_{1\perp}\cdot k_{2\perp} 
-M^2 x_3 - Mm_N x_1 x_2\right)
}
{\sqrt{x_1 x_2 x_3}}
\frac{\alpha_3}{\sqrt{3}}
\, \phi_{N_a}(1,2,3)\\
&-
\frac{4 \left(
x_1 k_{2\perp}^2
+ (x_1 + x_3) k_{1\perp}\cdot k_{2\perp} 
+M^2 x_1 + Mm_N x_2 x_3
\right)
}
{\sqrt{x_1 x_2 x_3}}\frac{\alpha_3}{\sqrt{3}}\, \phi_{N_a}(2,3,1)\\
&+
\frac{
4 M(M+x_2m_N)(x_1+x_3)}{\sqrt{x_1 x_2 x_3}}
\, \frac{\alpha_3}{\sqrt{3}}\phi_{N_a}(3,1,2)
\end{aligned}
\end{equation}

\begin{equation}
\begin{aligned}
\psi^{(2)}_N(1,2,3)=&
\frac{2(x_1 + x_3)}{\sqrt{x_1 x_2 x_3}}\alpha_1 \, \phi_{N_s}(3,1,2)+\frac{4(M - m_N x_3)}{m_N \sqrt{x_1 x_2 x_3}}\left(\alpha_2-\frac{\alpha_3}{\sqrt{3}}\right)\phi_{N_a}(1,2,3)\\
&+\frac{2(M - m_N x_1)}{m_N \sqrt{x_1 x_2 x_3}} \left(\alpha_2-\frac{\alpha_3}{\sqrt{3}}\right)\, \phi_{N_a}(2,3,1)-\frac{8 \, (x_1 + x_3)}{\sqrt{x_1 x_2 x_3}}\frac{\alpha_3}{\sqrt{3}}\phi_{N_a}(1,2,3)\\
&-
\frac{4(x_1 + x_3)}{\sqrt{x_1 x_2 x_3}}\frac{\alpha_3}{\sqrt{3}}\phi_{N_a}(2,3,1)
\end{aligned}
\end{equation}

\begin{equation}
    \begin{aligned}
\psi^{(3)}_N(1,2,3)=&-
\frac{
2(x_2M + m_N x_1x_2)
}{\sqrt{x_1 x_2 x_3}}
\alpha_1\,\phi_{N_s}(2,3,1)
-\frac{4\left(k_{2\perp}^2+M^2 
- m_N^2 x_1 x_2
\right)}{m_N \sqrt{x_1 x_2 x_3}}\left(\alpha_2-\frac{\alpha_3}{\sqrt{3}}\right)\phi_{N_a}(1,2,3)
\\
&+\frac{2M\left(M -m_N x_1\right)}
{m_N \sqrt{x_1 x_2 x_3}}\left(\alpha_2-\frac{\alpha_3}{\sqrt{3}}\right)\phi_{N_a}(2,3,1)+\frac{2k_{2\perp}^2}{m_N \sqrt{x_1 x_2 x_3}}
\left(\alpha_2-\frac{\alpha_3}{\sqrt{3}}\right)\phi_{N_a}(3,1,2)
\\
&
+\frac{8
m_N x_1 x_2}{\sqrt{x_1 x_2 x_3}}\frac{\alpha_3}{\sqrt{3}}\phi_{N_a}(1,2,3)-\frac{4 M x_1}{\sqrt{x_1 x_2 x_3}}\frac{\alpha_3}{\sqrt{3}}\phi_{N_a}(2,3,1)
\\[6pt]
&
+\frac{4\left(
 x_1M+ m_N x_1 x_2
\right)}{\sqrt{x_1 x_2 x_3}}\frac{\alpha_3}{\sqrt{3}}\phi_{N_a}(3,1,2)
\\[6pt]
\end{aligned}
\end{equation}

\begin{equation}
    \begin{aligned}
\psi^{(4)}_N(1,2,3)=&-
\frac{
2(x_2+x_3)\left(
M+ m_N x_1
\right)
}{\sqrt{x_1 x_2 x_3}}
\alpha_1\,\phi_{N_s}(2,3,1)-
\frac{
2M\left(
x_1 +x_3
\right)
}{\sqrt{x_1 x_2 x_3}}
\alpha_1\,\phi_{N_s}(3,1,2)\\
&
+\frac{4\left(
k_{1\perp}^2 + 2\, k_{1\perp} \cdot k_{2\perp}-M^2 + m_N^2 x_1 x_2
\right)}{m_N \sqrt{x_1 x_2 x_3}}\left(\alpha_2-\frac{\alpha_3}{\sqrt{3}}\right)\phi_{N_a}(1,2,3)
\\
&
-\frac{2\left(k_{1\perp}^2 + 2\, k_{1\perp} \cdot k_{2\perp}+M^2 - m_N^2 x_1 x_3
\right)}{m_N \sqrt{x_1 x_2 x_3}}\left(\alpha_2-\frac{\alpha_3}{\sqrt{3}}\right)\phi_{N_a}(3,1,2)
\\
&
+\frac{4(Mx_1 + m_N (x_1x_2 + x_1x_3))}{\sqrt{x_1 x_2 x_3}}\frac{\alpha_3}{\sqrt{3}}[2\phi_{N_a}(1,2,3)+\phi_{N_a}(3,1,2)]
\\[6pt]
\end{aligned}
\end{equation}

\begin{equation}
\begin{aligned}
    \psi^{(5)}_N(1,2,3)=&-\frac{2x_3(M+x_1m_N)}{\sqrt{x1 x2 x3}}\alpha_1\phi_{N_s}(2,3,1)-\frac{2M (M-x_2 m_N)}{m_N \sqrt{x_1 x_2 x_3}}\left(\alpha_2-\frac{\alpha_3}{\sqrt{3}}\right)\phi_{N_a}(1,3,2)\\
    &+\frac{4M}{\sqrt{x_1 x_2 x_3}}\frac{\alpha_3}{\sqrt{3}}\phi_{N_a}(1,3,2)
 \end{aligned}
\end{equation}

\begin{equation}
\begin{aligned}
    \psi^{(6)}_N(1,2,3)=&-\frac{2x_2}{\sqrt{x_1x_2x_3}}\alpha_1\phi_{N_s}(2,3,1)- \frac{2M}{m_N \sqrt{x_1 x_2 x_3}}\left(\alpha_2-\frac{\alpha_3}{\sqrt{3}}\right)\phi_{N_a}(1,2,3)
\end{aligned}
\end{equation}

\begin{equation}
\begin{aligned}
    \psi^{(6)'}_N(1,2,3)=&-\frac{2M}{m_N \sqrt{x_1 x_2 x_3}}\left(\alpha_2-\frac{\alpha_3}{\sqrt{3}}\right)\phi_{N_a}(1,2,3)
\end{aligned}
\end{equation}

\end{widetext}

\subsection{Details of the derivation of $\psi_N^{(2)}(1,2,3)$ in the Light-Front Basis}
To illustrate the procedure, we detail here the extraction of the
scalar contribution leading to $\psi_N^{(2)}(1,2,3)$. Indeed, 
the coefficient $\psi_N^{(2)}(1,2,3)$ arises from the projection of the covariant light-front nucleon wave function onto the helicity Fock basis and is defined as the coefficient multiplying the transverse antisymmetric structure $i\epsilon^{\alpha\beta} k_{1\perp}^\alpha k_{2\perp}^\beta$ in the decomposition of the three-quark state.

The starting point is the covariant light-front wave function
\begin{equation}
\begin{aligned}
\Phi_N \sim& 
\phi_{Ns}\,
\big(\bar u_{s_1}\gamma_5 C \bar u_{s_2}^T\big)\,
\bar u_{s_3}\Gamma_s u_N
\\
&+
\phi_{Na}\,
\big(\bar u_{s_1}\gamma^\mu C \bar u_{s_2}^T\big)\,
\bar u_{s_3}\Gamma_{a\mu} u_N,
\end{aligned}
\end{equation}
where $\phi_{Ns}$ and $\phi_{Na}$ denote the scalar and axial-vector diquark momentum wave functions, respectively, and $\Gamma_s$, $\Gamma_{a}^\mu$ encode the effective quark-diquark couplings fixed by the Faddeev equation. The coefficients $\alpha_1,\alpha_2,\alpha_3$ enter through the decomposition of $\Gamma$ in Eq.~\eqref{vertex_N}.

To obtain the light-front wave function in the canonical Fock basis, the Dirac spinors are replaced by light-front spinors $u_s(k)$ and the bilinears are evaluated explicitly. This generates helicity-dependent structures together with kinematic factors involving $x_i$, $k_{i\perp}$, and the constituent mass $M$. The resulting expression is then reorganized into a basis of independent spin-flavor operators, such as
\begin{equation}
u^\dagger_{\alpha\uparrow}(1)\Big(u^\dagger_{\beta\downarrow}(2)d^\dagger_{\gamma\uparrow}(3) - d^\dagger_{\beta\downarrow}(2)u^\dagger_{\gamma\uparrow}(3)\Big),
\end{equation}
with accompanying scalar momentum coefficients.
The function $\psi_N^{(2)}$ is identified as the coefficient multiplying the structure
\begin{equation}
i\epsilon^{\alpha\beta} k_{1\perp}^\alpha k_{2\perp}^\beta
= \frac{1}{2i}\left(k_{1R}k_{2L} - k_{1L}k_{2R}\right),
\end{equation}
which carries total light-front orbital angular momentum $l_z=0$. This term originates from the interference of transverse momentum components in the light-front spinors, primarily from the axial-vector diquark contribution, while the scalar diquark yields a simpler kinematic structure.

The scalar diquark contribution generates the term proportional to $\alpha_1\,\phi_{Ns}$, arising from the $\gamma_5$ Dirac structure without additional transverse tensor contractions. In contrast, the axial-vector diquark contribution produces the remaining terms proportional to $\alpha_2$ and $\alpha_3$, through contractions of the form
$
\big(\bar u \gamma^\mu C \bar u^T\big)\,\big(\bar u \Gamma_{a\mu} u_N\big)$,
which yield nontrivial combinations of transverse momenta and kinematic factors such as $(M - m_N x_i)/m_N$. These factors reflect the relativistic spin structure induced by the transverse (Melosh) rotation between canonical spin and light-front helicity.
The combination  of the arguments $(1,2,3)$, $(2,3,1)$, and $(3,1,2)$ in $\phi_{Ns}$ and $\phi_{Na}$ follows from the required $S_3$ permutation symmetry of the three-quark state, corresponding to different choices of the spectator quark in the quark-diquark decomposition.

\subsection{Delta baryon}
For simplicity, we consider the delta-baryon $\Delta^{++}$ as an illustrative example. The leading Fock-space component of the proton wave function is given by

\begin{widetext}

\begin{equation}
\begin{aligned}
|\Delta^{++}, \lambda = 3/2\rangle^{l_z=0}
&= \int d[123]\,
\left(
\psi^{(1)}_\Delta(1,2,3)
+ i \epsilon^{\alpha\beta} k_{1\alpha} k_{2\beta}\,\psi^{(2)}_\Delta(1,2,3)
\right)
\nonumber \\
&\quad \times
\frac{\epsilon^{\alpha\beta\gamma}}{\sqrt{N_c!}}\,
u^\dagger_{\alpha\uparrow}(1)\,
u^\dagger_{\beta\uparrow}(2)\,
u^\dagger_{\gamma\uparrow}(3)
|0\rangle ,
\\[1em]
|\Delta^{++}, \lambda= 3/2\rangle^{l_z=1}
&= \int d[123]\,
k_{1R}\,\psi^{(3)}_\Delta(1,2,3)
\nonumber \\
&\quad \times
\frac{\epsilon^{\alpha\beta\gamma}}{\sqrt{N_c!}}\,
u^\dagger_{\alpha\uparrow}(1)\,
u^\dagger_{\beta\uparrow}(2)\,
u^\dagger_{\gamma\downarrow}(3)
|0\rangle ,
\\[1em]
|\Delta^{++}, \lambda= 3/2\rangle^{l_z=2}
&= \int d[123]\,
\left(
k_{1R} k_{2R}\,\psi^{(4)}_\Delta(1,2,3)
+ k_{2R} k_{3R}\,\psi^{(5)}_\Delta(1,2,3)
\right)
\nonumber \\
&\quad \times
\frac{\epsilon^{\alpha\beta\gamma}}{\sqrt{N_c!}}\,
u^\dagger_{\alpha\uparrow}(1)\,
u^\dagger_{\beta\downarrow}(2)\,
u^\dagger_{\gamma\downarrow}(3)
|0\rangle ,
\\[1em]
|\Delta^{++}, \lambda= 3/2\rangle^{l_z=3}
&= \int d[123]\,
k_{1R}^2 k_{2R}\,
\psi^{(6)}_\Delta(1,2,3)
\nonumber \\
&\quad \times
\frac{\epsilon^{\alpha\beta\gamma}}{\sqrt{N_c!}}\,
u^\dagger_{\alpha\downarrow}(1)\,
u^\dagger_{\beta\downarrow}(2)\,
u^\dagger_{\gamma\downarrow}(3)
|0\rangle .
\end{aligned}
\end{equation}
and

\begin{equation}
\begin{aligned}
|\Delta^{++}, \lambda = 1/2\rangle^{l_z=0}
&= \int d[123]\,
\left(
\psi^{(1)}_\Delta(1,2,3)
+ i \epsilon^{\alpha\beta} k_{1\alpha} k_{2\beta}\,\psi^{(2)}_\Delta(1,2,3)
\right)
\nonumber \\
&\quad \times
\frac{\epsilon^{\alpha\beta\gamma}}{\sqrt{N_c!}}\,
u^\dagger_{\alpha\uparrow}(1)\,
u^\dagger_{\beta\uparrow}(2)\,
u^\dagger_{\gamma\downarrow}(3)
|0\rangle ,
\\[1em]
|\Delta^{++}, \lambda = 1/2\rangle^{l_z=1}
&= \int d[123]\,
k_{2R}\,\psi^{(3)}_\Delta(1,2,3)
\nonumber \\
&\quad \times
\frac{\epsilon^{\alpha\beta\gamma}}{\sqrt{N_c!}}\,
u^\dagger_{\alpha\uparrow}(1)\,
u^\dagger_{\beta\downarrow}(2)\,
u^\dagger_{\gamma\downarrow}(3)
|0\rangle ,
\\[1em]
|\Delta^{++}, \lambda = 1/2\rangle^{l_z=-1}
&= \int d[123]\,
k_{2L}\,\psi^{(4)}_\Delta(1,2,3)
\nonumber \\
&\quad \times
\frac{\epsilon^{\alpha\beta\gamma}}{\sqrt{N_c!}}\,
u^\dagger_{\alpha\uparrow}(1)\,
u^\dagger_{\beta\uparrow}(2)\,
u^\dagger_{\gamma\uparrow}(3)
|0\rangle ,
\\[1em]
|\Delta^{++}, \lambda = 1/2\rangle^{l_z=2}
&= \int d[123]\,
k_{1R} k_{2R}\,\psi^{(5)}_\Delta(1,2,3)
\nonumber \\
&\quad \times
\frac{\epsilon^{\alpha\beta\gamma}}{\sqrt{N_c!}}\,
u^\dagger_{\alpha\downarrow}(1)\,
u^\dagger_{\beta\downarrow}(2)\,
u^\dagger_{\gamma\downarrow}(3)
|0\rangle .
\end{aligned}
\end{equation}

\end{widetext}

For helicity $\lambda=+3/2$, the wave functions in momentum space reads

\begin{equation}
    \psi^{(1)}_\Delta(1,2,3)=-\frac{\sqrt{2} M\, \bar{x}_3\, (M+ m_N x_3)}{\sqrt{x_1 x_2 x_3}}\phi_\Delta(1,2,3)
\end{equation}

\begin{equation}
    \psi^{(2)}_\Delta(1,2,3)=0
\end{equation}

\begin{equation}
\begin{aligned}
    &\psi^{(3)}_\Delta(1,2,3)=-\frac{2\sqrt{2}\, M\, (x_1 + x_2)}{\sqrt{x_1 x_2 x_3}}\phi_\Delta(1,2,3)\\
    &-\frac{2\sqrt{2}\, x_2\, (M+ m_N x_1)}{\sqrt{x_1 x_2 x_3}}\phi_\Delta(2,3,1)\\
    &-\frac{2\sqrt{2}\, x_1 \, (M+ m_N x_2)}{\sqrt{x_1 x_2 x_3}}\phi_\Delta(3,1,2)
\end{aligned}
\end{equation}

\begin{equation}
\begin{aligned}
    \psi^{(4)}_\Delta(1,2,3)=0
\end{aligned}
\end{equation}

\begin{equation}
\begin{aligned}
    \psi^{(5)}_\Delta(1,2,3)=&-\frac{\sqrt{2}\, x_1}{\sqrt{x_1 x_2 x_3}}\phi_\Delta(1,2,3)\\
    &-\frac{\sqrt{2}\, x_1}{\sqrt{x_1 x_2 x_3}}\phi_\Delta(3,1,2)
\end{aligned}
\end{equation}

\begin{equation}
\begin{aligned}
    \psi^{(6)}_\Delta(1,2,3)=0
\end{aligned}
\end{equation}

For helicity $\lambda=+1/2$, the wave functions in momentum space reads

\begin{widetext}
\begin{equation}
\begin{aligned}
    \psi_\Delta^{(1)}(1,2,3)=&-\sqrt{\frac{2}{3}}\frac{M\left(k_{3\perp}^2 + m_{\Delta}\bar{x}_3(M+m_{\Delta}x_3)\right)}{m_{\Delta}\sqrt{x_1x_2x_3}}\phi_\Delta(1,2,3)\\
    &-\sqrt{\frac{2}{3}}\frac{k_{1\perp}\cdot k_{3\perp}\left(M+m_{\Delta}(x_1-x_2)\right) + \left(M+m_{\Delta}x_1\right)\left(k_{3\perp}^2+M^2+m_{\Delta}^2x_2x_3\right)}{m_{\Delta}\sqrt{x_1x_2x_3}}\phi_\Delta(2,3,1)\\
    &-\sqrt{\frac{2}{3}}\frac{M^3+m_{\Delta}\left(k_{3\perp}^2x_1+M^2x_2\right)-k_{1\perp}\cdot k_{3\perp}\left(M-m_{\Delta}(x_1-x_2)\right)}{m_{\Delta}\sqrt{x_1x_2x_3}}\phi_\Delta(3,1,2)\\
    &-\sqrt{\frac{2}{3}}\frac{m_{\Delta}x_1x_3\left(M+m_{\Delta}x_2\right)}{\sqrt{x_1x_2x_3}}\phi_\Delta(3,1,2)
\end{aligned}
\end{equation}

\begin{equation}
\begin{aligned}
    \psi_\Delta^{(2)}(1,2,3)=
    &\sqrt{\frac{2}{3}}\frac{M+m_{\Delta}(x_1-x_2)}{m_{\Delta}\sqrt{x_1x_2x_3}}\phi_\Delta(2,3,1)-\sqrt{\frac{2}{3}}\frac{M-m_{\Delta}(x_1-x_2)}{m_{\Delta}\sqrt{x_1x_2x_3}}\phi_\Delta(3,1,2)
\end{aligned}
\end{equation}

\begin{equation}
\begin{aligned}
\psi^{(3)}_\Delta(1,2,3)=&\sqrt{\frac{8}{3}}\frac{
k_{3\perp}^2 + M m_{\Delta} x_1 + m^2_{\Delta} x_1x_3}{m_{\Delta}\, \sqrt{x_1 x_2 x_3}}\phi_\Delta(1,2,3)\\
&+\sqrt{\frac{8}{3}}\frac{
k_{3\perp}^2+M^2 + m_{\Delta}^2 x_1 x_3}{m_{\Delta}\, \sqrt{x_1 x_2 x_3}}\phi_\Delta(3,1,2)\\
&+\sqrt{\frac{8}{3}}\frac{M\, (M + m_\Delta x_1)}{m_{\Delta}\, \sqrt{x_1 x_2 x_3}}\phi_\Delta(2,3,1)
\end{aligned}
\end{equation}

\begin{equation}
\begin{aligned}
    \psi^{(4)}_\Delta(1,2,3)=\sqrt{\frac{2}{3}}\frac{M^2 +Mm_{\Delta}(x_2-\bar{x}_2)}{m_{\Delta}\, \sqrt{x_1 x_2 x_3}}\phi_\Delta(3,1,2)
\end{aligned}
\end{equation}

\end{widetext}

\begin{equation}
\begin{aligned}
    \psi^{(5)}_\Delta(1,2,3)=&-\sqrt{\frac{2}{3}}\frac{M}{m_{\Delta}\,\sqrt{x_1 x_2 x_3}}\phi_\Delta(3,1,2)\\
    &-\sqrt{\frac{2}{3}}\frac{M}{m_{\Delta}\,\sqrt{x_1 x_2 x_3}}\phi_\Delta(2,3,1)
\end{aligned}
\end{equation}

\section{Diquark correlation functions}
\label{vac_pol}

The diquark correlation function can be calculated by resumming the quark bubble chain in Fig.~\ref{fig:diquark_bubble}. Bubble diagrams in each channel can be evaluated by the following loop integrals.
\begin{widetext}
\begin{itemize}
    \item \textbf{scalar-scalar diquark loop}

\begin{equation}
\begin{aligned}
   \Pi_{SS}=& -2i\int\frac{d^4k}{(2\pi)^4}\mathrm{Tr}\left[S(k)S(k-P)\right]\mathcal{F}(k)\mathcal{F}(P-k)
\end{aligned}
\end{equation}

\item\textbf{vector-vector diquark loop}

\begin{equation}
\begin{aligned}
\Pi^{\mu\nu}_{VV}=& -2i\int\frac{d^4k}{(2\pi)^4}\mathrm{Tr}\left[S(k)\gamma^\mu S(k-P)\gamma^\nu\right]\mathcal{F}(k)\mathcal{F}(P-k)\\
=&-\Pi_{VV}(P^2)\left(g^{\mu\nu}-\frac{P^\mu P^\nu}{P^2}\right)-\delta\Pi_{VV}(P^2)\frac{P^\mu P^\nu}{P^2}
\end{aligned}
\end{equation}

\item\textbf{pseudoscalar-pseudoscalar diquark loop}
\begin{equation}
\begin{aligned}
   \Pi_{PP}=& -2i\int\frac{d^4k}{(2\pi)^4}\mathrm{Tr}\left[S(k)i\gamma^5S(k-P)i\gamma^5\right]\mathcal{F}(k)\mathcal{F}(P-k)
\end{aligned}
\end{equation}

\item\textbf{axial-pseudoscalar diquark loop}
\begin{equation}
\begin{aligned}
   \Pi^{\mu}_{PA}=& -2i\int\frac{d^4k}{(2\pi)^4}\mathrm{Tr}\left[S(k)i\gamma^5S(k-P)\gamma^\mu\gamma^5\right]\mathcal{F}(k)\mathcal{F}(P-k)\\
   =&i\Pi_{PA}(P^2)\frac{P^\mu}{\sqrt{P^2}}
\end{aligned}
\end{equation}
and
\begin{equation}
\begin{aligned}
   \Pi^{\mu}_{AP}=& -2i\int\frac{d^4k}{(2\pi)^4}\mathrm{Tr}\left[S(k)\gamma^\mu\gamma^5S(k-P)i\gamma^5\right]\mathcal{F}(k)\mathcal{F}(P-k)\\
   =&-i\Pi_{PA}(P^2)\frac{P^\mu}{\sqrt{P^2}}
\end{aligned}
\end{equation}

\item\textbf{axial-axial diquark loop}

\begin{equation}
\begin{aligned}
   \Pi^{\mu\nu}_{AA}=& -2i\int\frac{d^4k}{(2\pi)^4}\mathrm{Tr}\left[S(k)\gamma^\mu\gamma^5S(k-P)\gamma^\nu\gamma^5\right]\mathcal{F}(k)\mathcal{F}(P-k)\\
   =&-\Pi^{(t)}_{AA}(P^2)\left(g^{\mu\nu}-\frac{P^\mu P^\nu}{P^2}\right)-\Pi^{(l)}_{AA}(P^2)\frac{P^\mu P^\nu}{P^2}
\end{aligned}
\end{equation}

\end{itemize}
where the quark propagator is approximated as

\begin{equation}
    S(k)=\frac{i(\slashed{k}+M(k))}{k^2-M^2(k)}\simeq\frac{i(\slashed{k}+M)}{k^2-M^2}
\end{equation}
with the momentum dependent constituent mass
$M(k)=m+(M-m)\mathcal{F}(k)$ determined by the gap equation \cite{Liu:2026tuf}.
\begin{equation}
\label{eq:gap}
    M=m+2\left(\frac{G_{I}}{4N_c}+\frac{4G_{IA}}{N_c}\right)\left(\int \frac{d^4k}{(2\pi)^4}
\frac{4M(k)\mathcal{F}(k)}{k^2 + M^2}\right)
\end{equation}
The momentum dependence of $M(k)$ in the denominator is suppressed to avoid spurious unphysical singularity.
With the identity $2k\cdot(P-k)=P^2-k^2-(P-k)^2$, the quark bubble functions can be decomposed into three master integrals $I_1$, $I_2$, and $I_3$ which are defined by

\begin{align}
    &I_1(P^2)=\int\frac{d^4k}{(2\pi)^4}\frac{-i}{(k_+^2-M^2)(k_-^2-M^2)}\mathcal{F}(k_+)\mathcal{F}(k_-)\\
    &I_2(P^2)=\int\frac{d^4k}{(2\pi)^4}\frac{i}{k^2-M^2}\mathcal{F}(k)\mathcal{F}(P-k)\\
    &I_3(P^2)=\int\frac{d^4k}{(2\pi)^4}\left(\frac{4i(k\cdot P)^2/P^2}{(k_+^2-M^2)(k_-^2-M^2)}-\frac{i}{k_+^2-M^2}-\frac{i}{k_-^2-M^2}\right)\mathcal{F}(k_+)\mathcal{F}(k_-)
\end{align}
where $k_\pm=k\pm P/2$. The external-momentum dependence of the nonlocal vertices induces $I_3$ in the spin-1 channel. This certain integral $I_3$ characterizes the contribution
associated with the nontransverse, or Ward-identity violation. With this in mind, the quark bubble functions read
\begin{align}
&\Pi_{SS}
   =4(P^2-4M^2)I_1(P^2)+8I_2(P^2)\\[8pt] 
&\Pi_{PP}
   =4P^2I_1(P^2)+8I_2(P^2)\\[8pt] 
&\Pi_{VV}
   =\frac{8}{3}(P^2+2M^2)I_1(P^2)+\frac{16}{3}I_2(P^2)+\frac43I_3(P^2)\\[8pt]
&\Pi_{PA}=8M\sqrt{P^2}I_1(P^2) \\[8pt]
& \Pi^{(t)}_{AA}=\frac{8}{3}(P^2-4M^2)I_1(P^2)+\frac{16}{3}I_2(P^2)+\frac{4}{3}I_3(P^2)\\[8pt]
&\Pi^{(l)}_{AA} =
-16M^2I_1(P^2)-4I_3(P^2)\\[8pt]
&\delta\Pi_{VV}
   =-4I_3(P^2)
\end{align}

Due to the explicit $P$-dependence of the nonlocal form factors
$\mathcal{F}(k\pm P/2)$, the quark bubble diagram in the vector channel is not exactly transverse, so that
$P_\mu \Pi_{VV}^{\mu\nu}\propto I_3P^\nu\neq 0$. If left untreated, this nontransverse contribution would violate the Ward identity of the full vector--vector current correlator, potentially generating spurious longitudinal poles. In a fully consistent nonlocal formulation, such terms are canceled by additional nonlocal current contributions obtained by properly gauging the nonlocal form factor. However, for typical quark virtualities $k^2\sim \rho^{-2}$ and external momenta satisfying
\[
P^2=m_X^2<4M^2\ll \rho^{-2},
\]
the momentum dependence of the nonlocal vertex may be expanded in $P$, so that to leading order
\[
\mathcal{F}(k\pm P/2)\simeq \mathcal{F}(k)
\]
inside the quark bubble. In this limit, $I_3\rightarrow0$ and the Ward identity recovers.

Eventually, the amputated correlation functions (diquark propagators) around the mass pole in each diquark channel can be written in terms of these bubble functions.

\begin{equation}
\begin{aligned}
    D_p(P^2)&=\frac2{N_c}\frac{ig_{\bar3_P}}{1-g_{\bar3_P}\Pi_{SS}(P^2)}\nonumber\\[7pt]
    D_s(P^2)&=\frac2{N_c}\frac{i\left[g_{\bar3_S}-g_{\bar3_V}+g_{\bar3_S}g_{\bar3_V}\left(\Pi_{PP}-\Pi^{(l)}_{AA}\right)\right]}{(1-g_{\bar3_S}\Pi_{PP}(P^2))(1-g_{\bar3_V}\Pi^{(l)}_{AA})+g_{\bar3_S}g_{\bar3_V}\Pi^2_{PA}}\nonumber\\[7pt]
    D^{\mu\nu}_a(P^2)&=\frac2{N_c}\frac{ig_{\bar3_A}}{1-g_{\bar3_A}\Pi_{VV}}\left(-g^{\mu\nu}+\frac{P^\mu P^\nu}{P^2}\right)\nonumber\\[7pt]
%
%
    D^{\mu\nu}_v(P^2)&=\frac2{N_c}\frac{ig_{\bar3_V}}{1-g_{\bar3_V}\Pi^{(t)}_{AA}}\left(-g^{\mu\nu}+\frac{P^\mu P^\nu}{P^2}\right) \nonumber\\[7pt]
%
\end{aligned}
\end{equation}

\section{Solving the equation for baryons in static approximation}
\label{Baryon}

The color and isospin structure as a quark-diquark boundstate has been projected properly in Sec.~\ref{sec:baryon}. For the spin structure, in this Appendix we follow the calculation in \cite{Mineo:2002bg} and diagonalize the Faddeev kernel in \eqref{baryon_eq} for the nucleon and delta baryons using the static approximation for simplicity. 

\begin{figure*}
    \centering
    \includegraphics[width=1\linewidth]{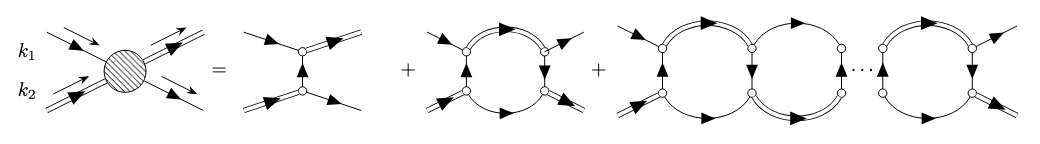}
    \caption{The Feynman diagrams for the quark-diquark bubble resummation in the baryon channels.}
    \label{fig:nucleon_bubble}
\end{figure*}

As shown in Fig.~\ref{fig:nucleon_bubble}, the nucleon channel can be viewed as a resummation of quark-diquark bubble in static approximation. Each quark-diquark bubble diagram are defined as

\begin{equation}
\label{Pi_N_Delta}
\Pi_N(P)
=
\begin{pmatrix}
\Pi_s(P) & 0 \\
0 & \Pi_a^{\mu\nu}(P)
\end{pmatrix},
\qquad
\Pi_\Delta(P)=\Pi_a^{\mu\nu}(P).
\end{equation}
where the loop integrals are defined by
\begin{equation}
\label{eq:Pi_diq}
\Pi_s(P)
=
i\int\frac{d^4k}{(2\pi)^4}
S(k)D_s(P-k)\mathcal{F}(k),
\qquad
\Pi_a^{\mu\nu}(P)
=
i\int\frac{d^4k}{(2\pi)^4}
S(k)D_a^{\mu\nu}(P-k)\mathcal{F}(k).
\end{equation}
As a simple proof of principle, we further adopt the pole approximation for the diquark propagators, with the diquark masses and effective quark–diquark couplings fixed by the two-body calculations presented in the preceding section.

\begin{equation}
D_s(k)\rightarrow \frac{-i\lambda_{qq_S}^2}{P^2-m_{qq_S}^2},
\qquad
D_a^{\mu\nu}(k)\rightarrow \frac{i\lambda_{qq_A}^2}{P^2-m_{qq_A}^2}\left(g^{\mu\nu}-\frac{k^\mu k^\nu}{m_{qq_A}^2}\right).
\end{equation}

The baryon vertex function given in \eqref{vertex_N} and \eqref{vertex_Delta} will be contructed in details. The diagonalization will be done in the rest frame of the baryon, and the solution will be boosted to a general frame.

\subsection{Nucleon}

Together with the Dirac index for the quark, the kernel $G_N$ in \eqref{G_N} is a $20 \times 20$ matrix, which can be reduced to $10\times10$ by projection to positive energy states. In rest frame, the projection is equivalent to the positive parity states. We directly diagonalize this kernel in the rest frame of the baryon by unitary transformations. It separates into a $6\times6$ block corresponding
to spin $1/2$ states (nucleon), and a $4 \times 4$ block for the spin $3/2$ states. 

In order to remove the Dirac matrices from the kernel $G_{N}\Pi_N$, we first apply the unitary transformation:

\begin{equation}
    G_N\rightarrow UG_N U=-\frac3M\begin{pmatrix}
        1 & -\sqrt{3} & -\sqrt{3} & -\sqrt{3} & -\sqrt{3} \\
        -\sqrt{3} & -1 &  1 &  1 &  1 \\
        -\sqrt{3} &  1 & -1 &  1 &  1 \\
        -\sqrt{3} &  1 &  1 & -1 &  1\\
        -\sqrt{3} &  1 &  1 &  1 & -1 \\
    \end{pmatrix}
\end{equation}
For simplicity, we have set the effective quark-diquark coupling to 1 and restore them at the end. The quark-diquark bubble function $\Pi_N$ in the rest frame after the same unitary transformation reads

\begin{equation}
    \Pi_N\rightarrow U^\dagger\Pi_N U^\dagger=\begin{pmatrix}
        \Pi^+_0 & 0 & 0 & 0 & 0 \\
        0 & \Pi^+_1 & \Pi^+_2 & \Pi^+_2 & \Pi^+_2\\
        0 & \Pi^+_2 & \Pi^+_3 & 0 & 0 \\
        0 & \Pi^+_2 & 0 & \Pi^+_3 & 0\\
        0 & \Pi^+_2 & 0 & 0 & \Pi^+_3 \\
    \end{pmatrix}\frac{1-\gamma^0}2+\begin{pmatrix}
        \Pi^-_0 & 0 & 0 & 0 & 0 \\
        0 & \Pi^-_1 & \Pi^-_2 & \Pi^-_2 & \Pi^-_2\\
        0 & \Pi^-_2 & \Pi^-_3 & 0 & 0 \\
        0 & \Pi^-_2 & 0 & \Pi^-_3 & 0\\
        0 & \Pi^-_2 & 0 & 0 & \Pi^-_3 \\
    \end{pmatrix}\frac{1+\gamma^0}2
\end{equation}
where unitary matrix is defined as 
$U=\mathrm{diag}(i\gamma^5,-\gamma^0,\gamma^1,\gamma^2,\gamma^3)$. 
The scalar part of the functions reads 
\begin{align}
    &\Pi^\pm_0(P^2)=-i\lambda^2_{qq_S}\int\frac{d^4k}{(2\pi)^4}\frac{\mp k^0-M}{k^2-M^2}\frac{1}{(P-k)^2-m^2_{qq_S}}\mathcal{F}(k)\\
    &\Pi^\pm_1(P^2)=-i\lambda^2_{qq_A}\int\frac{d^4k}{(2\pi)^4}\frac{\pm k^0-M}{k^2-M^2}\frac{1}{(P-k)^2-m^2_{qq_A}}\left(1-\frac{(P^0-k^0)^2}{m^2_{qq_A}}\right)\mathcal{F}(k)\\
    &\Pi^\pm_2(P^2)=-i\lambda^2_{qq_A}\int\frac{d^4k}{(2\pi)^4}\frac{\pm k^i}{k^2-M^2}\frac{1}{(P-k)^2-m^2_{qq_A}}\frac{(P-k)^0k^i}{m^2_{qq_A}}\mathcal{F}(k)\\
    &\Pi^\pm_3(P^2)=-i\lambda^2_{qq_A}\int\frac{d^4k}{(2\pi)^4}\frac{\mp k^0-M}{k^2-M^2}\frac{1}{(P-k)^2-m^2_{qq_A}}\left(1+\frac{(k^i)^2}{m^2_{qq_A}}\right)\mathcal{F}(k)
\end{align}

For the simplicity, we can solve the Faddeev equation in \eqref{baryon_eq} in rest frame and boost the solution to the general frame. The nucleon vertex function in the rest frame is defined as
\begin{equation}
\label{vertex_N2}
    \Gamma_N(S)=\begin{pmatrix}
        -i\alpha_1u_N(S) \\[4pt]
        \alpha_2\gamma^5u_N(S) \\[4pt]
        \alpha_3\sum_{\lambda,S'}\left\langle1\lambda;\frac12S'\big|\frac12S\right\rangle\varepsilon^1_\lambda u_N(S') \\[4pt]
        \alpha_3\sum_{\lambda,S'}\left\langle1\lambda;\frac12S'\big|\frac12S\right\rangle\varepsilon^2_\lambda u_N(S') \\[4pt]
        \alpha_3\sum_{\lambda,S'}\left\langle1\lambda;\frac12S'\big|\frac12S\right\rangle\varepsilon^3_\lambda u_N(S')
        
    \end{pmatrix}\rightarrow U\Gamma_N(S)=\begin{pmatrix}
        \alpha_1\gamma^5u_N(S) \\[4pt]
        \alpha_2\gamma^5u_N(S) \\[4pt]
        \alpha_3\sum_{\lambda,S'}\left\langle1\lambda;\frac12S'\big|\frac12S\right\rangle\gamma^1\varepsilon^1_\lambda u_N(S') \\[4pt]
        \alpha_3\sum_{\lambda,S'}\left\langle1\lambda;\frac12S'\big|\frac12S\right\rangle\gamma^2\varepsilon^2_\lambda u_N(S') \\[4pt]
        \alpha_3\sum_{\lambda,S'}\left\langle1\lambda;\frac12S'\big|\frac12S\right\rangle\gamma^3\varepsilon^3_\lambda u_N(S') 
    \end{pmatrix}
\end{equation}
where the nucleon spinor in rest frame is defined as
\begin{equation}
    u_N(S)=\sqrt{m_N}\left(\begin{array}{c}
        \chi_S  \\
        \chi_S
    \end{array}\right)
\end{equation}
The last three terms are known as Clebsch Gordan projection $1\otimes\frac12\rightarrow\frac12$ for the vector–spinor combination. With the basis convention $\varepsilon^\mu_{\pm 1} = \left(0, \mp\frac{1}{\sqrt{2}}, -\frac{i}{\sqrt{2}}, 0 \right)$ and $\varepsilon^\mu_{0} = (0,0,0,1)$, they can be expressed in Lorentz covariant form as

\begin{equation}
    \sum_{\lambda, S'} 
\left\langle 1\lambda;\, \frac{1}{2} S' \,\middle|\, \frac{1}{2}, S \right\rangle 
\, \varepsilon^\mu_{\lambda}\, u_N(P,S')
= -\frac{1}{\sqrt{3}}\, \left(\gamma^\mu+\frac{P^\mu}{m_N} \right)\gamma^5\, u_N\!\left(P,S\right)
\end{equation}
with Wigner-Eckart theorem defined as

\begin{equation}
    \left\langle 1\lambda;\frac{1}{2}s' \middle| \frac{1}{2}s \right\rangle
=
\frac{(-1)^{\lambda+1}}{\sqrt{3}}
\left(\sigma^{[1]}_{-\lambda}\right)_{s,s'}
\end{equation}
and the spherical basis for Pauli matrices $\sigma^{[1]}_{-1}=-\sigma^-$, $\sigma^{[1]}_{0}=\sigma_3$, and $\sigma^{[1]}_{1}=\sigma^+$.

With this in mind, the vertex function in this basis is reducible. We can reduce the spinor by projecting into the irreducible basis spin-$1/2$ and spin-$3/2$. In this basis, it is clear to see the nucleon is a spin-$1/2$ couple to the scalar diquark and axial vector diquark.

\begin{equation}
    \Gamma_N(S)\rightarrow R U\Gamma_N(S)=\begin{pmatrix}
        \alpha_1 \\
        \alpha_2 \\
        \alpha_3 \\
        0 \\
        0 \\
    \end{pmatrix}\gamma^5u_N(S)
\end{equation}
with the normalization condition  $\alpha_1^2+\alpha_2^2+\alpha_3^2=1$.
The corresponding unitary matrix is defined as
\begin{equation}
    R=\mathrm{diag}(\left(\begin{array}{cc}
        1 & 0 \\
        0 & 1
    \end{array}\right),\left(\begin{array}{ccc}
        1/\sqrt{3} & 1/\sqrt{3} & 1/\sqrt{3} \\
        0 & -1/\sqrt{2} & 1/\sqrt{2} \\
        -2/\sqrt{6} & 1/\sqrt{6} & 1/\sqrt{6}
    \end{array}\right))
\end{equation}

The matrix transform the Faddeev kernel into block form 

\begin{equation}
\label{N_kernel}
    G_N\rightarrow RUG_NU R^\dagger=-\frac3M\left(\begin{array}{ccc|cc}
      1 & -\sqrt{3} & -3 & 0 & 0 \\
        -\sqrt{3} & -1 & \sqrt{3} & 0 & 0 \\
        -3 & \sqrt{3} & 1 & 0 & 0 \\
        \hline 
        0 & 0 & 0 & -2 & 0\\
        0 & 0 & 0 & 0 & -2 \\
    \end{array}\right)
\end{equation}
and
\begin{equation}
    \Pi_N\rightarrow RU^\dagger\Pi_NU^\dagger R^\dagger\frac{1-\gamma^0}2=\left(\begin{array}{ccc|cc}
      \Pi^+_0 & 0 & 0 & 0 & 0 \\
        0 & \Pi^+_1 & \sqrt{3}\Pi^+_2 & 0 & 0 \\
        0 & \sqrt{3}\Pi^+_2 & \Pi^+_3 & 0 & 0 \\
    \hline
        0 & 0 & 0 & \Pi^+_3 & 0\\
        0 & 0 & 0 & 0 & \Pi^+_3 \\
    \end{array}\right)\frac{1-\gamma^0}2
\end{equation}

It is natural to understand that the coefficient in rest frame stands for three components in the nucleon state: i) scalar diquark state couple to a quark ii) time component of the quark-axial-diquark state couple to a quark iii) spatial component of the axial-diquark state couple to a quark. The homogeneous equation for the 3 coefficients $\alpha_1, \alpha_2, \alpha_3$ leads to the eigenvalue equation for the nucleon mass. 

\begin{equation}
    -\frac3M\left(\begin{array}{ccc}
      1 & -\sqrt{3} & -3  \\
        -\sqrt{3} & -1 & \sqrt{3}  \\
        -3 & \sqrt{3} & 1 
    \end{array}\right)\left(\begin{array}{ccc}
      \Pi^+_0 & 0 & 0 \\
        0 & \Pi^+_1 & \sqrt{3}\Pi^+_2 \\
        0 & \sqrt{3}\Pi^+_2 & \Pi^+_3 \\
    \end{array}\right)\begin{pmatrix}
        \alpha_1 \\
        \alpha_2 \\
        \alpha_3 \\
    \end{pmatrix}=\begin{pmatrix}
        \alpha_1 \\
        \alpha_2 \\
        \alpha_3 \\
    \end{pmatrix}
\end{equation}

The largest eigenvalue determines the nucleon mass. If we neglect the mixing from axial vector diquark, the bound state equation reduces to a simple form

\begin{equation}
    -\frac3M \Pi^+_0(P^2=m^2_N)=1
\end{equation}

The covariant form of the nucleon vertex function can be obtained by boosting. The result reads

\begin{equation}
\begin{aligned}
    \Gamma_N(P,S)=&\begin{pmatrix}
        -i\alpha_1u_N(P,S) \\[4pt]
        \alpha_2\frac{P^\mu}{m_N}\gamma^5u_N(P,S) +\alpha_3\sum_{\lambda,s'}\left\langle1\lambda;\frac12S'\big|\frac12S\right\rangle\varepsilon^\mu_\lambda(P) u_N(P,S')
    \end{pmatrix}\\
    =&\begin{pmatrix}
        -i\alpha_1u_N(S) \\[4pt]
        \left(\alpha_2-\frac1{\sqrt{3}}\alpha_3\right)\frac{P^\mu}{m_N}\gamma^5u_N(P,S) -\frac{1}{\sqrt{3}}\alpha_3\gamma^\mu\gamma^5 u_N(P,S)
    \end{pmatrix}
\end{aligned}
\end{equation}
with the normalization condition
\begin{equation}
    \frac{1}{2P^0}\Gamma_N^\dagger(P,S)\Gamma_N(P,S)=\alpha_1^2+\alpha_2^2+\alpha_3^2=1
\end{equation}

\subsection{Delta baryon}
For the $\Delta$ baryon, the $16 \times 16$ matrix of the Faddeev kernel $G_\Delta$ (see \eqref{G_Delta}) can be reduced to $8 \times 8$ matrix with positive parity projection. The direct diagonalization in the rest frame of the baryon separates the kernel into a $4\times4$ block corresponding
to spin $1/2$ states (nucleon), and a $4 \times 4$ block for the spin $3/2$ states. In order to remove the Dirac matrices, we first apply the unitary transformation:
\begin{equation}
    G_\Delta\rightarrow UG_\Delta U=\frac6M
    \begin{pmatrix}
         -1 & 1 & 1 & 1 \\
         1 & -1 & 1 & 1 \\
         1 & 1 & -1 & 1\\
         1 & 1 & 1 & -1 \\
    \end{pmatrix}
\end{equation}

\begin{equation}
    \Pi_\Delta\rightarrow U^\dagger\Pi_\Delta U^\dagger=
    \begin{pmatrix}
         \Pi_1^{+} & \Pi_2^{+} & \Pi_2^{+} & \Pi_2^{+} \\
         \Pi_2^{+} & \Pi_3^{+} & 0 & 0 \\
         \Pi_2^{+} & 0 & \Pi_3^{+} & 0\\
         \Pi_2^{+} & 0 & 0 & \Pi_3^{+} \\
    \end{pmatrix}\frac{1-\gamma^0}2+
    \begin{pmatrix}
         \Pi_1^{-} & \Pi_2^{-} & \Pi_2^{-} & \Pi_2^{-} \\
         \Pi_2^{-} & \Pi_3^{-} & 0 & 0 \\
         \Pi_2^{-} & 0 & \Pi_3^{-} & 0\\
         \Pi_2^{-} & 0 & 0 & \Pi_3^{-} \\
    \end{pmatrix}\frac{1+\gamma^0}2
\end{equation}
where $U=-i\,\mathrm{diag}(-\gamma^0,\gamma^1,\gamma^2,\gamma^3)$

In the rest frame, the vertex function for spin $\frac32$ is given by an expression. 
\begin{equation}
\label{vertex_Delta2}
    \Gamma_\Delta(S)=\begin{pmatrix}
        0 \\[4pt]
        \sum_{\lambda,s'}\left\langle1\lambda;\frac12s'\big|\frac32S\right\rangle\varepsilon^1_\lambda u_{s'}\\[4pt]
        \sum_{\lambda,s'}\left\langle1\lambda;\frac12s'\big|\frac32S\right\rangle\varepsilon^2_\lambda u_{s'}\\[4pt]
        \sum_{\lambda,s'}\left\langle1\lambda;\frac12s'\big|\frac32S\right\rangle\varepsilon^3_\lambda u_{s'}
        
    \end{pmatrix}\rightarrow iU\Gamma_\Delta(S)=\begin{pmatrix}
        0 \\[4pt]
        
        \sum_{\lambda,s'}\left\langle1\lambda;\frac12s'\big|\frac32S\right\rangle\gamma^1\varepsilon^1_\lambda u_{s'} \\[4pt]
        \sum_{\lambda,s'}\left\langle1\lambda;\frac12s'\big|\frac32S\right\rangle\gamma^2\varepsilon^2_\lambda u_{s'}\\[4pt]
        \sum_{\lambda,s'}\left\langle1\lambda;\frac12s'\big|\frac32S\right\rangle\gamma^3\varepsilon^3_\lambda u_{s'}
    \end{pmatrix}
\end{equation}
where the Dirac spinor in the rest frame is defined as
\begin{equation}
    u_{s}=\sqrt{m_\Delta}\left(\begin{array}{c}
        \chi_s  \\
        \chi_s
    \end{array}\right)
\end{equation}
with the polarization basis convention 
\begin{align}
    \varepsilon^\mu_\pm=\mp(0,1,\pm i,0)/\sqrt{2}  && \varepsilon^\mu_0=(0,0,0,1)
\end{align}

The last three terms are known as Clebsch Gordan projection $1\otimes\frac12\rightarrow\frac32$ for the vector–spinor combination. The expression is similar to the last three components in \eqref{vertex_N2}, but with the diquark and quark spins coupled to total spin $3/2$ instead of $1/2$, giving a standard Rarita-Schwinger spinor form. By boosting, the covariant expression for Rarita–Schwinger spinor form reads

\begin{equation}
\begin{aligned}
\label{eq:RS_spinor}
&u^\mu_\Delta(P,S)=\sum_{\lambda,s'}\left\langle1\lambda;\frac12s'\big|\frac32S\right\rangle\varepsilon^\mu_\lambda(P) u_{s'}(P)
\end{aligned}
\end{equation}
where the spinor satisfies the projection

\begin{equation}
    u^\mu_\Delta(P,S)=
    \left(g_{\mu\nu}
- \frac{1}{3}\,\gamma_\mu \gamma_\nu
- \frac{1}{3 m_\Delta}\left(\gamma_\mu P_\nu - \gamma_\nu P_\mu\right)
- \frac{2}{3 m_\Delta^2} P^\mu P^\nu\right)u^\mu_\Delta(P,S)
\end{equation}

The expression of the vertex function in  \eqref{vertex_Delta2} is reducible. By projecting into the irreducible basis corresponding to spin-$1/2$ and spin-$3/2$, it is clear to see that $\Delta$ is a spin-$3/2$ state.
\begin{equation}
    \Gamma_\Delta(S=+3/2)\rightarrow R U\Gamma_\Delta(S=+3/2)=\frac12\begin{pmatrix}
        0 \\
        0 \\
        1 \\
        -\sqrt{3} \\
    \end{pmatrix}\gamma^5u_\downarrow
\end{equation}
and
\begin{equation}
    \Gamma_\Delta(S=+1/2)\rightarrow R U\Gamma_\Delta(S=+1/2)=\frac1{2}\begin{pmatrix}
        0 \\
        0 \\
        -\sqrt{3} \\
        -1 \\
    \end{pmatrix}\gamma^5u_\uparrow
\end{equation}

The corresponding unitary matrix is defined as
\begin{equation}
    R=\mathrm{diag}(1,\left(\begin{array}{ccc}
        1/\sqrt{3} & 1/\sqrt{3} & 1/\sqrt{3} \\
        0 & 1/\sqrt{2} & -1/\sqrt{2} \\
        -2/\sqrt{6} & 1/\sqrt{6} & 1/\sqrt{6}
    \end{array}\right))
\end{equation}

The unitary transform convert the Faddeev kernel into block form:
\begin{equation}
    G_\Delta\rightarrow RUG_\Delta U R^\dagger=\frac6M\left(\begin{array}{cc|cc}
         -1 & \sqrt{3} & 0 & 0 \\
         \sqrt{3} & 1 & 0 & 0 \\
        \hline 
         0 & 0 & -2 & 0\\
         0 & 0 & 0 & -2 \\
    \end{array}\right)
\end{equation}
and
\begin{equation}
    \Pi_\Delta\rightarrow RU\Pi_\Delta U R^\dagger\frac{1-\gamma^0}2=\left(\begin{array}{cc|cc}
         \Pi^+_1 & \sqrt{3}\Pi^+_2 & 0 & 0 \\
         \sqrt{3}\Pi^+_2 & \Pi^+_3 & 0 & 0 \\
    \hline
         0 & 0 & \Pi^+_3 & 0\\
         0 & 0 & 0 & \Pi^+_3 \\
    \end{array}\right)\frac{1-\gamma^0}2
\end{equation}
The diagonalization of the isospin-$3/2$ kernel proceeds in the same way by applying the unitary transformations as above. The positive parity part of the kernel then separates into a $2 \times 2$ block for $J=1/2$ (corresponding to the coupling of the time component and the space components of the axial vector diquark with the quark, respectively) and a diagonal $2 \times 2$ block
for $J=3/2$. The latter one has the same form as the lower $2 \times 2$ block of the kernel \eqref{N_kernel}, but with the factor $-2$ replaced by $4$ (from the isospin structure)

Now the delta baryon mass is determined by 
\begin{equation}
    -\frac{12}M\Pi^+_3(P^2=m^2_\Delta)=1
\end{equation}
\end{widetext}

\bibliography{ILM}

\end{document}